\documentclass[prb,twocolumn,superscriptaddress,showpacs,aps,amssymb,floatfix,reprint]{revtex4-1}

\usepackage{amsmath,amssymb,bm}
\usepackage{graphicx}
\usepackage{epstopdf}
\usepackage{latexsym}
\usepackage{subfigure}
\usepackage{color}
\usepackage{multirow}
\usepackage{hhline}
\usepackage{comment}
\usepackage{tabularx}

\begin{document}
\title{Phase diagram of weakly coupled Heisenberg spin chains subject to a uniform Dzyaloshinskii-Moriya interaction}
\author{Wen Jin and Oleg A. Starykh}
\affiliation{Department of Physics and Astronomy, University of Utah, Salt Lake City, Utah 84112, USA}

\begin{abstract} \label{abstract}
Motivated by recent experiments on spin chain materials K$_2$CuSO$_4$Cl$_2$ and K$_2$CuSO$_4$Br$_2$, we theoretically investigate 
the problem of weakly coupled spin chains (chain exchange $J$, interchain $J'$) subject to a {\em staggered between chains}, but {\em uniform}
within a given chain,
Dzyaloshinskii-Moriya (DM) interaction of magnitude $D$. In the experimentally relevant limit $J' \ll D \ll J$ of strong DM interaction
the spins on the neighboring chains are forced to rotate in opposite directions, effectively resulting in a cancelation of 
the interchain interaction between components of spins in the plane normal to the vector ${\bm D}$. This has the effect of promoting 
two-dimensional collinear spin density wave (SDW) state, which preserves U(1) symmetry of rotations about the $\bm D$-axis. 
We also investigate response of this interesting system to an external magnetic field ${\bm h}$ and obtain the $h-D$ phase
diagrams for the two important configurations, ${\bm h} \parallel {\bm D}$ and ${\bm h} \perp {\bm D}$. 

\end{abstract}

\maketitle
\section{Introduction}\label{sec:introduction}

Many interesting quantum magnets are characterized by significant spatial anisotropy of the exchange interaction pattern and
often can be understood as being built from one-dimensional spin chains. Several recent examples of these include 
triangular antiferromagnets Cs$_2$CuCl$_4$ \cite{coldea2003} and Cs$_2$CuBr$_4$ \cite{Ono2005,Ono2005b,takano2009}, actively investigated for their fractionalized spinon continuum
and pronounced 1/3 magnetization plateau, correspondingly, and high-field candidate spin nematic materials such as LiCuVO$_4$ \cite{mourigal2012,svistov2014}
and PbCuSO$_4$(OH)$_2$ \cite{linarite2016,povarov2016}. 

Quasi-one-dimensional nature of this class of materials is responsible for the hierarchy of temperature/energy scales when
at high temperature, relative to the weak inter-chain exchange $J'$, the material exhibits mainly one-dimensional physics 
with little correlations between spins from different chains. Upon further cooling the inter-chain interactions become 
important and determine the ultimate ground state type of order that is realized below the ordering temperature $T_c \sim J'$\cite{schulz1996}.
If the interchain interaction is geometrically frustrated, as for example happens in triangular \cite{olegandbalents} and kagome \cite{Schnyder2008} lattices, the ordering temperature
may be further suppressed below the intuitive mean-field $T_c \sim J'$ estimate. 

In the present work we describe novel mechanism of frustrating inter-chain spin exchange. We show that spin chains with strong 
uniform Dzyaloshinskii-Moriya (DM) anisotropic exchange interaction, orientation of the DM vector of which is however staggered
between the chains, are too characterized by strongly reduced ordering temperature.

Our work is strongly motivated by two new interesting materials - K$_2$CuSO$_4$Cl$_2$ and K$_2$CuSO$_4$Br$_2$~\cite{Halg2014,smirnov,Halg'sphdthesis} - which are 
described by Hamiltonian \eqref{eq:H_0} representing weakly coupled spin chains (chain exchange $J$, inter-chain exchange $J'$, and $J'\ll J$) perturbed by the 
uniform within the chain, but staggered between chains, Dzyaloshinskii-Moriya (DM) anisotropic exchange interaction of magnitude $D$, 
as shown in Fig.~\ref{fig:lattice}. (Similar DM geometry is also realized in a spin-ladder material (C$_7$H$_{10}$N)$_2$CuBr$_4$.\cite{glazkov2015})
Despite close structural similarity, the two materials are characterized by different $h-T$ phase diagrams in the 
situation when magnetic field $\bm h$ is applied along the DM axis $\bm D$ of the material. Our objective here is to provide 
theoretical explanation of those phase diagrams, and find reasons for their differences. 
We also extend analysis to another special field configuration, when magnetic field is perpendicular to the DM vector.

Individual spin chains with uniform \cite{Gangadharaiah2008,Halg2014,Garate2010,smirnov} and staggered \cite{dender} DM interactions 
respond differently to the magnetic field. In the latter case it leads to the opening of significant spin gap \cite{oshikawa1999} while in the former the (much smaller) 
gap opens up only in the ${\bm h} \perp {\bm D}$ geometry \cite{Gangadharaiah2008,Garate2010}.
We show below that this difference persists in the presence of the weak inter-chain interaction and is responsible for a very different set
of the ordered states for the uniform DM problem in comparison with the staggered DM one \cite{sato2004}.

The plan of the paper is as follows.
In Sec.~\ref{sec:hamiltonian}, we introduce the pertinent spin chain model. Focusing on the low-energy physics, we attack the problem with the help of bosonization in Sec.~\ref{subsec:bosonization}. We examine the phase diagram of the model for the two special magnetic field orientations, 
${\bm h} \parallel {\bm D}$, Sec.~\ref{sec:parallel} and Sec.~\ref{sec:CMF}, and ${\bm h} \perp {\bm D}$, Sec.~\ref{sec:perpendicular}.

Throughout the paper we find competition between transverse cone-like orders and longitudinal spin density wave (SDW) ones. Here by the cone order we mean the order
that develops in the plane perpendicular to the external magnetic field. Combined with finite magnetization, this order can be visualized as the one where spins lie on the surface
of the cone whose axis is oriented along the magnetic field. The longitudinal SDW order is quite different - spins order in the direction of the  magnetic field.
Magnitude of the local magnetic moment is position dependent, which makes the resultant modulated pattern quite similar to a charge density wave order often found in itinerant
electron systems. 

In Sec.~\ref{sec:parallel}, by means of the renormalization group (RG) analysis, we find  a single commensurate cone state 
(magnetic order develops in the plane transverse to ${\bm h}$) for weak DM interaction ($D\ll J'$). In the opposite, and novel, case of strong DM interaction 
($D\gg J'$ but still $D \ll J$)  the inter-chain coupling is strongly frustrated and the cone state is destroyed. 
Instead, a collinear longitudinal spin density wave emerges as the ground state of the system of weakly coupled spin chains. 

We next show how quantum fluctuations generate a transverse spin exchange between {\em next-nearest} (NN) chains, which competes with the SDW order. The resultant cone-like order, denoted as coneNN, is found to develop above a critical magnetic field $h_c\sim J'$. The coneNN order is a juxtaposition of the two separate cone orders, formed by spins of {\em even} and {\em odd} chains correspondingly.
Owing to the opposite direction of DM axis on even/odd chains, spins making up even/odd cones appear to rotate in {\em opposite} directions. These RG-based findings are supported by the 
chain mean-field (CMF) calculations in Sec. \ref{sec:CMF}, where we compute and compare ordering temperature of various two-dimensional instabilities.

Turning to the ${\bm h} \perp {\bm D}$ arrangement in Sec.~\ref{sec:perpendicular}, we carry out chiral rotation of spin currents which reduces the problem
to that in the effective magnetic field the magnitude of which is given by the $\sqrt{h^2 + D^2}$. Subsequent RG analysis leads to detailed $h-D$ phase diagram
which harbors three different orders: two commensurate SDWs along and perpendicular to DM vector, respectively, 
and a \textit{distorted-cone} state (elliptic spiral structure). We find that in the experimentally relevant limit $D\ll J$, the phase transition between two different SDWs 
happens at $h_c\sim 0.23\pi J$, which is independent of $D$ and is of a spin-flop kind. The distorted-cone phase requires unrealistically large DM interaction $D\sim J$
and is separated from the SDW by a boundary at $h/D\simeq1.5$, which matches well with the classical prediction \cite{Garate2010}. 

We conclude the manuscript with a brief summary and a discussion of the relevance of our results to ongoing experimental studies of  K$_2$CuSO$_4$Br$_2$
and related materials. Numerous technical details of our analysis are presented in Appendices. 
\section{Hamiltonian}
\label{sec:hamiltonian}

\begin{center}
\begin{figure}[!tbp]
\includegraphics[width=0.8\columnwidth]{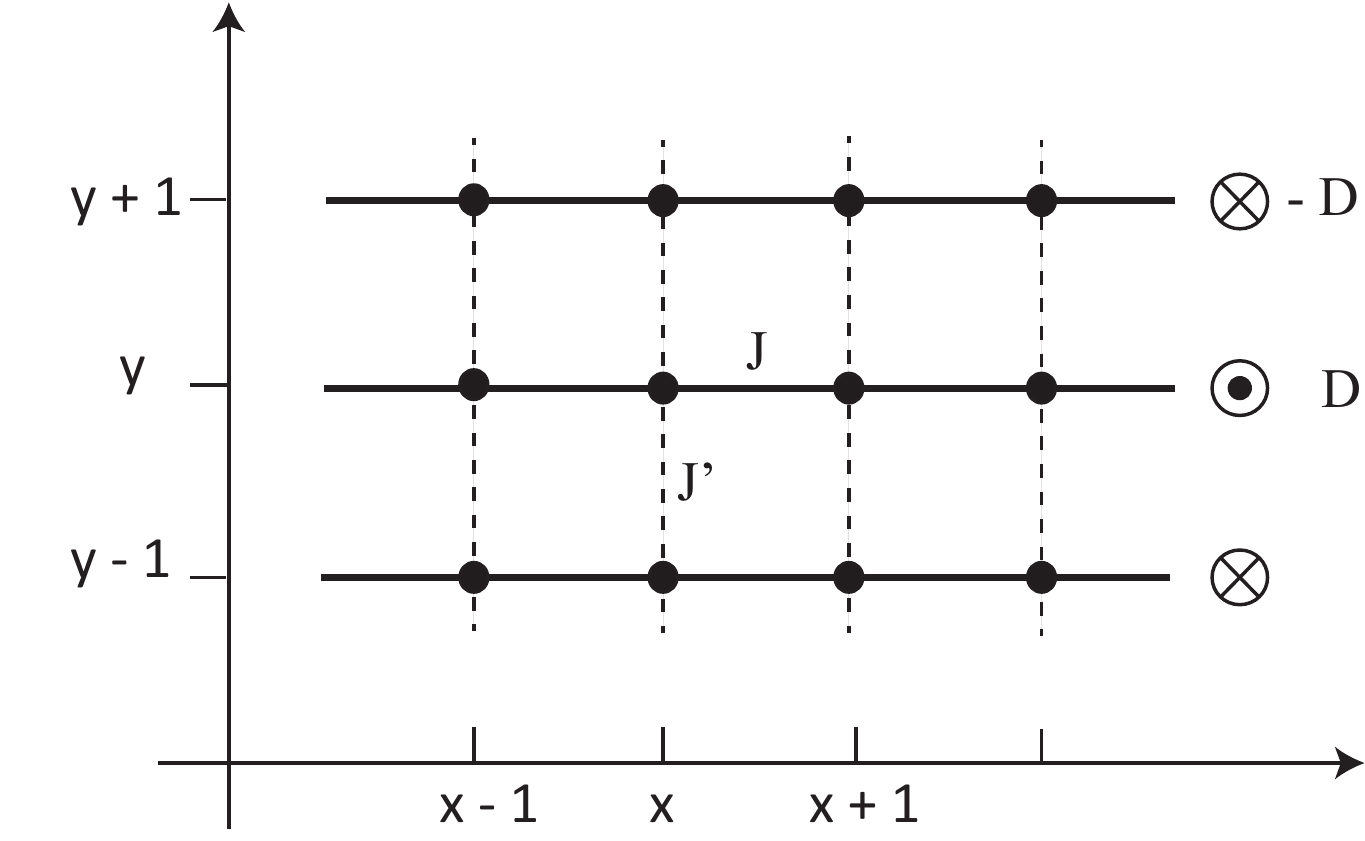}
\caption{Geometry of the problem. Intra-chain bonds $J$ (thick lines along $\hat{x}$), inter-chain bonds $J'$ (dashed lines along $\hat{y}$), and $J'\ll J$. 
DM vectors on neighboring chain have opposite direction, pointing either into or out of the page.} 
\label{fig:lattice}
\end{figure}
\end{center}

We consider weakly coupled antiferromagnetic Heisenberg spin-$1/2$ chains subject to a uniform Dzyaloshinskii-Moriya (DM) interaction and an external magnetic field. 
The system is described by the following Hamiltonian,
\begin{equation}
\begin{split}
{\cal H}&=\sum_{x,y}[J{\bm S}_{x,y}\cdot{\bm S} _{x+1,y} + J'{\bm S}_{x,y}\cdot{\bm S} _{x,y+1}]\\
&\quad +{\bm D}\cdot\sum_{x,y}(-1)^{y}{\bm S}_{x,y}\times {\bm S}_{x+1,y}-{\bm h}\cdot \sum_{x,y}{\bm S}_{x,y},\\
\end{split}
\label{eq:H_0}
\end{equation}
where ${\bm S}_{x,y}$ is the spin-$1/2$ operator at position $x$ of $y$-th chain. $J$ and $J'$ denote isotropic intra- and inter-chain antiferromagnetic exchange couplings 
as shown in Fig.~\ref{fig:lattice}, 
and we account for interactions between nearest neighbors only. The inter-chain exchange is weak, of the order of $J'\sim 10^{-2} J$. 
DM interaction \cite{Dzyalo,moriya} is parameterized by the DM vector ${\bm D}=D\hat{z}$, direction of which is staggered between adjacent chains -- note the factor $(-1)^y$ in \eqref{eq:H_0}.
Importantly, within a given $y$-th chain vector ${\bm D}$ is uniform. $\bm h$ is an external magnetic field.
\subsection{Lattice rotation of spins}
\label{subsec:lattice-rot}
DM interaction in Eq.~\eqref{eq:H_0} can be gauged away by a position-dependent rotation of spins about $\hat z$ axis \cite{Perk1976,shekhtman1992,oshikawa1999,bocquet2001},
\begin{equation}
S_{x,y}^+\to \tilde S_{x,y}^{+}e^{i\alpha_y x},\quad S_{x,y}^{z}\to \tilde S_{x,y}^{z},
\end{equation}
where the rotation angle $\alpha_y=\arctan[(-1)^y D/J]$ for the $y$-th chain changes sign between even and odd chains.
In our work, we consider $D\ll J$, which is the limit relevant for real materials\cite{Halg2014, smirnov, Dmitrienko2014}, therefore the rotation angle $\alpha_y$ is small. 
After the rotation Hamiltonian \eqref{eq:H_0} reads
\begin{equation}
\begin{split}
\tilde{\cal H}&=\sum_{x,y}[\frac{\tilde J}{2} ( \tilde{S}^{+}_{x,y} \tilde{S}^{-}_{x+1,y}+{\rm h.c.})+J \tilde {S}^{z}_{x,y}\tilde{S}_{x+1,y}^{z}]\\
&\quad +\sum_{x,y}[\frac{J'}{2}(\tilde S_{x,y}^{+}\tilde S_{x,y+1}^{-}e^{2i\alpha_y x}+{\rm h.c.})+J' \tilde {S}_{x,y}^{z} \tilde{S}_{x,y+1}^{z}]\\
&\quad -\frac{h_x}{2}\sum_{x,y} (\tilde S_{x,y}^{+}e^{i\alpha_y x}+h.c.)-h_z\sum_{x,y} \tilde S_{x,y}^{z}.
\label{eq:ttH_0}
\end{split}
\end{equation}
where $\tilde J=\sqrt{J^2+D^2}$ describes the transverse component of  exchange interaction for the obtained XXZ chain. 
Observe that the transverse component of the {\em inter-chain} interaction, $J' e^{2i\alpha_y x}$, 
is oscillating function of the chain coordinate $x$.

It is intuitively clear that for sufficiently fast oscillation (that is, for sufficiently large $|\alpha_y|$) this term
must ``average out'' and disappear from the Hamiltonian. Our detailed calculations, reported below, fully confirm this intuition.

\subsection{Determination of the DM vector by ESR experiments}
\label{sec:esr}
The DM vector $\bm D$ can be characterized by the electronic spin resonance (ESR) measurements\cite{Halg2014,p,smirnov}. 
In a magnetic field $\bm h\parallel \bm D$, two resonance lines (ESR doublet) are observed at resonance frequencies $\nu_{\pm}$,
\begin{equation}
2\pi\hbar\nu_{\pm}=|g\mu_Bh\pm\frac{\pi}{2}D|.
\label{eq:esr-split}
\end{equation}
This ESR doublet is only observable for magnetic field having a component along $\bm D$, thus this property can be used to determined the direction of $\bm D$. 
In another limiting case $\bm h\perp \bm D$, the resonance occurs at the ``gapped" frequency
\begin{equation}
2\pi\hbar\nu=\sqrt{(g\mu_Bh)^2+(\frac{\pi}{2}D)^2}.
\label{eq:esr-gap}
\end{equation}
This gap provides an alternative way to obtain the amplitude $D$. (The lineshape and the temperature dependence of the width of the resonance
were studied in Refs.\onlinecite{karimi2011} and \onlinecite{Furuya2016}, Appendix D, correspondingly.)
In the case of K$_2$CuSO$_4$Br$_2$ several ESR measurements \cite{Halg2014,smirnov}
have consistently predicted $D_{\rm Br} \approx 0.28$ K. In K$_2$CuSO$_4$Cl$_2$ the DM interaction is smaller. Recent experiment \cite{smirnov2016}
estimates it to be $D_{\rm Cl} \approx 0.11$ K.
With regards to other parameters of the microscopic Hamiltonian, the intra-chain exchange $J$ has been estimated\cite{Halg2014} as $J_{\rm Cl} = 3.1$ K and $J_{\rm Br} = 20.5$ K.
Inter-chain interaction $J'$ is most difficult to estimate. Appendix~\ref{app:j'} describes fit of our CMF calculations of the ordering temperatures to experimental
values which allows us to estimate inter-chain exchanges as $J'_{\rm Cl} = 0.08$ K and $J'_{\rm Br} = 0.09$ K.
Thus the ratio $D/J'$ is about $1.3$ for K$_2$CuSO$_4$Cl$_2$ and $3.1$ for K$_2$CuSO$_4$Br$_2$ respectively. This, according to our investigation, places these
two materials into two distinct limits of weak and strong DM interaction, respectively.

\subsection{Bosonization: low-energy field theory}
\label{subsec:bosonization}

In the low-energy continuum limit the spin operator is represented by\cite{Gangadharaiah2008},
\begin{equation}
{\bm S}_{x,y}\to{\bm J}_{yL}(x)+{\bm J}_{yR}(x)+(-1)^{x/a}{\bm N}_y(x),
\label{eq:spin}
\end{equation}
where $a$ is the lattice spacing, and continuous space coordinate is introduced via $x=na$, with $n$ an integer. 
${\bm J}_{yL}(x)$ and ${\bm J}_{yR}(x)$, are the uniform left and right spin currents, and ${\bm N}_y(x)$ is the staggered magnetization. 
These fields can be conveniently expressed in terms of abelian bosonic fields $(\phi_y(x),\theta_y(x))$,
\begin{equation}
\begin{gathered}
J_{yR}^{+}=\frac{1}{2\pi a}e^{-i\sqrt{2\pi}(\phi_y-\theta_y)},\,
J_{yR}^{z}=\frac{1}{2\sqrt{2\pi}}(\partial_x\phi_y-\partial_x\theta_y),\\
J_{yL}^{+}=\frac{1}{2\pi a}e^{i\sqrt{2\pi}(\phi_y+\theta_y)},\;\,\;
J_{yL}^{z}=\frac{1}{2\sqrt{2\pi}}(\partial_x\phi_y+\partial_x\theta_y).\\
\end{gathered}
\label{eq:J}
\end{equation}
and
\begin{equation}
 {\bm N}_y= A(-\sin[\sqrt{2\pi}\theta_y],\; \cos[\sqrt{2\pi}\theta_y],\;-\sin[\sqrt{2\pi}\phi_y]).
 \label{eq:N}
 \end{equation}
Here, $A\equiv \gamma/(\pi a)$, and $\gamma=\langle \cos(\sqrt{2\pi}\varphi_{\rho})\rangle \sim O(1)$ is 
determined by gapped charged modes of the chain.

The above parameterization, applied to the Hamiltonian \eqref{eq:H_0}, produces the following continuum Hamiltonian \cite{Gangadharaiah2008,Schnyder2008,Garate2010}
\begin{equation}
{\cal H}=\sum_y [{\cal H}_0  + {\cal V}  + {\cal H}_{\rm bs}+{\cal H}_{\rm inter}],
\label{system_1}
\end{equation}
where
\begin{equation}
\begin{gathered}
{\cal H}_0=\frac{2\pi v}{3}\int \mathrm{d}x ({\bm {J}_{yR} }\cdot{\bm {J}_{yR} }+{\bm {J}_{yL} }\cdot{\bm {J}_{y L} }),\\
{\cal V}=-h_z\int \mathrm{d}x (J_{yR}^{z}+ J_{yL}^{z})-h_x\int \mathrm{d}x(J_{yR}^{x}+J_{yL}^{x})\\
\quad +(-1)^y \tilde{D} \int \mathrm{d}x(J_{yR}^{z}-J_{yL}^{z}),\\
{\cal H}_{\rm bs}=-g_{\rm bs}\int \mathrm{d}x [ J_{yR}^xJ_{yL}^x+J_{yR}^yJ_{yL}^y+(1+\lambda)J_{yR}^zJ_{yL}^z ],\\
{\cal H}_{\rm inter}=J'\int \mathrm{d}x{\bm N}_{y}\cdot{\bm N}_{y+1},\\
\end{gathered}
\label{system_2}
\end{equation}
where $v\simeq J\pi a/2$ is the spin velocity and $\tilde{D}=D(1+2\gamma^2)/\pi \approx D$.  $\cal V$ contains the second line of Eq.~\eqref{eq:H_0},
it collects all vector-like perturbations of the bare chain Hamiltonian ${\cal H}_0$.
${\cal H}_{\rm bs}$ describes residual backscattering interaction between right- and left-moving spin modes of the chain,
its coupling is estimated as $g_{\rm bs} \approx 0.23 \times (2\pi v)$, see Ref.~\onlinecite{Garate2010} for details.
An important DM-induced anisotropy parameter $\lambda$ is given, according to Ref.~\onlinecite{Garate2010} (see Eq. (B2) there), by
\begin{equation}
\lambda=c' \frac{D^2}{J^2}, ~\text{where}~  c'=\big(\frac{2\sqrt{2}v}{g_{\rm bs}}\big)^2 \approx 3.83.
\label{eq:lambda}
\end{equation}
 The inter-chain interaction is described by ${\cal H}_{\rm inter}$, in which we kept the most relevant, in renormalization group sense, contribution,
 ${\bm S}_{x,y}\cdot{\bm S}_{x,y+1}\to{\bm N}_{y}(x)\cdot{\bm N}_{y+1}(x)$. 
 
Now we examine phase diagram of the system described by Eq.~\eqref{system_1} and Eq.~\eqref{system_2} under two different field configurations, with 
external magnetic field ${\bm h}$ placed parallel, Sec.~\ref{sec:parallel}, and perpendicular, Sec.~\ref{sec:perpendicular}, to the DM vector ${\bm D}$. 

\begin{figure}[!tbp]
 \includegraphics[width=0.8\columnwidth]{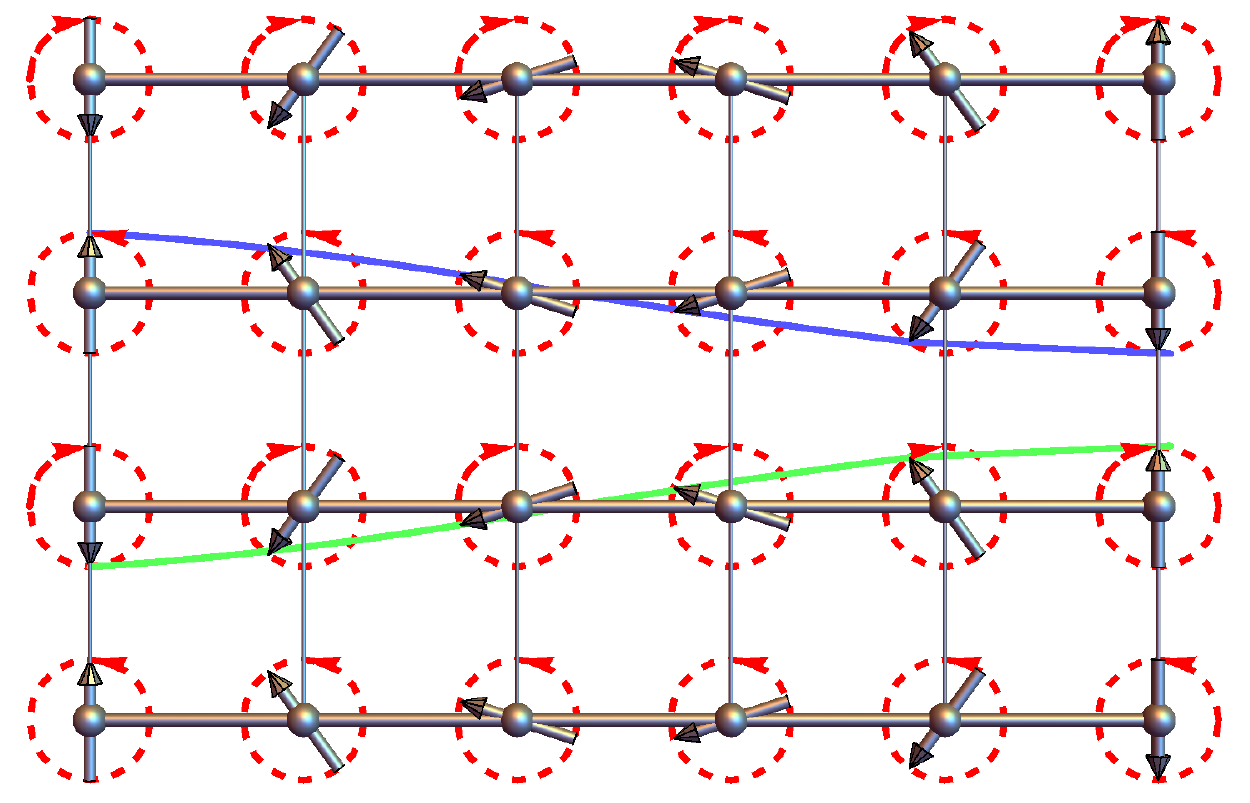}
 \caption{(Color online) Staggered magnetization $\bm N$ of the coneNN state from Section~\ref{sec:coneNN}. $\bm h\parallel \bm D$ and spins are ordered in the transverse to $\bm h$ plane. 
 Red circles with arrows indicates the precession direction of spins, as one moves along each chain. 
 Note that the arrows' direction alternates between consecutive chains, owing to the staggering of DM vector. 
 Blue and green curves visualize relative orientation of spin on neighboring chains which oscillates from parallel to anti-parallel as one moves along the chain
 leading to the cancellation of exchange interaction between nearest chains.}
 \label{fig:cone}
\end{figure}

\section{Key ideas of RG and CMF}
\label{sec:overview}

Our work describes an extended study of a novel mechanism of frustrating inter-chain exchange interaction in a system of weakly coupled spin-1/2 chains.
This section summarizes key ideas of the two main theoretical techniques - renormalization group (RG) and chain mean-field theory (CMF) - that are used in 
the paper. 

We assume that all interchain couplings are weak. RG proceeds by integrating short-distance modes (small distance $x$ or large momentum $k_x$) 
and by progressively reducing the large momentum cutoff from its bare value $\Lambda \sim 1/a$, which is of the order of the inverse lattice spacing $a$ (which we take to be $O(1)$),
to $\Lambda_\ell = \Lambda e^{-\ell}$,
where $\ell \in (0,\infty)$ is the logarithmic RG scale. Correspondingly, the minimal real space scale increases as $a e^\ell$.
Various interaction couplings $\gamma_i$, which enter the Hamiltonian as ${\cal H} = {\cal H}_0 + \sum_i \int dx \gamma_i {\cal O}^i_y(x) {\cal O}^i_{y+1}(x)$, see \eqref{system_2},
where ${\cal O}^i_y$ represent the $y$-th chain operator $J^a_y$ in \eqref{eq:J} or $N^a_y$ in \eqref{eq:N}, get renormalized (flow) during this procedure.
This renormalization is described by the {\em perturbative} RG flow equation of the dimensionless coupling \cite{oleg_cmf} $\tilde{\gamma}_i = \gamma_i/(v \Lambda_\ell^2)$ 
\begin{equation}
\frac{d \tilde{\gamma}_i}{d \ell} = (2 - 2\Delta_i) \tilde{\gamma}_i 
\label{eq:basics}
\end{equation}
Here $\Delta_i$ is the scaling dimension of the operator ${\cal O}^i_y$, which in the case of {\em relevant} operator \eqref{eq:N}, can be represented as $\Delta_i = 1/2 + O(y)$,
where $y$ stands for the dimensionless {\em marginal} coupling. For the marginal operator, say ${\cal O}^k_y$, the scaling dimension is close to 1,
$\Delta_k = 1 + O(y)$, and as a result the flow of the marginal operator obeys $d y/d\ell \sim y^2$. 
(See \eqref{eq:rg2} below for the specific example of both of these features.) Dimensionless coupling constants of the relevant operators increase with $\ell$. RG flow
need to be stopped at the RG scale $\ell^*$ at which the {\em first} coupling, say $\tilde{\gamma}_j$, reaches the value $C \sim O(1)$ of order $1$. 
According to \eqref{eq:basics} $\ell^*$ can be estimated as $\ell^* = \ln[C/\tilde{\gamma}_j(\ell=0)]/(2-2\Delta_j)$.
The length scale $\xi = a e^{\ell^*}$ defines the correlation length above which the system needs to be treated as two (or three) dimensional.  
The type of the developed two-dimensional order is determined by the most relevant operator ${\cal O}^j_y$ the coupling constant of which has reached $C \sim O(1)$ first.
Its expectation value can be estimated as $\langle {\cal O}^j\rangle \sim \xi^{-\Delta_j}$ and therefore, using $\tilde{\gamma}_j(\ell=0) = \gamma_j/(v \Lambda_{\ell=0}^2)$ and $\Lambda_{\ell=0} \sim O(1)$,
we obtain
\begin{equation}
\langle {\cal O}^j\rangle \sim \xi^{-\Delta_j} = \Big(\frac{\gamma_j}{C v}\Big)^{\Delta_j/(2-2\Delta_j)} .
\label{eq:orderparam}
\end{equation}
This discussion makes it clear that perturbative RG procedure is inherently uncertain since both the equation \eqref{eq:basics} and the ``strong-coupling value'' estimate $C$ 
are based on the perturbation expansion in terms of the coupling constants $\gamma_i$. Moreover, in the case of the competition between the two orders,
associated with operators ${\cal O}^j$ and ${\cal O}^i$ correspondingly, the transition from the one order to another can only be {\em estimated} from the condition
$\ell^*_j = \ell^*_i$.

This approximate treatment becomes more complicated when some of the interactions acquire coordinate-dependent oscillating factor, symbolically 
$\int dx \gamma_i {\cal O}^i_y(x) {\cal O}^i_{y+1}(x) e^{i f x}$. Such a dependence is caused by external magnetic field and/or DM interactions, 
see for example equations \eqref{shift:J-N} and \eqref{H'} below. Perturbative RG calculation is still possible, see for example Sec.4.2.3 of Giamarchi book \cite{Giamarchi}
for its detailed description, but becomes technically challenging. At the same time the key effect of the oscillating term $e^{i f x}$ can be understood with the help
of much simpler qualitative consideration outlined, for example, in Ref.~\onlinecite{oshikawa1999} and in Sec.18.IV of Gogolin {\em et al} book \cite{Gogolin}. 
Oscillation becomes noticeable on the spatial scale $x \sim 1/f$ which has
to be compared with the running RG scale $a e^\ell$. As a result, RG flow can be separated into two stages. During the first stage $0 \leq \ell \leq \ell_{\rm osc} = \ln(1/f)$
oscillating factor $e^{i f x}$ can be approximated by 1, i.e. it does not influence the RG flow. At this stage all RG equations can be well approximated by their
zero-$f$ form. During the second stage $\ell_{\rm osc} \leq \ell \leq \ell^*$ and the product $f x$ is not small anymore. The factor $e^{i f x}$ produces sign-changing 
integrand. Provided that the coupling constant of that term remain small (which is the essence of the condition $\ell \leq \ell^*$), the 
integration over $x$ removes such an oscillating interaction term from the Hamiltonian altogether.

This is the strategy we assume in this paper. It is clearly far from being exact but it is an exceedingly good approximation in the two important limits:
the small-$f$ limit when $\ell_{\rm osc} \gg \ell^*$ and the external field/DM interaction is not important at all, and in the large-$f$ limit when $\ell_{\rm osc} \ll \ell^*$
and the oscillations are so fast that corresponding interactions average to zero. In-between these two clear limits the proposed {\em two-stage} scheme \cite{oshikawa1999} 
provides for a physically sensible interpolation. 

Perturbative RG procedure outlined above is great for understanding relative relevance of competing interchain interactions and for approximate 
understanding of the role of the field and DM induced oscillations. Its inherent ambiguity makes one to look for a more quantitative description which
matches RG at the scaling level but also allows to account for the numerical factors associated with various interaction terms at the better than
logarithmic accuracy level. Such description is provided by the chain mean-field (CMF) theory proposed in Ref.~\onlinecite{schulz1996} and numerically tested 
for the system of weakly coupled chains in Refs.~\onlinecite{Sandvik1999,Yasuda2005}. In CMF, interchain interactions are approximated by a self-consistent Weiss fields
introduction of which reduces the coupled-chains problem to an effective single-chain one of the sine-Gordon kind, which is understood extremely well \cite{schulz1996,Lukyanov1997}. 
As described in Section~\ref{sec:CMF} and 
Appendix~\ref{app:cmf} below, this approximation allows one to calculate critical temperature $T_i$ of the order associated with operator ${\cal O}^i$.
The order with the highest $T_i$ is assumed to be dominant. As mentioned above, at the scaling level CMF theory matches the RG procedure and
the highest $T_i$ corresponds to the order with the shortest $\ell^*_i$. The benefit of CMF approach consists in the ability to account for the field-dependent 
scaling dimensions of various chain operators in a more systematic and uniform way as we detail below.

\section{Parallel configuration, ${\bm h} \parallel{\bm D} $}
\label{sec:parallel}

When the external magnetic field is parallel to DM vector ${\bm D}$ along $\hat{z}$, $h_z=h$ and $h_x=0$. In this configuration it is convenient
to use Abelian bosonization \eqref{eq:J}, by expressing spin currents in $\cal V$ of Eq.~\eqref{system_2} in terms of fields $(\phi_y,\theta_y)$, 
\begin{equation}
\begin{gathered}
{\cal H}_0=\frac{v}{2}\int \mathrm{d}x [(\partial_x{\phi_y})^2+(\partial_x{\theta_y})^2], \;{\cal V}={\cal H}_{\rm Z}+{\cal H}_{\rm DM},\\
{\cal H}_{\rm Z}=-\frac{h}{\sqrt{2\pi}}\int \mathrm{d}x \partial_x \phi_y,\\
{\cal H}_{\rm DM}=-(-1)^y\frac{D}{\sqrt{2\pi}}\int \mathrm{d}x \partial_x \theta_y,
\end{gathered}
\label{eq:abelian-H}
\end{equation}
where ${\cal H}_Z$ and ${\cal H}_{\rm DM}$ are the Zeeman and DM interactions, respectively.
Evidently, these linear terms can be {\em absorbed} into ${\cal H}_0$ by shifting fields $\phi_y$ and $\theta_y$ appropriately,
\begin{equation}
\begin{gathered}
 \phi_y=\tilde{\phi}_y+\frac{t_{\phi}}{\sqrt{2\pi}}x, \quad t_{\phi} \equiv \frac{h}{v},\\
 \theta_y= \tilde{\theta}_y + (-1)^y\frac{t_{\theta}}{\sqrt{2\pi}} x = \tilde{\theta}_y + \frac{t^y_{\theta}}{\sqrt{2\pi}} x ,\\ 
 \quad t^y_{\theta} \equiv (-1)^y t_\theta = (-1)^y\frac{D}{v}.
\end{gathered}
\label{shift1}
\end{equation}
Note that $t_{\theta}^y$ depends on the parity of the chain index $y$, and it is just the continuum version of the angle $\alpha_y$ in Sec.~\ref{subsec:lattice-rot}. 

As a result of the shifts, the spin currents and the staggered magnetization are modified as 
\begin{equation}
\begin{gathered}
J_{yR}^{+}\to \tilde{J}_{yR}^{+}e^{-i(t_\phi-t_{\theta}^y)x},\quad
J_{yL}^{+}\to \tilde{J}_{yL}^{+}e^{i(t_\phi+t_{\theta}^y)x},\\
J_{yR}^{z}\to \tilde{J}_{yR}^{z}+\frac{(t_\phi-t_{\theta}^y)}{4\pi},\quad
J_{yL}^{z}\to \tilde{J}_{yL}^{z}+\frac{(t_\phi+t_{\theta}^y)}{4\pi},\\
N^{+}_y \to \tilde{N}^{+}_y e^{it_{\theta}^yx},\quad
N^{z}_y \to -A\sin[\sqrt{2\pi}\tilde{\phi}_y+t_\phi x].\\
\end{gathered}
\label{shift:J-N}
\end{equation}
It is important to observe here that {\em tilded} operators in \eqref{shift:J-N} are obtained from the original ones \eqref{eq:J} and \eqref{eq:N} by replacing
original $\phi_y$ and $\theta_y$ with their {\em tilded} versions $\tilde{\phi}_y$ and $\tilde{\theta}_y$.
Note also that the shift introduces oscillating position-dependent factors to transverse components of ${\bm J}_y$ and ${\bm N}_y$. 
The Hamiltonian now reads
\begin{equation}
{\cal H}_{\rm chain}=\tilde{\cal H}_0+\tilde{\cal H}_{\rm bs}+\tilde{\cal H}_{\rm inter},
\label{eq:15}
\end{equation}
where $\tilde{\cal H}_0$ retains its quadratic form \eqref{eq:abelian-H} in terms of tilded fields. 
It is perturbed by backscattering $\tilde{H}_{\rm bs}$ and inter-chain $\tilde{H}_{\rm inter}$ interactions, which now read
\begin{eqnarray}
\tilde{\cal H}_{\rm bs}&=&\int \mathrm{d}x \big\{\pi vy_B\big(\tilde{J}_{yR}^{+}\tilde{J}_{yL}^{-}e^{-i2t_\phi x}+{\rm h.c.}\big) \nonumber\\
&&+ 2\pi v y_z \tilde{J}_{yR}^{z}\tilde{J}_{y, R}^{z}\big\},
\label{H'1}
\end{eqnarray}
and $\tilde{\cal H}_{\rm inter} = {\cal H}_{\rm cone} + {\cal H}_{\rm sdw}$, where
\begin{equation}
\begin{gathered}
{\cal H}_{\rm cone}=\pi v A^2 g_{\theta} \int \mathrm{d}x \big(e^{i[\sqrt{2\pi}(\tilde{\theta}_y-\tilde{\theta}_{y+1})+2t_{\theta}^{y}x]}+{\rm h.c.}\big),\\
{\cal H}_{\rm sdw}=
\pi v A^2\int  \mathrm{d}x \Big\{ {g}_{\phi }\big(e^{i\sqrt{2\pi}(\tilde{\phi}_y-\tilde{\phi}_{y+1})}+{\rm h.c.}\big)\\
\quad\quad-\tilde{g}_{\phi}\big(e^{i[\sqrt{2\pi}(\tilde{\phi}_y+\tilde{\phi}_{y+1})+2t_\phi x]}+{\rm h.c.}\big)\Big\}.\\
\end{gathered}
\label{H'}
\end{equation}
${\cal H}_{\rm cone}$ and ${\cal H}_{\rm sdw}$ are the transverse and longitudinal (with respect to the $z$-axis) components of inter-chain interaction respectively. 
Their effect consists in promoting two-dimensional ordered cone and SDW state, correspondingly.
Small terms resulting from the additive shifts in $J_{R/L}^z$ in \eqref{shift:J-N} have been neglected. 
Table~\ref{table:term1} describes which inter-chain interactions produce which state.
\begin{table}[!tbp]
	\begin{center}
		{\renewcommand{\arraystretch}{1.2}%
		\begin{tabular}{c| c c c}
			\hline\hline
			Interaction & $\quad$Coupling$\quad$ & $\quad$Coupling$\quad$  & $\quad$Induced$\quad$\\
			term        & operator & constant  & state\\
			\hline
			${\cal H}_{\rm cone}$ & $N^+_y N^-_{y+1} $ & $g_{\theta}$ & cone \\[0.6ex]
			${\cal H}_{\rm sdw}$ & $N^z_y N^z_{y+1}$ & $g_{z}$ & SDW \\  [0.6ex]
			${\cal H}_{\rm NN}$ & $N^+_y N^-_{y+2} $ & $G_{\theta}$ & coneNN \\[0.6ex]
			\hline\hline
		\end{tabular}}
	\end{center}
	\caption{Three relevant perturbations from interchain interaction ${\cal H}_{\rm cone}$, ${\cal H}_{\rm sdw}$ in Eq.~\eqref{H'}  and ${\cal H}_{\rm NN}$ in Eq.~\eqref{2nd neighbour}, 
		their operator forms, associated coupling constants and types of the ordered states they induce.}
	\label{table:term1}
\end{table}

In writing the above we introduced several running coupling constants
\begin{equation}
\begin{gathered}
y_B=\frac{1}{2}(y_x+y_y),  \quad y_B(0)=-\frac{g_{\rm bs}}{2\pi v},\\
g_{\theta}=\frac{1}{2}(g_x+g_y), \quad g_\theta(0)=\frac{J'}{2\pi v}, \\
g_{\phi}=\tilde{g}_{\phi}=\frac{1}{2}g_z, \quad g_z(0) = \frac{J'}{2\pi v},\\
\label{eq:ini2}
\end{gathered}
\end{equation}
initial values of which follow from 
\begin{equation}
\begin{gathered}
y_x(0)=y_y(0)=-\frac{g_{\rm bs}}{2\pi v},\;\;\;
y_z(0)=-\frac{g_{\rm bs}}{2\pi v}(1+\lambda),\\
g_x(0)=g_y(0)=g_z(0)=\frac{J'}{2\pi v}.
\end{gathered}
\label{eq:ini1}
\end{equation}
Observe that DM interaction produces an effective anisotropy $\lambda=c'(D/J)^2>0$ which leads to $|y_z(0)|>|y_{x,y}(0)|$. 

Next we need to identify the most-relevant coupling in perturbation $H'=\tilde{\cal H}_{\rm bs}+\tilde{\cal H}_{\rm inter}$, which is accomplished by the renormalization group (RG) analysis.  

\begin{figure}[!tbp]
\centering
 \includegraphics[width=0.75\columnwidth]{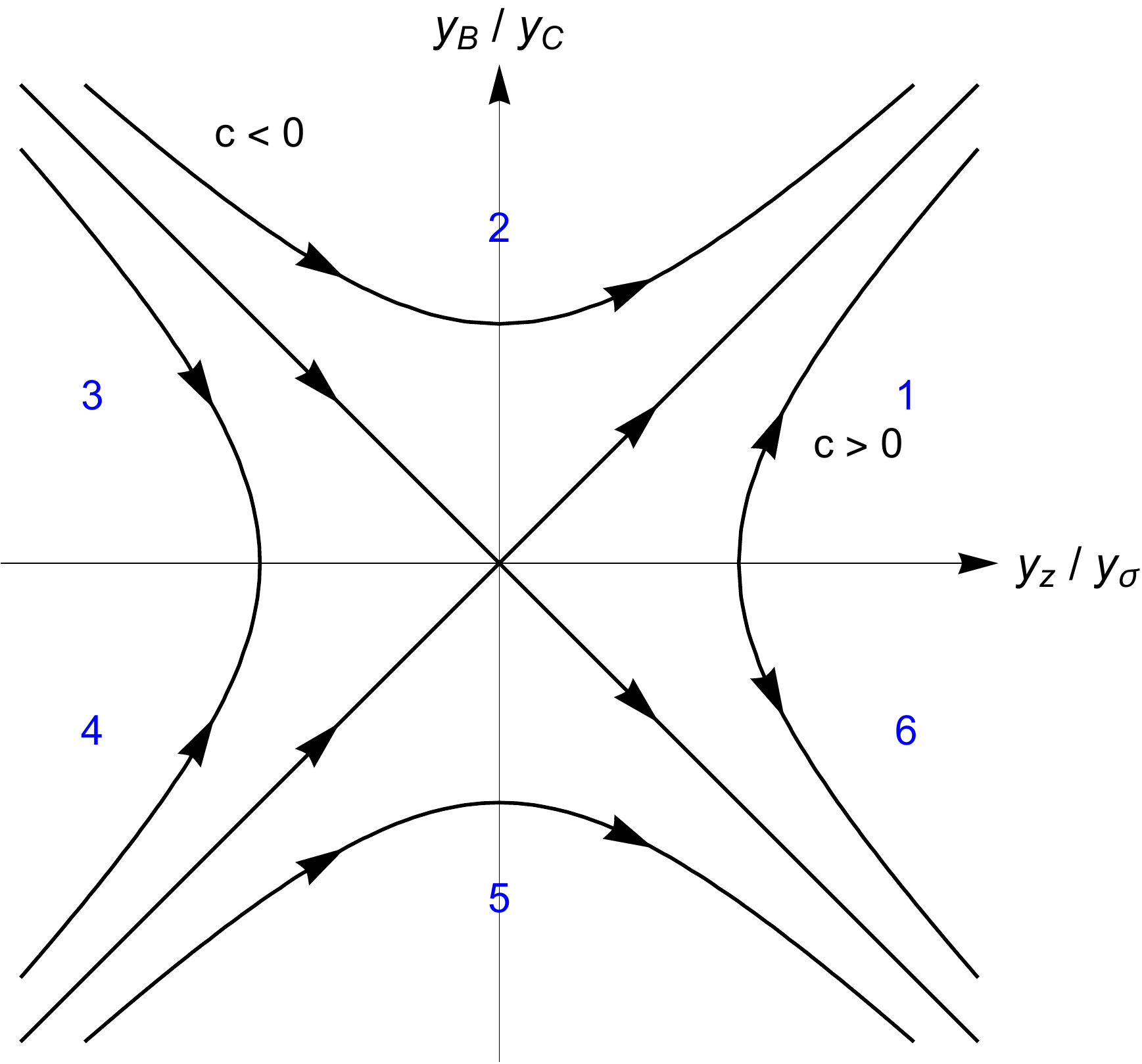}
 \caption{(Color online) Solution of Kosterlitz-Thouless (KT) equations (first line of \eqref{eq:rg2}). 
 Five sectors of the flow are divided according to the initial conditions. For example, in sector 3: $y_{z/\sigma}(0)<0,y_{B/C}(0)>0$ and $C>0$.}
 \label{fig:kt}
\end{figure}

\subsection{Renormalization group (RG) analysis}
\label{subsec:rg_1}
According to standard RG arguments, the low energy properties of the system are determined by the couplings which renormalize to dimensionless values of order one first.
We derived RG equations for various coupling constants with the help of operator product expansion (OPE) technique \cite{fradkin} (see Appendix~\ref{app:ope} for  details),

\begin{equation}
\begin{gathered}
\frac{d y_B}{d \ell}=y_By_z,\;\;\;
\frac{d y_z}{d \ell}=y_B^2,\\
\frac{d g_\theta}{d \ell}=g_{\theta}(1-\frac{1}{2}y_z),\;\;\;
\frac{dg_z}{d\ell}=g_z(1+\frac{1}{2}(y_z - 2 y_B))  .\\
\end{gathered}
\label{eq:rg2}
\end{equation}
The first two equations in Eq.~\eqref{eq:rg2} are the well-known Kosterlitz-Thouless (KT) equations for the marginal backscattering couplings $y_{B,z}$ in \eqref{H'1}.
They admit analytic solution which is illustrated in Fig.~\ref{fig:kt}.
Initial conditions \eqref{eq:ini2}, \eqref{eq:ini1} correspond to $y_B<0, y_z<0$ and $C=y_z(\ell)^2-y_B(\ell)^2>0$, which places the KT flow
in sector $4$ in Fig.~\ref{fig:kt}. Physically, this corresponds to DM-induced easy-plane anisotropy ($\lambda >0$) which, if acting alone, would drive
the chain into a critical LL state. 

This marginally-irrelevant flow of $y_{B,z}$ is, however, interrupted by the exponentially fast growth of the inter- chain interactions $g_{\theta,\phi}$
which, according to \eqref{eq:rg2}, reach strong coupling limit at $\ell_{\rm inter} \approx \ln(2\pi v/J')$.
This growth describes development of the two-dimensional magnetic order in the system of 
weakly coupled chains. As a result, we are allowed to treat chain backscattering $y_{B,z}$, which barely changes on the scale of $\ell_{\rm inter}$, 
as a weak correction to the relevant inter-chain interaction.
This is the physical content of the second line of RG equations in \eqref{eq:rg2}.

DM interaction and magnetic field strongly perturb RG flow \eqref{eq:rg2} via coordinate-dependent factors $e^{i 2t_{\theta}^y x}$  and $e^{i 2t_\phi x}$, 
rapid oscillations of which become significant once running RG scale $\ell$ becomes greater than $\ell_{\theta} (\ell_\phi)$, where
\begin{equation}
\begin{gathered}
\ell_{\theta}=\ln(\frac{1}{a_0t_\theta}) = \ln(\frac{v}{D a_0}), \: \ell_{\phi}=\ln(\frac{1}{a_0t_\phi}) = \ln(\frac{v}{h a_0}). \\
\end{gathered}
\label{length1}
\end{equation}
These oscillations have the effect of nullifying, or averaging out, corresponding interaction terms in the Hamiltonian, provided that the corresponding 
coupling constants remain small at RG scales $\ell_{\theta,\phi}$.
The affected terms are ${\cal H}_{\rm cone}$ and $\tilde{g}_\phi$ term in ${\cal H}_{\rm sdw}$, respectively.
Also affected is backscattering $y_B$ term in \eqref{H'1}.
The short-distance cut-off $a_0$ that appears in \eqref{length1} is determined by the initial value of the backscattering $g_{\rm bs}(0) = 0.23 \times (2\pi v)$,
see Ref.~\onlinecite{Garate2010} for detailed explanation of this point. 

In accordance with general discussion in Sec.~\ref{sec:overview}, we define $\ell^*$ as an RG scale
at which the most relevant coupling constant $g$ reaches value of $1$, namely $|g(\ell^*)|=1$. For interchain couplings, we find that 
$\ell^*$ is close to $\ell_{\rm inter}\approx \ln(2\pi v/J')$ introduced below Eq.~\eqref{eq:rg2}, and this is noted in the caption of Figures~\ref{fig:flow1}, 
\ref{fig:nn1} and Figures~\ref{fig:flow01} - \ref{fig:flow03}.

Magnetic field induced oscillations in ${\cal H}_{\rm sdw}$ are well-known and describe magnetization-induced shift of longitudinal spin modes 
from the zero wave vector. In addition, magnetic field works to increase scaling dimension of $N^z$ field, from 1/2 at 
zero magnetization $M=0$ to 1 at full polarization $M=1/2$, see Table~\ref{table:scaling_d}, making the $N^z$ field less relevant. 
Typically, this makes ${\cal H}_{\rm sdw}$ term less important
than ${\cal H}_{\rm cone}$ one, which is build out of transverse spin operators which become more relevant with the field 
(the corresponding scaling dimension of which becomes smaller with the field, it changes from 1/2 at  $M=0$ to 1/4 at $M=1/2$).

\begin{table}[!tbp]
	\begin{center}
		{\renewcommand{\arraystretch}{1.2}%
		\begin{tabular}{c| c c c}
			\hline\hline
			Operator    &$\quad \Delta\quad$  & $\quad$M=0$\quad$ & $\quad$M=1/2$\quad$ \\ [0.6ex] 
			\hline
			$N^z \quad $ & $\pi/ \beta^2$ & 1/2 & 1 \\  [0.6ex]
			$N^+ \quad $   & $\pi R^2$ & 1/2 & 1/4 \\
			\hline\hline
		\end{tabular}}
	\end{center}
	\caption{Scaling dimensions $\Delta$ of longitudinal and transverse  components for staggered magnetization ${\bm N}$ vs magnetization $M$. }
	\label{table:scaling_d}
\end{table}

In our problem, however, the prevalence of the cone state is much less certain due to the presence of the built-in DM-induced
oscillations in ${\cal H}_{\rm cone}$ \eqref{H'}, originating from the staggered geometry of DM interaction. As a result, one needs to distinguish
the cases of weak and strong DM interaction, which in the current case should be compared with the inter-chain exchange interaction $J'$.

\subsection{Weak DM interaction, $D\ll J'$}
\label{sub:weak_DM_para}

\begin{figure}[!tbp]
	\includegraphics[width=0.8\columnwidth]{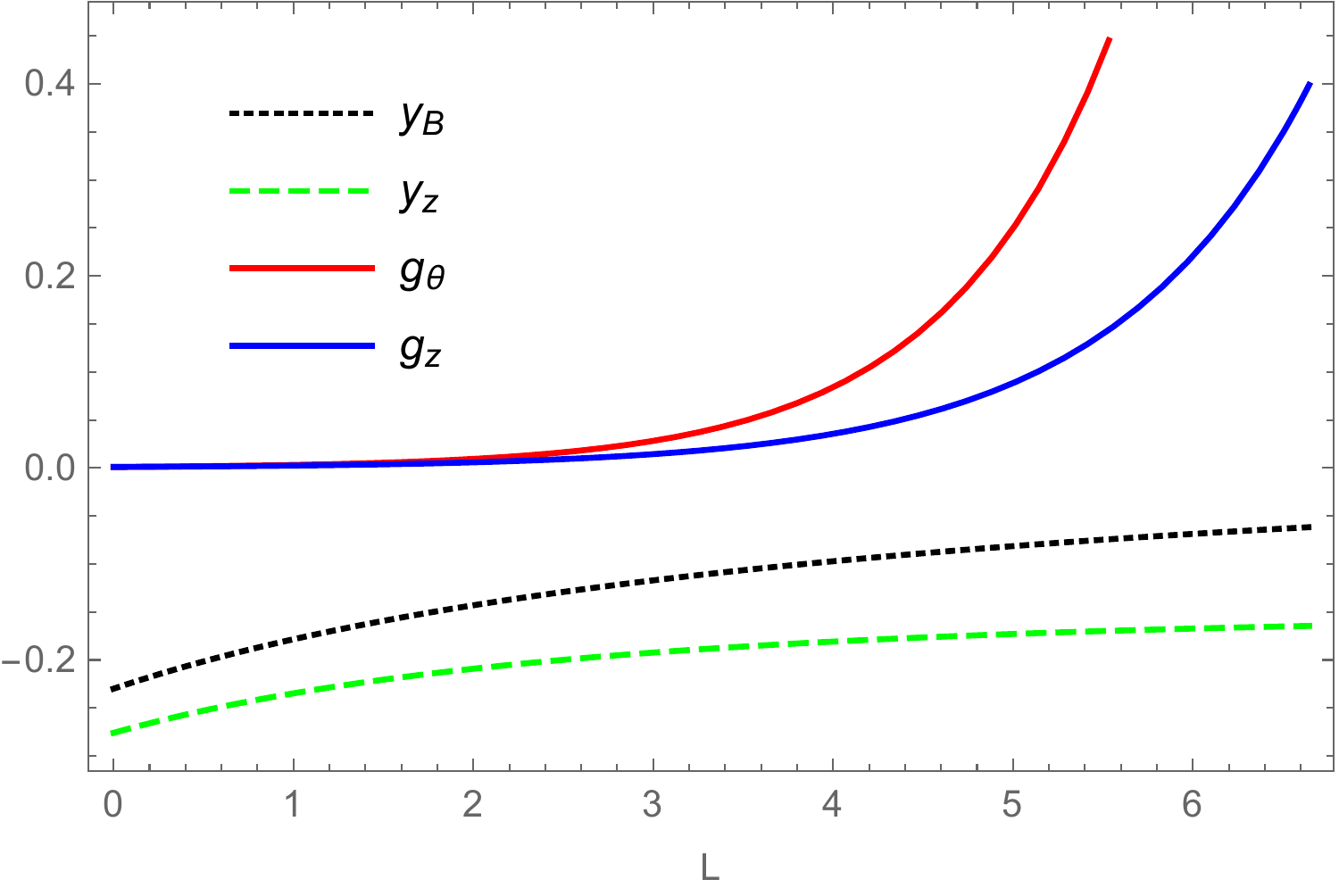}
	\caption{(Color online) Typical RG flow of the coupling constants for weak DM interaction and ${\bm h}\parallel {\bm D}$, $h_x=0$. $D=1\times10^{-4}J$, $g_{\rm bs}/(2\pi v)=0.23$, $J'/(2\pi v)=0.001$,
		$h_z/D=1$ and $\lambda=0.2$. Here $\ell_{\rm inter}\simeq 6.9$, $\ell_{\phi}=\ell_{\theta}\simeq 6.6$. The dominant coupling is $g_\theta $ (red solid line), and $g_\theta(\ell^*)=1$ at $\ell^*\simeq6.3$. }
	\label{fig:flow1}
\end{figure}

First, we consider the case of weak DM interaction, $D\ll J'$. This means $\ell_\theta > \ell_{\rm inter}$,  the integrand of ${\cal H}_{\rm cone}$ oscillates slowly
so that the factor $e^{i 2t_{\theta}^y x}$ does not affect the RG flow. As discussed in Appendix~\ref{app:ope}, backscattering terms break the symmetry between $g_\theta$ and $g_z$,
$g_{\theta}(\ell) > g_z(\ell)$.  As a result, inter-chain interaction ${\cal H}_{\rm cone}$ reaches strong coupling before ${\cal H}_{\rm sdw}$ and
the ground state realizes the cone phase. Typical RG flow of coupling constants for this case is shown in Fig.~\ref{fig:flow1}. 

Minimization of the argument of cosine in ${\cal H}_{\rm cone}$ requires that 
$\sqrt{2\pi}(\tilde{\theta}_y-\tilde{\theta}_{y+1})+2t_{\theta}^{y}x = \pi$. This is solved by requiring 
$\tilde{\theta}_y(x) = \hat{\theta} - (-1)^y t_\theta x/\sqrt{2\pi} - \sqrt{\pi/2} ~y$, where $\hat{\theta}$ is position-independent constant which describes orientation of the staggered magnetization $N^+_y(x) \sim (-1)^y i e^{i\sqrt{2\pi} \hat{\theta}}$ in the plane perpendicular to the magnetic field.

Observe that the obtained solution describes a {\em commensurate cone} configuration.
The original shift \eqref{shift1} is compensated by the opposite shift needed to minimize the $\tilde{\theta}$ configuration.
As a result the obtained cone state is commensurate along the chain direction: $N^+_y$ is uniform along the chain direction which
means the spin configuration is actually staggered, $S^+_y(x) \sim (-1)^x N^+_y$, see \eqref{eq:spin}. Note also that $N^+_y$ is staggered between chains
(so as to minimize the antiferromagnetic inter-chain exchange $J' >0$), so that in fact $S^+_y(x)$ realizes the standard N\'eel configuration. Thus ground state spin configuration of 
the cone phase is described by
\begin{eqnarray}
\langle {\bm S}_y(x) \rangle &&= M {\bf z} + (-1)^{x + y} \Psi_{\rm cone} (-\sin[\sqrt{2\pi}\hat{\theta}] {\bf x} + \nonumber\\
&&+\cos[\sqrt{2\pi}\hat{\theta}] {\bf y}).
\label{eq:cone}
\end{eqnarray}
Here $\Psi_{\rm cone}$ denotes the magnitude of the order parameter at the scale $\ell^*$. According to \eqref{eq:orderparam} and using equations \eqref{eq:N} and \eqref{eq:ini2}, 
it can be estimated as $\Psi_{\rm cone} = \gamma/(\pi a) \sqrt{g_\theta} \propto (J'/v)^{1/2}$. The square-root dependence of the order parameter on the inter-chain exchange $J'$
is a well-known feature of weakly coupled chain problems \cite{schulz1996}. CMF theory, which we introduce in the next section, can too be used to calculate the
cone order parameter. This is described in Appendix~\ref{app:order_parameter} and its dependence on magnetization $M$, at a fixed $J'/v$ ratio, is illustrated in Fig.~\ref{fig:orderCl}.
Note that its dependence on $M$ occurs via $M$-dependence of scaling dimensions and other parameters in the Hamiltonian which are not easy to capture 
with the help of the RG procedure.

\subsection{Strong DM interaction, $D>J'$}\label{sub:strong_DM_para}
\label{strong_DM_para}
\subsubsection{SDW order}
Now we turn to a less trivial case of strong DM interaction, when $D \gg J'$. Here $\ell_\theta <\ell_{\rm inter}$, which simply eliminates ${\cal H}_{\rm cone}$
from the competition, and from the Hamiltonian. The physical reasoning is that strong DM interaction introduces strong frustration to the transverse 
inter-chain interaction, which oscillates rapidly and averages to zero. As a result, the only inter-chain interaction that survives in this situation
is ${\cal H}_{\rm sdw}$,  Eq.\eqref{H'}, which establishes two-dimensional longitudinal SDW order. 

Two types of SDW ordering are possible. The first - {\em commensurate} SDW order - realizes in low magnetic field $h \leq h_{\rm c-ic} \sim O(J')$
when spatial oscillations due to $t_\phi x$ term in $N^z_y$ operator \eqref{shift:J-N} are not important. This is the regime of $\ell_\phi \gg \ell_{\rm inter}$,
when both $g_\phi$ and $\tilde{g}_\phi$ terms in the SDW inter-chain interaction ${\cal H}_{\rm sdw}$ in \eqref{H'} contribute equally.
In a close similarity to the commensurate cone state discussed above, the $\tilde{\phi}$ configuration here is minimized by 
$\tilde{\phi}_y(x) = \hat{\phi} - t_\phi x/\sqrt{2\pi} - \sqrt{\pi/2} ~y$. Here the global constant $\hat{\varphi}$ is determined by the requirement 
that $\sin[\sqrt{2\pi} \hat{\phi}] = \pm 1$, corresponding to a maximum possible magnitude of  $N^z_y \sim (-1)^y \sin[\sqrt{2\pi} \hat{\phi}]$.
Therefore $\hat{\phi} = \hat{\phi}_k = \sqrt{\pi/2} (k + 1/2)$, where $k=0, 1$. This describes the situation of the commensurate longitudinal SDW order
which is pinned to the lattice, $N^z_y \sim (-1)^y (-1)^k$. Changing $k \to k \pm 1$ corresponds to a discrete translation of the SDW order by one lattice spacing.
In terms of spins this too is a N\'eel-like order, but it is collinear one along the magnetic field axis, 
\begin{equation}
\langle {\bm S}_{x,y} \rangle = (M + \Psi_{\rm sdw-c} (-1)^{x+y} (-1)^k){\bf z}.
\label{eq:sdw-comm}
\end{equation}

Increasing the field beyond $h_{\rm c-ic}$ un-pins the SDW ordering from the lattice and transforms spin configuration into collinear {\em incommensurate} SDW.
Technical details of this are described in the Appendix \ref{app:cmf} and here we focus on the physics of this commensurate-incommensurate (C-IC) transition.
Increasing $h$ makes $\ell_\phi$ smaller and at $\ell_\phi \approx \ell_{\rm inter}$ oscillating $e^{i 2 t_\phi x}$ factor in the $\tilde{g}_\phi$ term in \eqref{H'}
becomes very strong and `washes out' that piece of the ${\cal H}_{\rm sdw}$ Hamiltonian. The remaining, $g_\phi$, part of ${\cal H}_{\rm sdw}$
continues to be the only relevant inter-chain interaction and flows to the strong coupling. Therefore now $\sqrt{2\pi}(\tilde{\phi}_y-\tilde{\phi}_{y+1}) = \pi$
which is solved by $\tilde{\phi}_y = \hat{\phi} - \sqrt{\pi/2} ~y$. As a result the shift \eqref{shift1} remains intact and one finds 
incommensurate SDW ordering with 
\begin{equation}
\langle {\bm S}_y(x)\rangle \sim (M + \Psi_{\rm sdw-ic} (-1)^{x+y} \sin[\sqrt{2\pi} \hat{\phi} + h x/v]) {\bf z}.
\label{eq:sdw-incomm}
\end{equation}
The magnitude of the SDW order parameter $\Psi_{\rm sdw-ic}$ in this equation is calculated in Appendix~\ref{app:order_parameter} and its dependence on magnetization 
$M$, at a fixed $J'/v$ ratio, is illustrated in Fig.~\ref{fig:orderBr}. Note that unlike the cone order, the SDW one weakens with increasing $M$.

\begin{figure}[!tbp]
	\includegraphics[width=0.8\columnwidth]{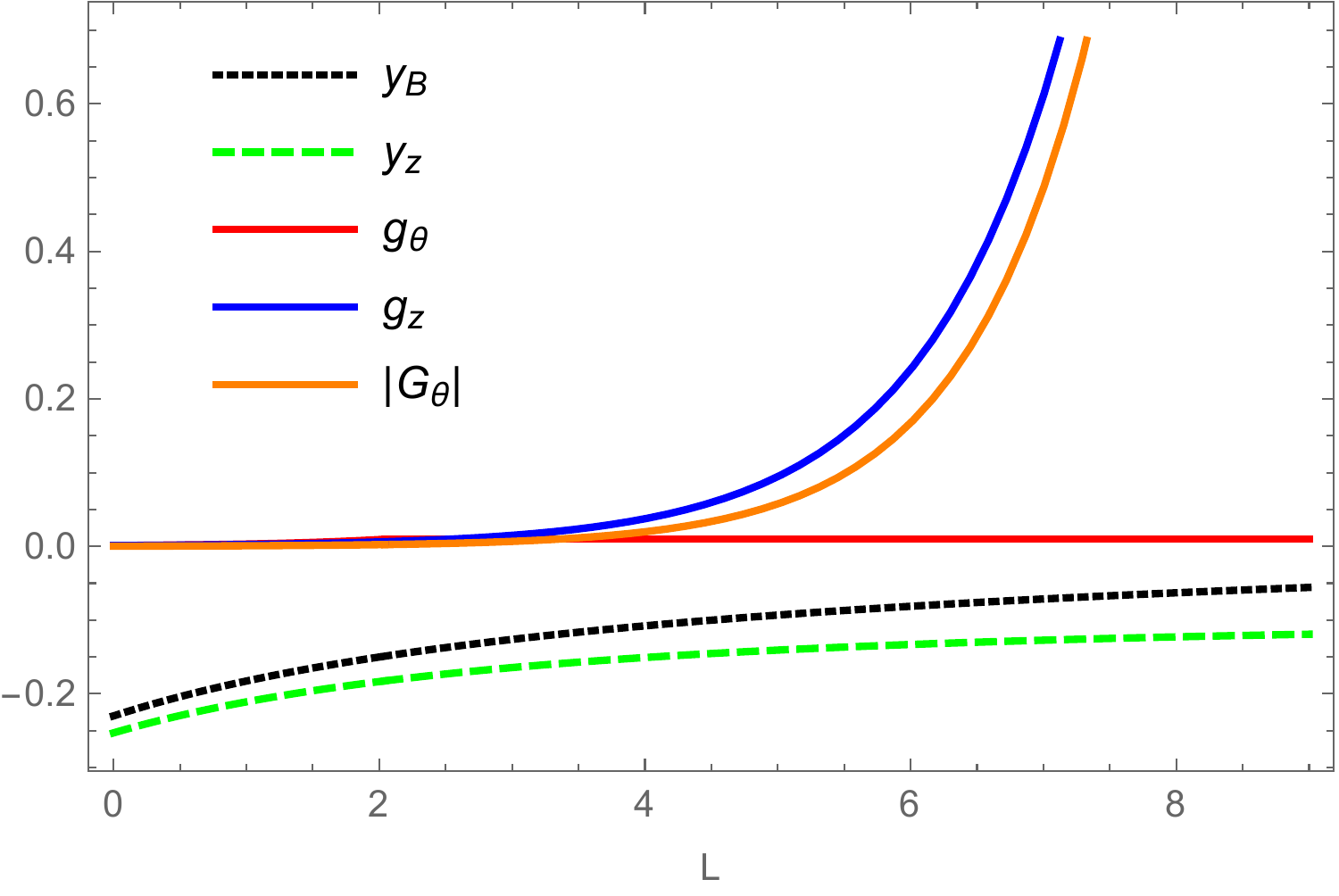}
	\caption{(Color online) RG flow of the coupling constants for strong DM interaction and ${\bm h}\parallel {\bm D}$, $h_x=0$. The case of low magnetic field  $h_z/D=0.005$. 
		$D=0.01J$, $g_{\rm bs}/(2\pi v)=0.23$, $J'/(2\pi v)=0.001$ and $\lambda=0.1$. Here $\ell_{\rm inter}\simeq 6.9$, $\ell_{\phi}\simeq7.4$, $\ell_{\theta}\simeq2$ and $g_{\theta}$ keeps as a constant after $\ell>\ell_\theta$, due to the rapid spatial oscillation. The dominant coupling is $g_z$ (blue solid line), and $g_z(\ell^*)=1$ at $\ell^*\simeq7.5$.}
	\label{fig:nn1}
\end{figure}

The global phase $\hat{\phi} \in (0,\sqrt{2\pi})$
is not pinned to any particular value - it describes emergent translational $U(1)$ symmetry of the `high-field' limit of the SDW Hamiltonian [Eq.\eqref{H'} without $\tilde{g}_\phi$ term], 
which does not depend on the value of $\hat{\phi}$. Spontaneous selection of some particular $\hat{\phi}$ corresponds to a spontaneous breaking
of the translational symmetry. The resulting incommensurate SDW order is characterized by the emergence of Goldstone-like longitudinal fluctuations, 
{\em phasons}. Recent discussion of some aspects of this physics can be found in Ref.~\onlinecite{Starykh2014}.

\subsubsection{Next-nearest chains cone order}
\label{sec:coneNN}

\begin{figure}[!tbp]
	\includegraphics[width=0.8\columnwidth]{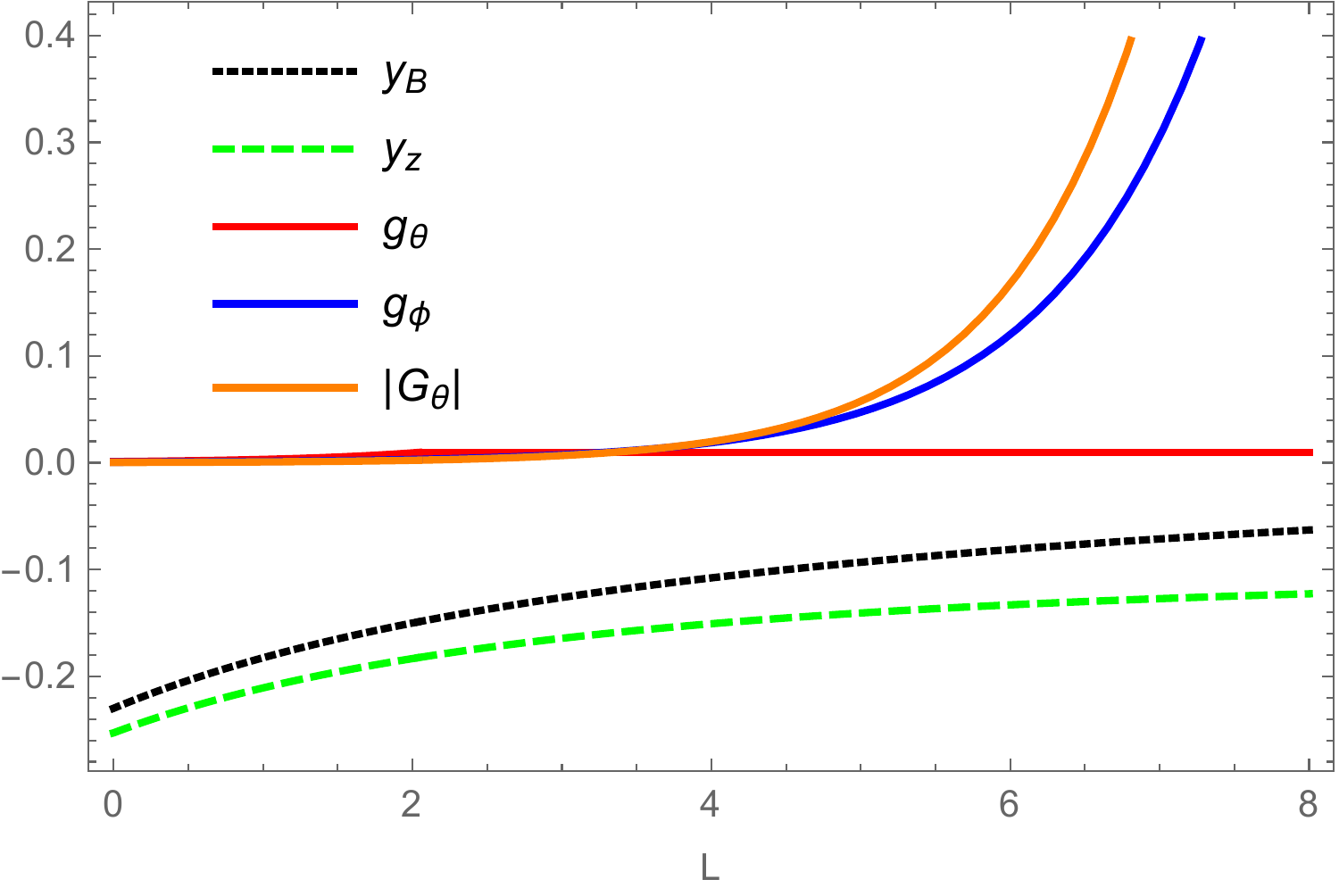}
	\caption{(Color online) Typical flow of the coupling constants for strong DM interaction and ${\bm h}\parallel {\bm D}$, $h_x=0$. This is the case of relatively high magnetic field $h_z/D=5$. 
		$D=0.01J$, $g_{\rm bs}/(2\pi v)=0.23$,  $J'/(2\pi v)=0.001$ and $\lambda=0.1$. Here $\ell_{\rm inter}\simeq 6.9$, $\ell_{\varphi}\simeq0.4$, $\ell_{\theta}\simeq2$. The dominant coupling is $G_\theta$ (orange solid line), and $|G_\theta(\ell^*)|=1$ at $\ell^*\simeq7.7$.}
	\label{fig:nn2}
\end{figure}
The above SDW-only arguments, however, do not take into account a possibility of a cone-like interaction between more distant chains. 
Even though such interactions are absent from the
lattice Hamiltonian \eqref{eq:H_0}, they can (and will) be generated by quantum fluctuations at low energies, as long as they remain consistent with symmetries
of the lattice model \cite{olegandbalents}. The simplest of such interactions is given by the transverse inter-chain interaction between the 
{\em next-neighbor} (NN) chains ${\cal H}_{\rm NN}$, see Appendix~\ref{app:nn_chain} for the detailed derivation,
\begin{equation}
{\cal H}_{\rm NN} = 2\pi v G_{\theta} \sum_y \int \mathrm{d}x (\tilde{N}_{y}^{+} \tilde{N}_{y+2}^{-}+{\rm h.c.}).
\label{2nd neighbour}
\end{equation}
This is an indirect exchange, mediated by an intermediate chain ($y+1$), and therefore its exchange coupling 
can be estimated as $2\pi v G_{\theta} \sim (J')^2/(2\pi v) \ll J'$.
However the scaling dimension of this term ($\approx 1$ without the magnetic field) is the same as of the original cone interaction ${\cal H}_{\rm cone}$ and thus $G_\theta$ is
expected to grow exponentially fast. Importantly, ${\cal H}_{\rm NN}$ is free of the DM-induced oscillations because DM vectors ${\bm D}$ on chains $y$ and $(y+2)$ 
point in the {\em same} direction. That is, fields $\tilde{\theta}_y$ and $\tilde{\theta}_{y+2}$ {\em co-rotate}. 
This basic physical reason makes ${\cal H}_{\rm NN}$ a legitimate candidate for fluctuation-generated interchain exchange interaction
of the cone kind. Calculation in Appendix~\ref{app:nn_chain} gives the NN coupling constant
\begin{equation}
G_{\theta}=-\frac{\pi A_3^2}{4}f(\Delta_1)\frac{J'}{D}g_{\theta},\; f(\Delta_1)=t_{\theta}^{2\Delta_1-1}\frac{\Gamma(1-\Delta_1)}{\Gamma(\Delta_1)},
\label{eq:gNN}
\end{equation}
which depends on magnetic field via scaling dimension $\Delta_1$. At low fields $\Delta_1\approx 1/2$ and $f(1/2) \approx 1$.
Observe that $G_{\theta}$ describes ferromagnetic interaction and, contrary to naive perturbation theory expectation, has significant
magnitude: $2\pi v G_\theta \propto (J')^2/D \gg (J')^2/J$. RG equation for $G_{\theta}$ coincides with that of $g_\theta$,
\begin{equation}
    \frac{d G_\theta}{d \ell}=G_{\theta}(1-\frac{1}{2}y_z).
    \label{ini3}
\end{equation}
When $G_{\theta}$ reaches strong coupling first, the $\tilde{\theta}$ configuration is uniform, $\tilde{\theta}_y = \tilde{\theta}_{y+2} = \hat{\theta}_{\rm \nu = e/o}$,  
where index $\nu = {\rm e}$ for even $y$ and $\nu = {\rm o}$ for odd $y$ values and in general $\hat{\theta}_{\rm e} \neq \hat{\theta}_{\rm o}$.
At this level of approximation subsystems of {\em even} and {\em odd} chains decouple from each other. The obtained coneNN order is {\em incommensurate},
\begin{eqnarray}
\langle {\bm S}_{x,y} \rangle &&= M {\bf z} + (-1)^{x + y} \Psi_{\rm coneNN} \Big(-\sin[\sqrt{2\pi}\hat{\theta}_\nu + (-1)^y t_\theta x] {\bm x}  \nonumber\\
&&+\cos[\sqrt{2\pi}\hat{\theta}_\nu + (-1)^y t_\theta x] {\bm y}\Big), ~\nu = {\rm e}, {\rm o}.
\label{eq:coneNN}
\end{eqnarray}
The described situation is actually very similar to one discussed in Ref.~\onlinecite{oleg_cmf}, see section IV there, where spins in the neighboring layers are found
to counter-rotate, due to oppositely oriented DM vectors, and are not correlated with each other.

By a simple manipulation this spin ordering can also be represented as 
\begin{eqnarray}
&&\langle {\bm S}_{x,y} \rangle = M {\bf z} + \nonumber\\
&& + (-1)^{x + y} \Psi_{\rm coneNN} \Big( \cos[t_\theta x] \{-\sin[\sqrt{2\pi}\hat{\theta}_\nu] {\bf x} + \cos[\sqrt{2\pi}\hat{\theta}_\nu] {\bf y}\} \nonumber\\
&& - (-1)^y \sin[t_\theta x] \{\cos[\sqrt{2\pi}\hat{\theta}_\nu] {\bf x} + \sin[\sqrt{2\pi}\hat{\theta}_\nu]{\bf y}\} \Big) .
\label{eq:coneNN2}
\end{eqnarray}
Expressions inside curly brackets represent orthogonal unit vectors which are obtained from the orthogonal pair $({\bf x}, {\bf y})$
by the chain-parity dependent rotation by angle $\pm \sqrt{2\pi} \hat{\theta}_\nu$.

The magnitude of the coneNN order parameter is shown in Appendix~\ref{app:order_parameter}, Figure~\ref{fig:orderBr}, for a particular experimentally 
relevant ratio of $J'/J$.

\subsubsection{Competition between SDW and cone/coneNN orders}
Quantitative description of the competition between SDW and cone orders within RG framework represents a very difficult task.
This basically has to do with the fact that RG is not well suited for describing oscillating perturbations such as \eqref{H'} and \eqref{H'1}.
It is quite good at extracting the essential physics of the slow- and fast-oscillation limits, as described in sections \ref{sub:weak_DM_para} and 
\ref{sec:coneNN} above, but is not particularly useful in describing the intermediate regime $D \sim J'$ in which the change from one behavior
to the another takes place (see Ref.~\onlinecite{oshikawa1999} for the example of the RG study of the much simpler problem of a single spin-1/2
chain in the magnetic field).

Applied to the cone-SDW competition, one needs to compare effects due to the DM-induced oscillations with those due to the magnetic field induced ones.
Given that magnetic field makes cone terms more relevant and SDW ones less relevant, one can anticipate that even if the DM interaction is strong enough to destroy the cone phase
in small magnetic field, the cone can still prevail over the SDW phase at higher fields. Chain mean field approximation, described in the next section (and also in more
details in Appendix \ref{app:cmf}) indeed shows that the critical $D/J'$ ratio required for suppressing the cone phase increases with magnetization $M$.
Nonetheless, the ratio $D/J'$ is bounded: there exists sufficiently large $D$ (still of the order $J'$) above which the cone order becomes impossible for any $M$.

For $D$ greater than that we need to 
examine competition between ${\cal H}_{\rm sdw}$ and ${\cal H}_{\rm NN}$.
Approximating $A$ as $1/2$ here (see Ref.~\onlinecite{Essler2003}, transverse normalization factor $A_3$ is close to $1/2$ at small magnetization), 
we observe that  $|G_{\theta}|$ is about $J'/(4D)$ times smaller than $g_z$. 
However, in the presence of magnetic field $G_{\theta}$ becomes more relevant in RG sense (similar to its frustrated `parent' $g_\theta$), and 
grows much faster than SDW interaction $g_z$, which becomes less relevant with magnetic field. 
Therefore there should be a range of $J'/D$ such that $G_{\theta}(\ell)$ can compete with $g_z(\ell)$.
 
Such an example is shown in Fig.~\ref{fig:nn1} and Fig.~\ref{fig:nn2}, $D/J' \sim 1$ there. Fig.~\ref{fig:nn1} shows RG flow in low magnetic field $h_z/D=0.005$, 
when $g_z$ grows faster than $|G_{\theta}|$, resulting in the SDW state. However, in higher magnetic field $h_z/D=5$, which is still rather low in comparison with
$J$, $G_{\theta}$ turns to be the most relevant coupling constant. Hence the ground state changes to the coneNN one.

Details of this competition depend strongly on the magnitude of the magnetic field. At low field $h \leq h_{\rm c-ic}$ SDW is commensurate, while at 
higher field $h \geq h_{\rm c-ic}$ it turns incommensurate. Calculations reported in Appendix~\ref{app:cmf} find that $h_{\rm c-ic} \approx 1.4 J'$
which is sufficiently small value (the corresponding magnetization is very small as well, $M_{\rm c-ic} = h_{\rm c-ic}/(2\pi v) \approx 1.4 J'/(\pi^2 J) \ll 1$) , 
especially in the most interesting to us regime of strong DM, $D \gg J'$. 
Given that the critical temperature of the incommensurate SDW  order is lower than that of the commensurate one, see Fig.~\ref{fig:2sdw}, the SDW-coneNN
competition is most pronounced in the $h \geq h_{\rm c-ic}$ limit, on which we mostly focus in the section~\ref{sec:CMF} below.
 
 \begin{figure}[!tbp]
 	\includegraphics[width=0.8\columnwidth]{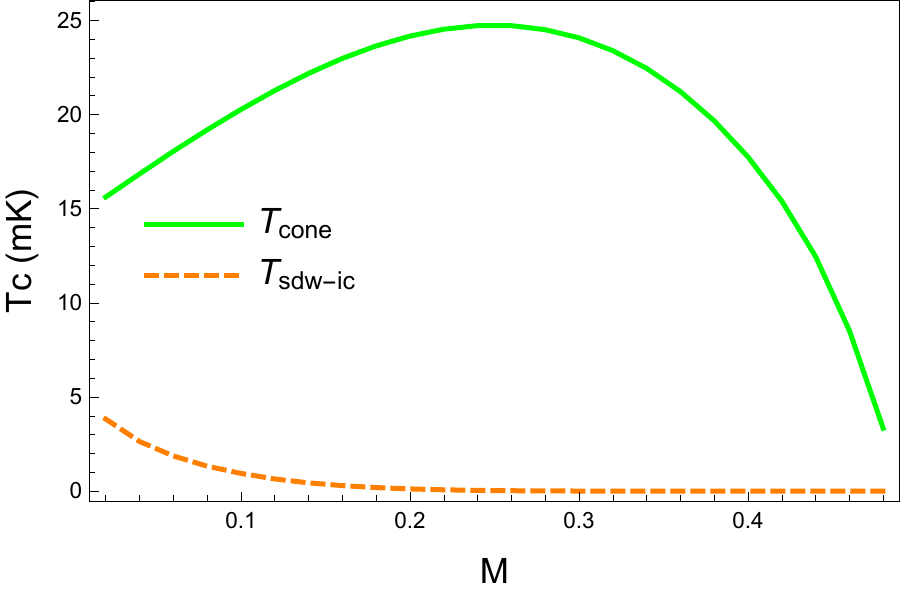}
 	\caption{(Color online) Ordering temperatures of the cone ($T_{\rm cone}$, green solid line) and incommensurate SDW ($T_{\rm sdw-ic}$, orange dashed line) states, vs. magnetization $M$, for the case
 		of  weak DM interaction. $J=1$ K, $J'=0.01$ K and $D=0.01$ K. Commensurate SDW state ($T_{\rm sdw-c}$) is characterized by $T_{\rm sdw-ic} < T_{\rm sdw-c} < T_{\rm cone}$ but
 		is present only is the very narrow magnetization interval $0 < M < M_{\rm c-ic} < 0.01$ and is not shown here.
 		The larger ordering temperature is dominant, thus the ground state is cone in the whole field/magnetization range.}
 	\label{fig:Tc}
 \end{figure}

\section{Chain Mean-field calculation}
\label{sec:CMF}

A more quantitative way to characterize DM-induced competition, described in the previous section with the help of qualitative RG arguments,
 is provided by the 
chain mean-field (CMF) approximation\cite{oleg_cmf} which allows one to calculate and compare critical temperatures for different
magnetic instabilities. The instability with maximal $T_c$ is assumed to describe the actual magnetic order. 
This calculation enables us to directly compare the resulting critical temperature of the dominant  instability to the experimental lambda peak in heat capacity measurements \cite{Halg2014}
and therefore to directly compare experimental and theoretical $h-T$ phase diagrams.
It provides one with a reasonable way to estimate the inter-chain exchange $J'$ of the material, as we describe in Appendix~\ref{app:j'}.
It also allows for a straightforward calculation of the microscopic order parameters, see Appendix~\ref{app:order_parameter} .

In applying CMF to our model, there are three inter-chain interactions in Eqns.~\eqref{H'} and ~\eqref{2nd neighbour} that need to be compared,
\begin{equation}
\begin{gathered}
{\cal H}_{\rm cone}=c_1\int  \mathrm{d}x \cos[\beta(\tilde{\theta}_{y} - \tilde{\theta}_{y+1})+2(-1)^yt_\theta x],\\
{\cal H}_{\rm sdw-ic}=c_2\int  \mathrm{d}x [ \cos\frac{2\pi}{\beta}(\tilde{\phi}_y - \tilde{\phi}_{y+1})],\\
{\cal H}_{\rm NN}=-c_3\int  \mathrm{d}x  \cos[\beta(\tilde{\theta}_{y} - \tilde{\theta}_{y+2})].\\
\end{gathered}
\label{eq:inter-cmf}
\end{equation}
In accordance with the discussion in the end of the previous section \ref{subsec:rg_1} we focus here on the $h \geq h_{\rm c-ic}$ regime and
neglect oscillating term $\tilde{g}_\phi$ in ${\cal H}_{\rm sdw}$. The amplitudes are
\begin{equation}
\begin{gathered} c_1=J'A_3^2$,\quad $c_2=J'A_1^2/2,\\
c_3=\frac{\pi}{4}\frac{J'^2}{D}A_3^4t_{\theta}^{2\Delta_1-1}\frac{\Gamma(1-\Delta_1)}{\Gamma(\Delta_1)}.
\end{gathered}
\label{eq:cmf_inter}
\end{equation}

\begin{figure}[!tbp]
	\includegraphics[width=0.8\columnwidth]{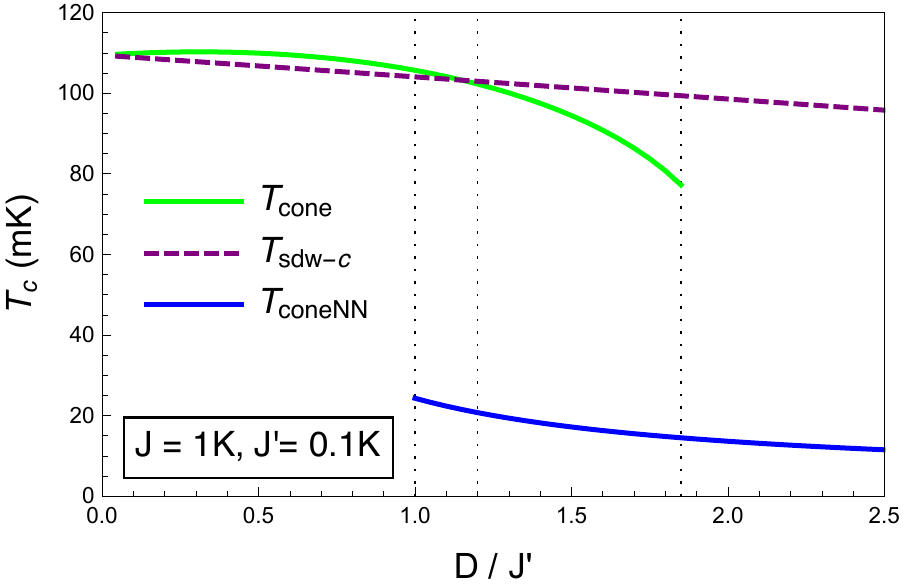}
	\caption{(Color online) Ordering temperatures of the cone state (green solid line), commensurate-SDW (purple dashed line) and coneNN (blue solid line) states as a function of $D/J'$ ratio, and
		in the limit of zero magnetic field, $M=0$. Here, $J=1$ K, $J'=0.1$ K. Note that solution for coneNN state has physical meaning in the limit $D/J'\gg 1$. 
		$T_{\rm sdw-c}$ overcomes $T_{\rm cone}$ at $D/J'\simeq1.2$ and solution for $T_{\rm cone}$ disappears at $D/J'\simeq 1.9$. See Section~\ref{sec:CMF} and Appendix \ref{app:cmf}.}
	\label{fig:three}
\end{figure}

CMF is designed for the analysis of the relevant perturbations and does not account for the marginal interactions, such as Eq.~\eqref{H'1}, directly.
However much of their effects can still be captured by adopting a more precise expression for the staggered magnetization, 
which encodes magnetic field dependence of the scaling dimensions
of transverse and longitudinal components via simple generalization of \eqref{eq:N},
\begin{equation}
{\bm N}_y(x)= (-A_3\sin[\beta\tilde{\theta}_y], A_3\cos[\beta\tilde{\theta}_y], -A_1\sin[\frac{2\pi}{\beta}\tilde{\phi}_y]).
\end{equation}
Here the magnetic field dependence of the scaling dimensions of transverse and longitudinal components of ${\bm N}$ is contained in the 
parameter $\beta =2 \pi R$, which in turn is related to the exactly known ``compactification radius" $R$ in the sine-Gordon (SG) model.
  At zero magnetization $M=h=0$, the SU(2) invariant
Heisenberg chain has $2\pi R^2=1$. In magnetic field, $\beta$ and $R$ decrease toward the limit $2\pi R^2=1/2$ as the chain approaches full polarization.
The amplitudes $A_1$ and $A_3$ have been determined numerically \cite{hikihara}.

Calculation of $T_c$ is standard and well-documented in Ref.~\onlinecite{oleg_cmf}, additional details are provided in Appendix~\ref{app:cmf}.

For weak DM interaction, we compare the ordering temperatures of ${\cal H}_{\rm cone}$ and ${\cal H}_{\rm sdw}$, and the $T_c$ for each state as a function of magnetization $M$ is 
shown in Fig.~\ref{fig:Tc}. For chosen parameters, critical temperature of the cone is always above that of the SDW, therefore the ground state is cone, in agreement with the 
RG analysis in Sec.~\ref{sub:weak_DM_para}. 
As magnetization increases, the transverse correlations are enhanced, and longitudinal ones are suppressed, 
resulting in a greater separation between the two critical temperatures.
At larger magnetization, $T_{\rm cone}$ also decreases, basically due to the Zeeman effect -- spins align more along the direction of the magnetic field, 
thereby reducing the magnitude of the transverse spin component.

\begin{figure}[!t]
	\includegraphics[width=0.8\columnwidth]{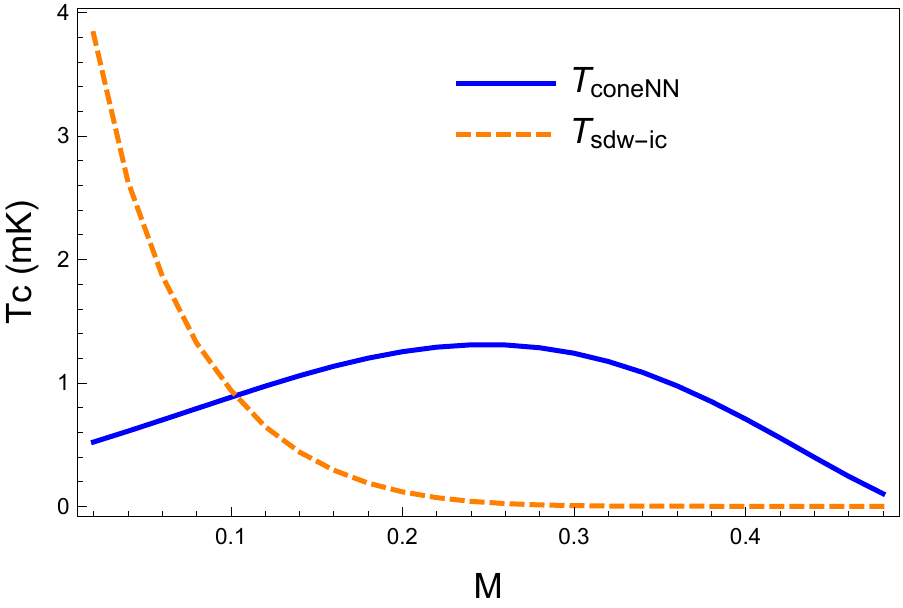}
	\caption{(Color online) Ordering temperatures of the incommensurate-SDW (orange dashed line) and cone2N (blue solid line) states, as a function of magnetization $M$, in the case of strong DM interaction. 
		$J=1$ K, $J'=0.01$ K and $D=0.1$ K. Two lines intersect at small magnetization $M\simeq 0.1$, above which the critical temperature of the cone-NN state overcomes that of the SDW one. }
	\label{fig:Tc_2}
\end{figure}

Increasing DM interaction frustrates ${\cal H}_{\rm cone}$ until, at some critical $D/J'$ value, its mean-field solution disappears completely, signifying the impossibility
of the standard cone state. This feature is described in much details in Appendices~\ref{app:cmf} and \ref{app:cic}.
Figure~\ref{fig:three} illustrates it.

With the cone state out of the picture, we now need 
 to consider the transverse NN-chain coupling ${\cal H}_{\rm NN}$ and its competition with the SDW state as magnetization increases from $0$ to the saturation at $M=0.5$.
 The result is shown in Fig.~\ref{fig:Tc_2}. In a small magnetic field (when $M\approx h/(2\pi v$)), 
 $T_{\rm sdw}$ is above $T_{\rm coneNN}$. As magnetization increases, the scaling dimensions get modified, and the two curves intersect, which indicates 
 a phase transition from the SDW to the cone-NN phase. 
 This result is fully consistent with our qualitative RG analysis in Sec.~\ref{strong_DM_para}.
 
 %
 
\section{Orthogonal configuration, ${\bm h}\perp {\bm D}$}
\label{sec:perpendicular}

When ${\bm h} \perp {\bm D}$, the system Hamiltonian is described by Eq.~\eqref{system_2} with $h_x=h$, and $h_z=0$. 
In order to treat both vector perturbations, $h$ and $D$, equally, we 
perform a chiral rotation of spin currents about the $\hat{y}$ axis,
\begin{equation}
{\bm J}_{y ,R/L}={\cal R}(\theta_{R/L}){\bm M}_{y ,R/L},
\label{chiral rotation}
\end{equation}
where ${\bm M}_{R/L}$ is spin current in the rotated frame, and $\cal R $ is the rotation matrix,
\begin{equation}
\begin{gathered}
{\cal R}(\theta_{R/L})=
\begin{pmatrix}
\cos\theta_{R/L} & 0 & \sin\theta_{R/L} \\
0 & 1 & 0 \\
-\sin\theta_{R/L} & 0 & \cos\theta_{R/L}.
\end{pmatrix},
\end{gathered}
\end{equation}
The general form of chiral rotation angles $\theta_{R/L}$ can be found in references \cite{Garate2010,Gangadharaiah2008}.
Here we apply it to our special ${\bm h}\perp {\bm D}$ case, which gives
\begin{equation}
\begin{gathered}
\theta_R=\frac{\pi}{2}+\theta_0^y,\; \theta_L=\frac{\pi}{2}-\theta_0^y,\; \theta_0^y\equiv(-1)^y \tan^{-1}[\frac{D}{h}].
\label{eq:theta0}
\end{gathered}
\end{equation}
The staggered nature of DM interaction is reflected in the $y$-dependence of the rotation angle $\theta_{R/L}$, via that of $\theta^y_0$, here 
(similar to $t^y_\theta$ and $\alpha_y$). The rotation does not affect ${\cal H}_0$ in Eq.\eqref{system_2} but transforms $\cal V$ into
\begin{equation}
\begin{split}
{\cal V}&=-\sqrt{D^2+h^2}\int \mathrm{d}x (M_{y, R}^z+M_{y, L}^z)\\
&=-\frac{\sqrt{D^2+h^2}}{\sqrt{2\pi}}\int \mathrm{d}x ~\partial_x\varphi_y.
\end{split}
\label{eq:effV}
\end{equation}
Here and below abelian fields in the rotated frame are denoted as $(\varphi_y,\vartheta_y)$ and spin current ${\bm M}_{R/L}$ is expressed 
in terms of them in the same way as ${\bm J}_{R/L}$ is in terms of original pair $(\phi_y,\theta_y)$ used for the $\bm h\parallel \bm D$ configuration in Sec.~\ref{sec:parallel}.

We see that in the rotated frame the spins are subject to an effective magnetic field $h_{\rm eff} = \sqrt{D^2+h^2}$ along $z$ axis.
The fact that $D$ and $h$ terms are treated equally here represents the major technical advantage of the chiral rotation transformation \eqref{chiral rotation}.
Importantly, $h_{\rm eff}$ is {\em finite} once $D\neq 0$, implying the presence of some oscillating terms in the Hamiltonian even in the absence
of external magnetic field. 
Being linear in derivative of $\varphi_y$, the term \eqref{eq:effV} is easily absorbed into ${\cal H}_0$, similar to what was done in \eqref{shift1}.
The parameters of this shift are
\begin{equation}
\begin{gathered}
t_{\varphi}=\frac{\sqrt{D^2+h^2}}{v} = \frac{h_{\rm eff}}{v},\quad t_{\vartheta}=0.
\end{gathered}
\label{s2}
\end{equation}
Observe that no shift of $\vartheta$ is required here.
The chiral rotation also transforms expressions for backscattering and inter-chain interactions, which we analyze next.

\begin{figure}[!tbp]
  \includegraphics[width=0.8\columnwidth]{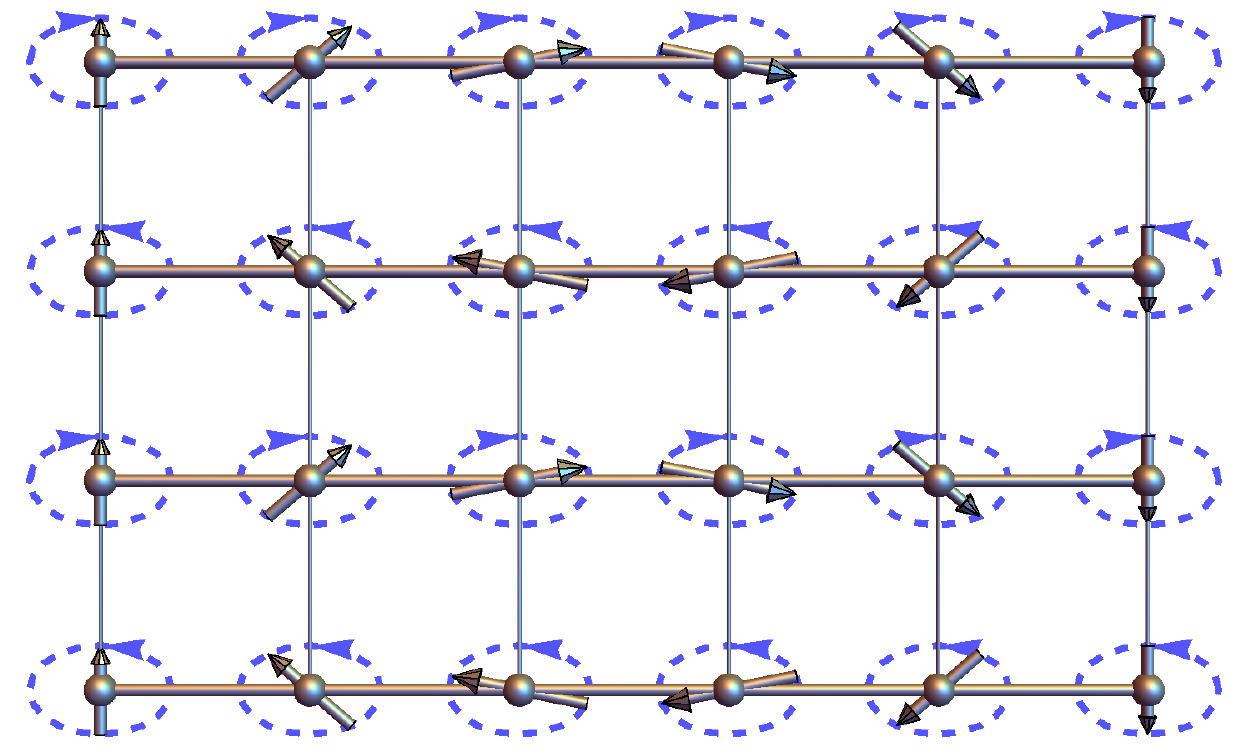}
   \caption{(Color online) Staggered magnetization in the \textit{distorted-cone} phase, Eq.\eqref{eq:dist-cone}, in the transverse to $\bm D$ plane. 
   This distortion is caused by magnetic field, the stronger the field the bigger the distortion. The opposite sense of spin precession in the neighboring chains is due to the staggered DM interaction.}
   \label{fig:distorted_cone}
\end{figure}
\subsection{Backscattering ${\cal H}_{\rm bs}$ }
\label{subsec:bs_perp}

Rotation \eqref{chiral rotation} of spin currents transforms backscattering Hamiltonian in \eqref{system_2} into
\begin{eqnarray}
{\cal H}_{\rm bs}&=&2\pi v\int \mathrm{d}x \Big[\sum_a y_a M_{y,R}^{a}M_{y,L}^{a} + \nonumber\\
&&+ y_A (M_{y,R}^z M_{y,L}^x-M_{y,R}^x M_{y,L}^z)\Big],
\end{eqnarray}
where $a=x,y,z$ and,
\begin{equation}
\begin{gathered}
y_x(0)=-\frac{g_{bs}}{2\pi v}\big[(1+\frac{\lambda}{2})\cos[2\theta_0^y] +\frac{\lambda}{2}\big],\\
 y_y(0)=-\frac{g_{bs}}{2\pi v},\\
y_z(0)=-\frac{g_{bs}}{2\pi v}\big[(1+\frac{\lambda}{2})\cos[2\theta_0^y] -\frac{\lambda}{2}\big],\\
y_A(0)=\frac{g_{bs}}{2\pi v}(1+\frac{\lambda}{2})\sin[2\theta_0^y].\qquad
\end{gathered}
\label{initial}
\end{equation}
Here $2\theta_0^y =  \theta_R-\theta_L$, see \eqref{eq:theta0}.
The subsequent shift of $\varphi$, which eliminates linear term \eqref{eq:effV}, 
\begin{equation}
 {\varphi_y}\to\varphi_y+\frac{t_{\varphi}}{\sqrt{2\pi}}x, 
 \label{eq:perp-shift}
 \end{equation}
produces the end result 
\begin{eqnarray}
&&{\cal H}_{\rm bs}\to {\cal H}_A+{\cal H}_B+{\cal H}_C+{\cal H}_{\sigma}, \\
&&{\cal H}_{A}=\pi v y_A\int \mathrm{d}x (M_{y,R}^{z}M_{y,L}^{+}e^{it_{\varphi}x}-M_{y,R}^{+}M_{y,L}^{z}e^{-it_{\varphi}x}+{\rm h.c.}),\nonumber\\
&&{\cal H}_{B}=\pi v y_B\int \mathrm{d}x(M_{y,R}^{+}M_{y,L}^{-}e^{-i2t_{\varphi}x}+{\rm h.c.}), \nonumber\\
&&{\cal H}_{C}=\pi v y_C\int \mathrm{d}x (M_{y,R}^{+}M_{y,L}^{+}+{\rm h.c.}), \nonumber\\
&&{\cal H}_{\sigma}=-2\pi v y_{\sigma}\int \mathrm{d}x M_{y,R}^{z}M_{y,L}^{z}, \nonumber
\label{eq:perp-BS}
\end{eqnarray}
where
\begin{equation}
y_C\equiv\frac{1}{2}(y_x-y_y), \;y_B\equiv\frac{1}{2}(y_x+y_y),\; y_{\sigma}\equiv-y_z.
\end{equation}

\begin{table}[!tbp]
\begin{center}
		{\renewcommand{\arraystretch}{1.4}%
\begin{tabular}{c| c c c}
 \hline
 \hline
Interaction & Coupling & Coupling  & Induced\\
term        & operator & constant  & state\\
 \hline
${\cal H}_{x}$ & $\quad$${\cal N}^x_{y} {\cal N}^x_{y+1} $$\quad$ & $\quad$$g_{x}$$\quad$ & SDW(z) \\[0.8ex]
${\cal H}_{y}$ & ${\cal N}^y_y {\cal N}^y_{y+1}$ & $g_{y}$ & SDW (y)  \\  [0.8ex]
${\cal H}_{\rm inter,\varphi}$ & $\cos[\sqrt{2\pi}(\varphi_{y}-\varphi_{y+1})]$ & $g_{\varphi_{1}}$ & \text{Distorted-cone} \\[0.8ex]
\hline
\hline
\end{tabular}}
\end{center}
\caption{When $\bm h\perp \bm D $, three relevant interchain interactions are ${\cal H}_{x}\propto {\cal N}^x_y{\cal N}^x_{y+1}$ , ${\cal H}_y\propto {\cal N}^y_y{\cal N}^y_{y+1}$ 
and ${\cal H}_{\rm inter,\varphi}$ in Hamiltonian~\eqref{eq:perp-inter} and \eqref{eq:perp-varphi}. The Table shows their operator forms
in the {\em rotated} frame, associated coupling constants and the ordered states they induce.}
\label{table:term2}
\end{table}

\subsection{Interchain interaction ${\cal H}_{\rm inter}$}
\label{subsec:inter_perp}
Under the rotation \eqref{chiral rotation} the staggered magnetization ${\bm N}_y$ in the original frame transforms, in terms of that in the rotated frame, ${\bm {\mathcal N}}_y$, as follows
\begin{equation}
{\bm N}_y(x)= ({\cal N}^{z}_y, \cos\theta_0^y{\cal N}^{y}_y +\sin\theta_0^y{\varepsilon}_y, -{\cal N}^{x}_y),
\label{eq:Nrotation}
\end{equation}
where 
\begin{equation}
\varepsilon_y=\frac{\gamma}{\pi a}\cos[\sqrt{2\pi}\varphi_y + t_\varphi x],
\end{equation}
is the {\em dimerization} operator in the rotated frame (while $\xi_y=\frac{\gamma}{\pi a}\cos[\sqrt{2\pi}\phi_y]$ is the dimerization in the original frame, see Appendix \ref{app:ope}). Observe that due to \eqref{eq:theta0} $\sin\theta_0^y$ actually oscillates in sign with the chain index $y$.
According to \eqref{eq:N}, 
\begin{equation}
{\bm {\mathcal N}}_y \propto (- \sin[\sqrt{2\pi}\vartheta_y], \cos[\sqrt{2\pi}\vartheta_y], -\sin[\sqrt{2\pi}\varphi_y + t_\varphi x]),
\label{eq:calN}
\end{equation}
where oscillatory $x$-dependence of ${\cal N}^{z}_y$ follows from the shift \eqref{eq:perp-shift}.
Relation \eqref{eq:Nrotation} can be obtained by connecting chiral rotation \eqref{chiral rotation} 
to the spinor rotation of Dirac fermions $\Psi_{R/L, s}$ ($s$ is the spin index) which are related to the spin current via, e.g., 
${\bm J}_R \sim \Psi^\dagger_{R, s} {\bm \sigma}_{s,s'} \Psi_{R, s'}$. The staggered magnetization is expressed in terms of these as
${\bm N} \sim \Psi^\dagger_{R, s} {\bm \sigma}_{s,s'} \Psi_{L, s'} + ({\rm L} \leftrightarrow {\rm R})$. 
Rotation of spinors $\Psi_{R/L}$ leads to \eqref{eq:Nrotation}.

Inter-chain interaction in terms of rotated operators reads
\begin{equation}
{\cal H}_{\rm inter}=2\pi v\sum_y \int \mathrm{d}x \Big[\sum_{a}g_a {\cal N}_y^a{\cal N}_{y+1}^a+g_E\varepsilon_{y}\varepsilon_{y+1}\Big].
\label{eq:perp-inter}
\end{equation}
The interchain couplings are
\begin{equation}
\begin{gathered}
g_x(0)=\frac{J'}{2\pi v}, \quad g_y(0)=\frac{J'}{2\pi v}\cos^2\theta_0^y,\\
g_z(0)=\frac{J'}{2\pi v}, \quad
g_E(0)=-\frac{J'}{2\pi v}\sin^2\theta_0^y,\\
\end{gathered}
\label{eq:ini_inter1}
\end{equation}
Two terms in \eqref{eq:perp-inter}, namely $g_z$ and $g_E$ ones, are expressed in terms of $\varphi$ field and therefore 
contain oscillating with position $x$ parts. In order to keep the presentation simple, we refrain here from
writing this dependence out explicitly. Beyond the oscillating RG scale $\ell_\varphi = -\ln[a_0 t_\varphi]$, introduced in Section~\ref{subsec:2steprg} below,
these two terms combine into 
\begin{equation}
\begin{gathered}
{\cal H}_{\rm inter,\varphi}=2\pi vA^2 \sum_y \int \mathrm{d}x ~g_{\varphi_1}\cos[\sqrt{2\pi}(\varphi_{y}-\varphi_{y+1})],\\
g_{\varphi_1}\equiv\frac{1}{2}(g_E+g_z), \quad g_{\varphi_1}(0)=\frac{J'}{4\pi v}\cos^2\theta_0^y.
\label{eq:perp-varphi}
\end{gathered}
\end{equation}
Interchain interactions \eqref{eq:perp-inter} (terms with $g_{x/y}$) and \eqref{eq:perp-varphi} are the most relevant perturbations.
Three parts of the inter-chain Hamiltonian (namely $g_x$, $g_y$ and $g_{\varphi_1}$ terms) and the ordered states they induce are summarized in Table~\ref{table:term2}.

As discussed previously, Eq.~\eqref{eq:effV}, as well as its consequence, Eq.\eqref{eq:perp-varphi}, 
implies an effective magnetic field along $z$ in the rotated frame. Recalling the effect of the magnetic field on the scaling 
dimensions of various operators, which was discussed in Sec.~\ref{sec:parallel} and ~\ref{sec:CMF}, 
we must conclude that this magnetic field will suppress the longitudinal ordering  
and enhance transverse ones. Therefore we expect $g_{x,y}$ terms in \eqref{eq:perp-inter} to be more relevant than 
$g_{\varphi_1}$ one.

\begin{table}
 \begin{center}
 	\centering
 	{\renewcommand{\arraystretch}{1.2}%
   \begin{tabular}{c|c|c|c|c|c}
\hline\hline
Region & I &  II  & III & IV  & V \\ 
\hline
$y_{C}(0)$ & $\quad+\quad$ & $\;\;+\;\;$ & $\;\;+\;\;$ & $\;\;+\;\;$ & $\quad-\quad$ \\
\hline
$y_{\sigma}(0)$ & $-$ & $-$ & $+$ & $+$ & $+$ \\
\hline
 C & $+$ & $-$ & $-$ & $+$ & $+$ \\
 \hline
Fastest& \multirow{2}{*}{$g_{\varphi_1}$ }& \multicolumn{3}{c|}{\multirow{2}{*}{$g_x$}}& \multirow{2}{*}{$g_y$} \\
growing& &\multicolumn{3}{c|}{} & \\
 \hline\hline
\end{tabular}}
 \caption{Signs of $y_C$, $y_{\sigma},C$ in different field regions for intermediate value of $\lambda$ of order $0.1$. 
 This table summarizes conditions the fastest growing coupling constant in RG system \eqref{eq:RG_tot3}.}
 \label{table:c}
\end{center}
\end{table}
\subsection{Two stage RG\cite{Garate2010, Gogolin}}
\label{subsec:2steprg}
RG flow of backscattering Hamiltonian \eqref{eq:perp-BS} is given by
\begin{equation}
\begin{gathered}
\frac{dy_x}{dl}=y_y y_z,\quad
\frac{dy_y}{dl}=y_x y_z+y_A^2,\\
\frac{dy_z}{dl}=y_x y_y,\quad
\frac{dy_A}{dl}=y_y y_A.\\
\end{gathered}
\label{eq:RG_bs}
\end{equation}
The interchain interaction \eqref{eq:perp-inter} changes as
\begin{equation}
\begin{split}
&\frac{dg_x}{dl}=g_x[1+\frac{1}{2}(y_x-y_y-y_z)],\\
&\frac{dg_y}{dl}=g_y[1+\frac{1}{2}(y_y-y_z-y_x)],\\
&\frac{dg_z}{dl}=g_z[1+\frac{1}{2}(y_z-y_x-y_y)],\\
&\frac{dg_E}{dl}=g_E[1+\frac{1}{2}(y_x+y_y+y_z)].\\
\end{split}
\label{eq:RG-inter}
\end{equation}
Similar to discussion around Eq.~\eqref{length1} for the $\bm h\parallel\bm D$ case, here too magnetic field induced oscillations $e^{i t_\varphi x}$ become
prominent beyond the RG scale
\begin{equation}
l_{\varphi}=-\log({a_0t_\varphi}).
\end{equation}

We find that for sufficiently strong DM interaction, approximately $D/J'>0.01$, the oscillating scale is shorter than the interchain one, $l_{\varphi} < l_{\rm inter}$. This means that the RG flow consists of two stages, $0 < l < l_{\varphi}$ and $l_{\varphi} < l <l_{\rm inter}$. During the first stage, $0 < l < l_{\varphi}$, full set of RG equations 
\eqref{eq:RG_bs} and \eqref{eq:RG-inter} needs to be analyzed. At this stage all of the couplings remain small.
During the second stage, for $l > l_{\varphi}$, strong oscillations in ${\cal H}_A$, ${\cal H}_B$, see \eqref{eq:perp-BS}, and in the `oscillating part' of \eqref{eq:perp-inter}
lead to the disappearance of these terms. 
\begin{figure}[t]
	\includegraphics[width=0.8\columnwidth]{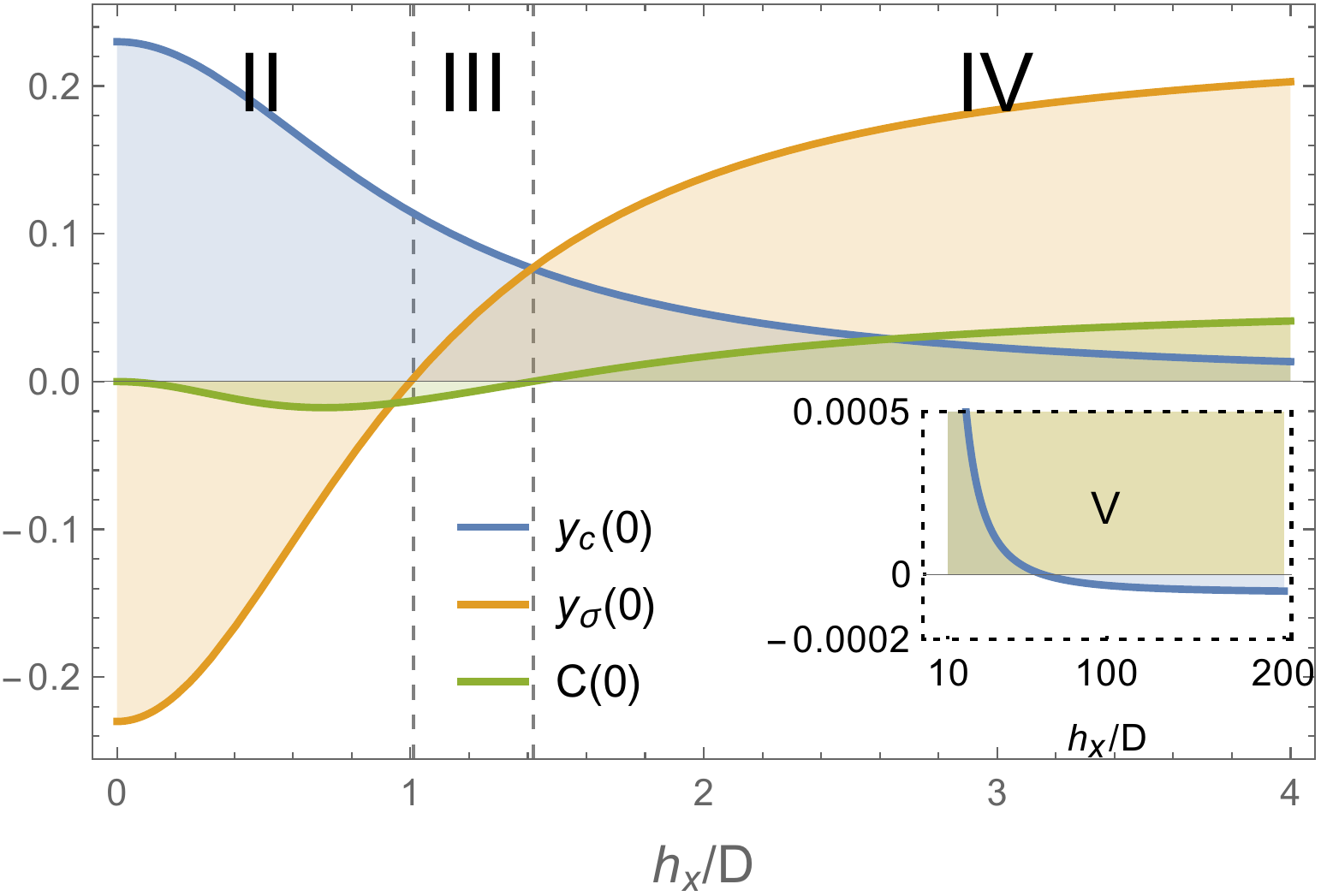}
	\caption{(Color online) $y_C(0)/\eta$, $y_{\sigma}(0)/\eta$ and $C/\eta$ in Eq.\eqref{yyc} as a function of the ratio $h_x/D$. Here we denote $\eta = {\cal G}_{\rm bs}/(2\pi v)$. 
		$\lambda=1\times10^{-4}$, and $D/J=\sqrt{\lambda/c'}\sim0.005$. Here only region II, III, IV in Table \ref{table:c} are present in low magnetic field. 
		The inset shows region V appearing when the ratio $h_x/D$ increases to about $50$, which indicates a phase transition from SDW(z) to SDW(y).}
	\label{fig:c1}
\end{figure}
Setting $y_A(l)=0$ and $y_B(l)=0$ reduces backscattering RG  to the Kosterlitz-Thouless (KT) equations
\begin{equation}
\frac{dy_C}{dl}=y_Cy_\sigma,\;\;\;
\frac{dy_\sigma}{dl}=y_C^2,\\
\label{eq:RG_tot2}
\end{equation}
analytic solution of which is illustrated in Fig.~\ref{fig:kt}. 
At the same time, interchain RG reduces to
\begin{equation}
\begin{gathered}
\frac{dg_x}{dl}=g_x(1+y_C+\frac{1}{2}y_{\sigma}),\\
\frac{dg_y}{dl}=g_y(1-y_C+\frac{1}{2}y_{\sigma}),\\
\frac{dg_{\varphi_1}}{dl}=g_{\varphi_1}(1-\frac{1}{2}y_{\sigma}).
\end{gathered}
\label{eq:RG_tot3}
\end{equation} 
Initial conditions for $y_C$, $y_{\sigma}$ and $g_{\varphi1}$ at the start of the 2nd RG stage are
\begin{equation}
\begin{gathered}
y_C(l_{\varphi})=\frac{1}{2}[y_x(l_{\varphi})-y_y(l_{\varphi})],\quad 
y_\sigma(l_{\varphi})=-y_z(l_{\varphi}),\\
g_{\varphi_1}(l_{\varphi})=\frac{1}{2}[g_E(l_{\varphi})+g_z(l_{\varphi})].\\
\end{gathered}
\end{equation}

\subsection{Types of two-dimensional order}
\label{sec:perp-phases}
In $\bm h\perp\bm D$ configuration, three competing interchain interactions $g_{x,y,\varphi_1}$ 
lead to three kinds of two-dimensional magnetic orders.
When $g_{x}$ (or $g_y$) is the most relevant coupling, one needs to minimize ${\cal N}^x_y {\cal N}^x_{y+1}$ (or ${\cal N}^y_y {\cal N}^y_{y+1}$), correspondingly.
It is clear that in both cases the appropriate component of $\bm {\mathcal N}$should be staggered as $(-1)^y$ between chains. In terms of $\vartheta_y$, this 
order is described by a simple $\vartheta_y = \sqrt{\pi/2} (y+1/2)$ (correspondingly, $\vartheta_y = \sqrt{\pi/2} y$) in the case of $g_x$ (correspondingly, $g_y$) relevance.
The resulting spin ordering is of commensurate SDW kind, which, according to \eqref{eq:Nrotation}, can be more informatively described as SDW(z) (correspondingly, SDW(y)) order
when the coupling $g_x$ (correspondingly, $g_y$) is the most relevant one:
\begin{eqnarray}
\begin{gathered}
\langle {\bm S}_{x,y} \rangle \sim M {\bf x} + (-1)^{x+y} \Psi_{\rm sdw(z)} {\bf z},\\
\langle {\bm S}_{x,y} \rangle \sim M {\bf x} + (-1)^{x+y} \frac{h}{\sqrt{h^2 + D^2}} \Psi_{\rm sdw(y)} {\bf y} .
\end{gathered}
\label{eq:SDWzy}
\end{eqnarray}
Note that uniform magnetization is along the direction of the external magnetic field $h_x$, see \eqref{system_2}, while the antiferromagnetically ordered component
is orthogonal to it. As noted at the end of section~\ref{subsec:inter_perp}, in the rotated frame effective field $h_{\rm eff}$ makes $g_{x,y}$ inter-chain interactions more relevant 
by reducing their scaling dimensions. 
Therefore, we expect that the critical temperatures of SDW(z) and SDW(y) orders will vary with magnetization $M$ similarly to that of the cone and coneNN phases,
see for example $T_{\rm coneNN}(M)$ in Fig.~\ref{fig:Tc_2}, which is indeed in semi-quantitative agreement with the  experiment\cite{Halg'sphdthesis}. 
Correspondingly, the magnetization dependence of the orders parameters $\Psi_{\rm sdw(z,y)}$ in \eqref{eq:SDWzy}, for a fixed $J'/J$, should look similar to that of cone and coneNN orders
in Appendix~\ref{app:order_parameter}.

\begin{figure}[!tbp]
	\includegraphics[width=0.8\columnwidth]{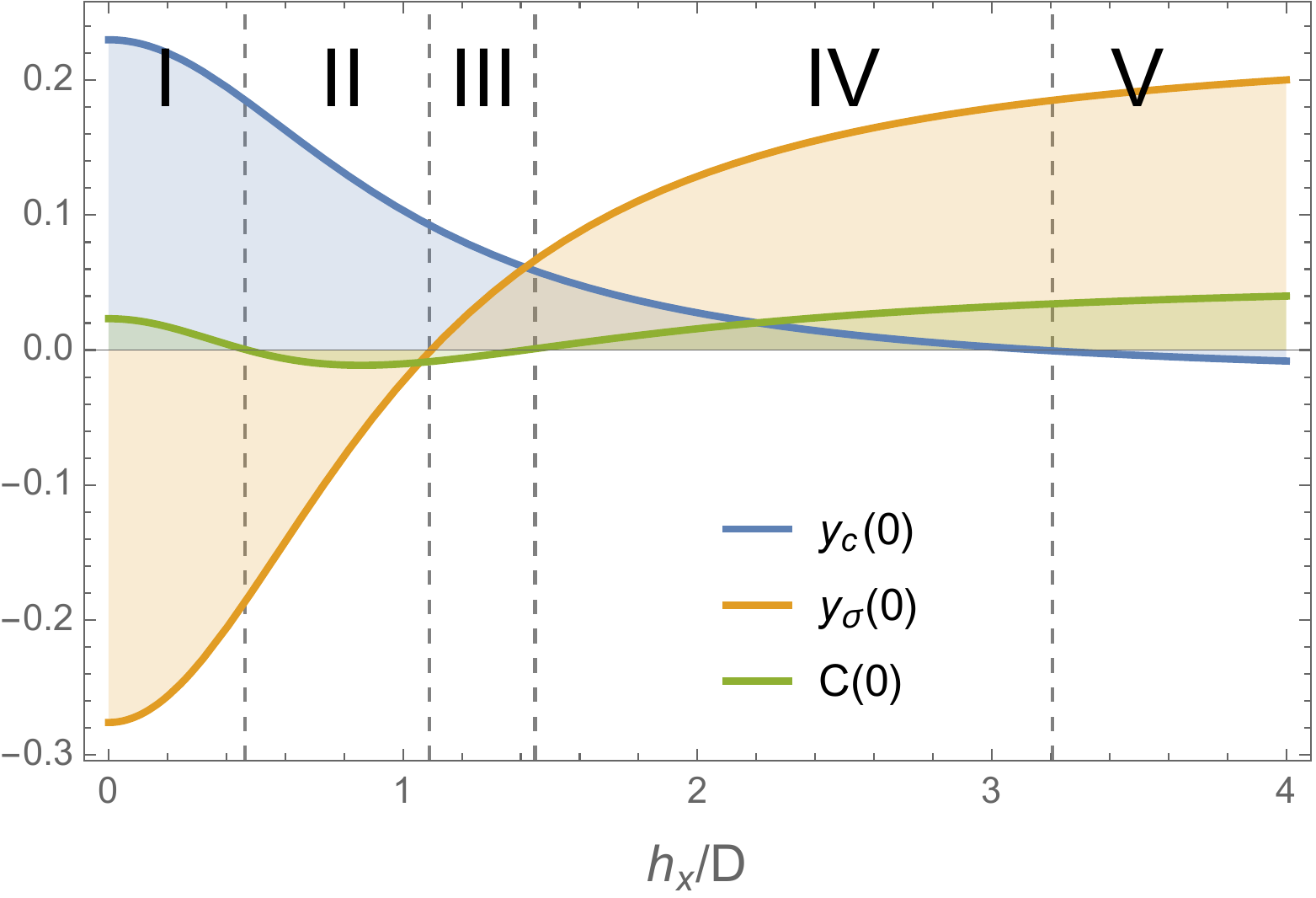}
	\caption{(Color online) Plot of $y_C(0)/\eta$, $y_{\sigma}(0)/\eta$ and $C/\eta$ in Eq.\eqref{yyc} versus the ratio $h_x/D$. Here
		$\eta = {\cal G}_{\rm bs}/(2\pi v)$. $\lambda=0.2$, and $D/J=\sqrt{\lambda /c'}\sim 0.23$. Here all five distinct regions from Table \ref{table:c} are present.}
	\label{fig:c2}
\end{figure}

When the most relevant coupling is $g_{\varphi_1}$, minimization of \eqref{eq:perp-varphi} leads to $\varphi_y = \sqrt{\pi/2} y + \hat{\varphi}$
so that the spin order is given by the incommensurate \textit{ distorted-cone} in the ${\bf x}-{\bf y}$ plane
\begin{eqnarray}
\langle {\bm S}_{x,y}  \rangle && \sim M {\bf x} + (-1)^{x+y} \Psi_{\rm dist-cone} \Big(\sin[\sqrt{2\pi} \hat{\varphi} + t_\varphi x] {\bf x} \nonumber\\
&& - \frac{(-1)^y D}{\sqrt{h^2 + D^2}} \cos[\sqrt{2\pi} \hat{\varphi} + t_\varphi x] {\bf y}\Big) .
\label{eq:dist-cone}
\end{eqnarray}
$N^{x/y}$ components of the staggered magnetization form an ellipse. We used \eqref{eq:theta0} in deriving this expression.
Notice that the spin pattern \eqref{eq:dist-cone} represents a rotated, by the chain-dependent angle, and then elliptically distorted version of the coneNN state \eqref{eq:coneNN2}.

\begin{figure}[!tbp]
	\includegraphics[width=0.8\columnwidth]{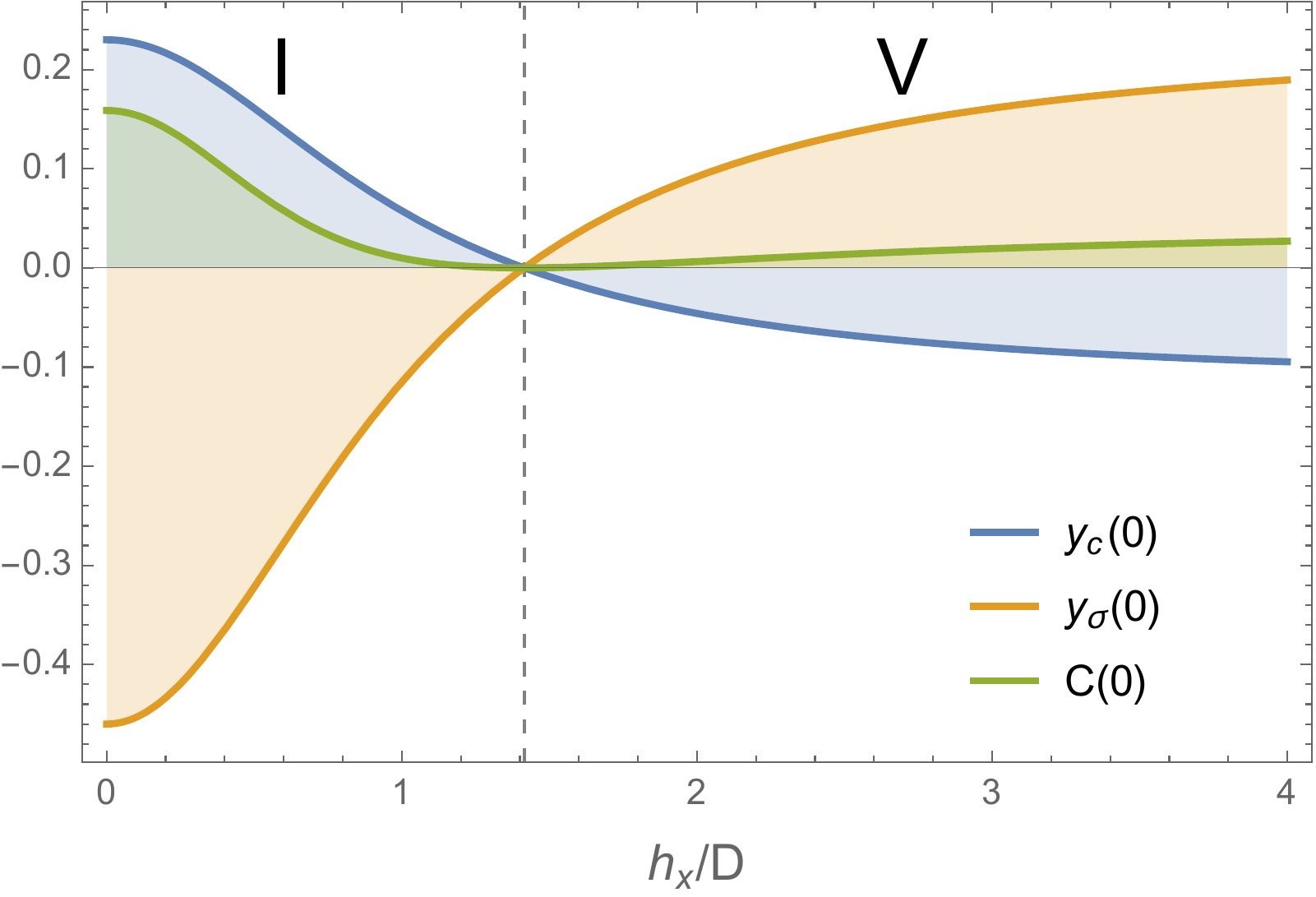}
	\caption{(Color online) $y_C(0)/\eta$, $y_{\sigma}(0)/\eta$ and $C/\eta$ in Eq.\eqref{yyc}  versus the ratio $h_x/D$,  and $\eta = {\cal G}_{\rm bs}/(2\pi v)$. 
	$\lambda=1$, and $D/J=\sqrt{\lambda/c'}\sim0.5$. Here region I and V from Table \ref{table:c} are present.}
	\label{fig:c3}
\end{figure}

\subsection{Distinguishing the most relevant interaction}
The above Eq.~\eqref{eq:RG_tot3} shows that the flow of inter-chain interactions is controlled by the signs of marginal couplings $y_C$ and $y_{\sigma}$, and their relative magnitude, 
which are determined by the initial condition in Eq.~\eqref{initial} as well as by their subsequent 1st stage flow. 
Given that DM-induced anisotropy $\lambda$ is very small, the effect of the 1st stage RG flow reduces to the overall renormalization
of the value of $g_{\rm bs}$. This really is a direct consequence of the assumed near-SU(2) symmetry of the backscattering Hamiltonian \eqref{eq:perp-BS},
which, in the absence of the field $h_{\rm eff}$ (which is the essence of the 1st stage RG where oscillating factors do not play any role, therefore $e^{i t_\varphi x} \to 1$), is just a rotated version 
of  the marginally-irrelevant interaction of spin currents $g_{\rm bs} {\bm J}_R \cdot {\bm J}_L$. Therefore the main effect of the 1st stage consists in the renormalization
$g_{\rm bs}(0) \to {\cal G}_{\rm bs} \equiv g_{\rm bs}(0)/(1-g_{\rm bs}(0) l_\varphi/(2\pi v))$, see Ref.~\onlinecite{Schnyder2008} for the discussion of 
a similar situation.

Thus,  initial values of backscattering couplings for the 2nd stage of the RG are
\begin{equation}
\begin{gathered}
y_C(l_\varphi)=-\frac{{\cal G}_{\rm bs}}{4\pi v}\big[(1+\frac{\lambda}{2})\cos[2\theta_0^y] -1+\frac{\lambda}{2}\big],\\
y_{\sigma}(l_\varphi)=\frac{{\cal G}_{\rm bs}}{2\pi v}\big[(1+\frac{\lambda}{2})\cos[2\theta_0^y] -\frac{\lambda}{2}\big],\\
C=y_{\sigma}(l)^2-y_C(l)^2=y_{\sigma}(l_\varphi)^2-y_C(l_\varphi)^2,\\
\end{gathered}
\label{yyc}
\end{equation}
Finite $h_{\rm eff}$ \eqref{s2} breaks spin-rotational symmetry 
and forces couplings $y_{C,\sigma}$ off the marginal diagonal directions in Fig.~\ref{fig:kt}.
Note that situations with significant $\lambda \sim O(1)$, such as shown in generalized phase diagrams in Fig.~\ref{fig:boundaryplot}, 
requires separate analysis with explicit numerical solution of the 1st stage equations \eqref{eq:RG_bs}.

Noting that $\cos[2\theta_0^y]=(h^2 - D^2)/(h^2 + D^2)$, we have identified 5 distinct regions with different signs of $y_{C,\sigma}$ and
integration constant $C$, which lead to different RG flows. The boundaries of these regions depend on $h/D$ and $\lambda$.
Expression for $C$ is approximated to $O(\lambda)$ accuracy because $\lambda \sim (D/J)^2\ll 1$.
The results are summarized in Table~\ref{table:c} which shows which interchain orders are promoted in different regions.
Several examples of $y_C(0)$, $y_{\sigma}(0)$, and $C$ vs. $h/D$, for three different values of $\lambda$, are shown as Fig.~\ref{fig:c1}, ~\ref{fig:c2}, ~\ref{fig:c3}.

Practically, $\lambda \sim 10^{-4}$ is very small, like in Fig.~\ref{fig:c1}. In low magnetic field one observes regions II, III and IV, all of which result in the two-dimensional 
commensurate SDW order along DM vector ($\hat{z}$). At large $h/D$ values ($ >50$, see the inset in the same figure), the region V appears, leading to 
a commensurate SDW order along $\hat{y}$-axis, orthogonal to the DM vector. This indicates a spin-flop phase transition where spins change their direction suddenly.
The actual value of the corresponding critical magnetic field $h_{\rm flop}$ does not have to be very high, and is experimentally accessible for most material. 
For instance, for $D=0.01J$ we get $h_{\rm flop}\sim50 D=0.5J$. 

\begin{figure}[t]
	\includegraphics[width=0.8\columnwidth]{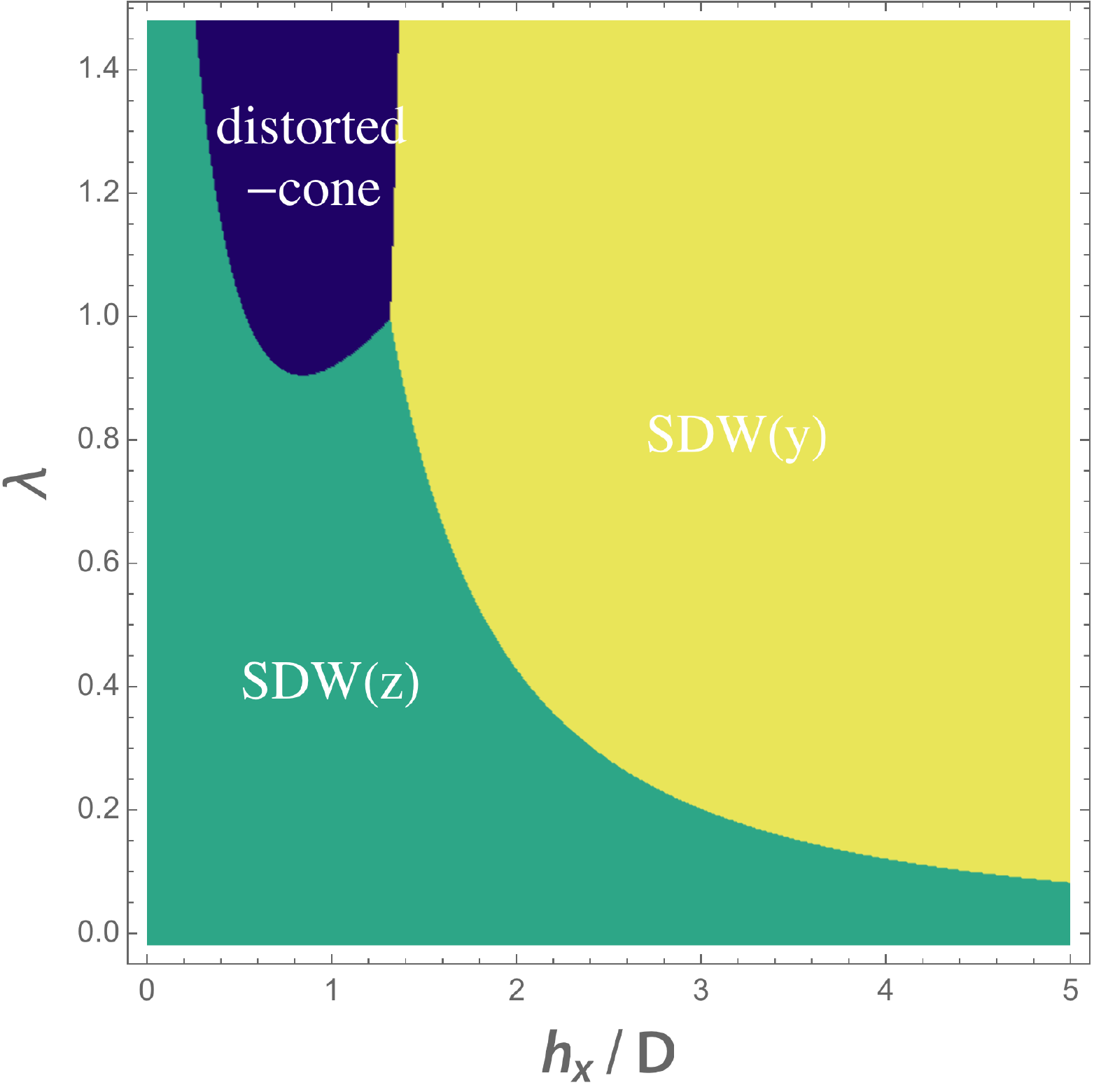}
	\caption{(Color online) Phase diagram for the case of ${\bm h} \perp {\bm D}$, $h_z=0$. 
		Here $g_{\rm bs}=0.23\times 2\pi v$, $J'=10^{-3}\times  2\pi v$ and $D=0.01J$. We vary $\lambda$ and $h_x$, and treat $\lambda$ as independent from $D$ parameter.
		At large $\lambda$ there is a phase transition from the distorted-cone to SDW(y) state. At small $\lambda$ the SDW(z) and SDW(y) phases are separated by the
		transition line which approaches $\lambda=0$ as $h_x/D\to \infty$. }
	\label{fig:pd2}
\end{figure}

In Fig.~\ref{fig:c2}, all 5 different regions are present, and we expect two phase transitions to be present. As magnetic field increases from zero the system transits from the distorted-cone
to the SDW(z), and then to the SDW(y). However, small initial value of $g_{\varphi_1}\propto\cos^2[\theta^y_0] \sim h^2/D^2$ at low field prevents it from reaching strong coupling limit.
Instead, coupling $g_x$ gets there first. As a result, the distorted-cone phase is not realized at low magnetic field. This feature of the RG flow is evident in the phase 
diagrams in Fig.~\ref{fig:pd2} and Fig.~\ref{fig:boundaryplot}, in which the distorted-cone state is present only in the strong DM limit of $D \sim O(1)$.
We therefore conclude that the distorted-cone phase is unlikely to realize in real materials with small $D/J$ ratio.

\subsection{Phase diagram}
\label{subsec:phase_diag_perp}

The ground state of the two-dimensional system is determined by the fastest growing coupling constant of \eqref{eq:RG_tot3}.
For $\lambda$ not vanishingly small (practically, for $\lambda > 0.01$) we numerically solve both the 1st step, Eq.~\eqref{eq:RG_bs},~\eqref{eq:RG-inter},
and the 2nd step, Eq.~\eqref{eq:RG_tot2} and ~\eqref{eq:RG_tot3}, RG equations. The $\lambda-h/D$ phase diagram is shown in Fig.~\ref{fig:pd2}. For small $\lambda$, which for a moment is treated as an independent parameter, there is a phase transition from 
SDW(z) to SDW(y) at large ratio of $h_x/D$, and the line separating the two states tends to be horizontal as $h_x/D\to \infty$. 
The distorted-cone state appears only at unrealistically large $\lambda$. It transforms to SDW(y) at $h_x/D\simeq 1.5$, 
for any $\lambda > 1$. This can be understood from Eq.~\eqref{yyc} and Table~\ref{table:c}: in order to change the sign of $y_C(0)$ and 
$y_\sigma(0)$ at the same time, one needs $1+\lambda>2/\lambda$, which implies $\lambda>1$. 
The distorted-cone-SDW(y) transition is of incommensurate-commensurate kind in agreement with the classical analysis prediction in Ref.~\onlinecite{Garate2010}.

\begin{figure}[!tbp]
	\includegraphics[width=0.8\columnwidth]{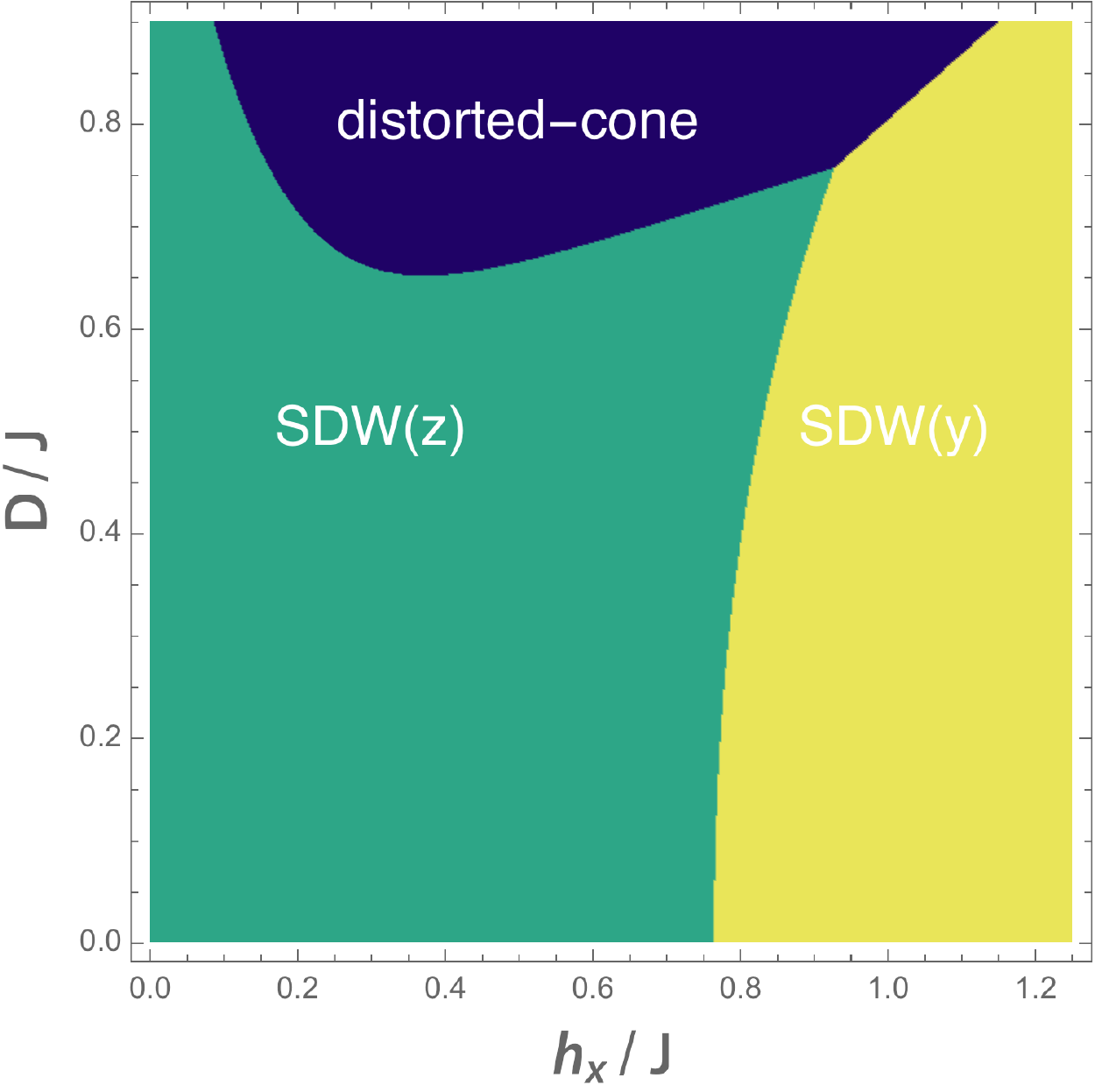}
	\caption{(Color online) $h-D$ phase diagram for the case of ${\bm h} \perp {\bm D}$, $h_z=0$. 
		Here $\lambda\approx 3.8 (D/J)^2$, see  Eq.\eqref{eq:lambda}, and $g_{\rm bs}=0.23 \times 2\pi v$ and $J'=10^{-3} \times 2\pi v$. 
		For small $D/J$, the critical field separating SDW(z) to SDW(y) phases is given by $h_x\simeq0.23\pi$. 
		The line separating distorted-cone and SDW(y) phases is described by $h_x/D\simeq1.5$. }
	\label{fig:boundaryplot}
\end{figure}

It is easy to see that stronger DM interaction leads to a more stable SDW(z).
Indeed, stronger DMI shortens the RG scale $l_{\varphi}$ thereby extending the 2nd stage RG flow which favors $g_x$ process.

\begin{figure}[!tbp]
	\includegraphics[width=0.8\columnwidth]{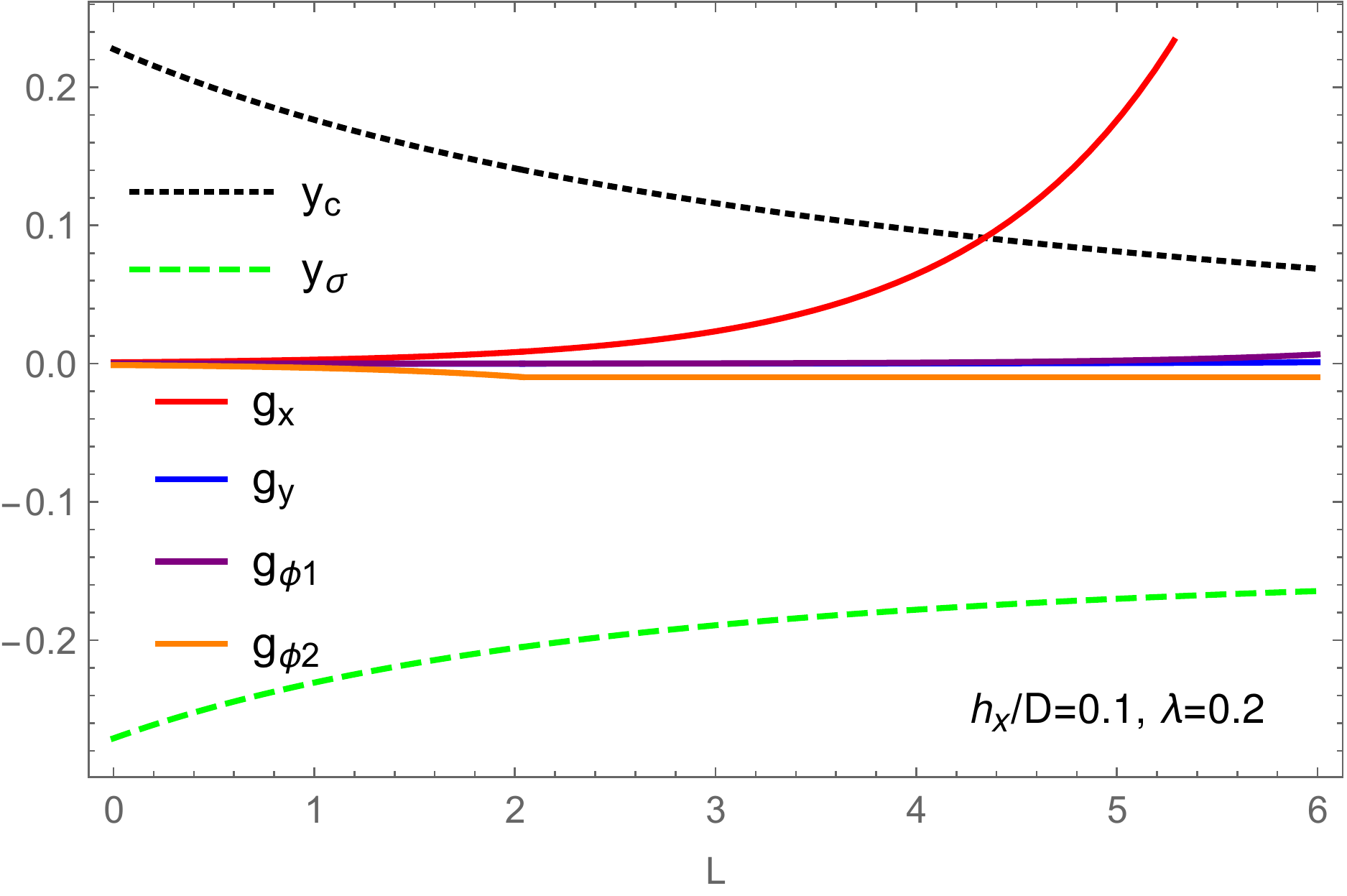}
	\caption{(Color online) Typical flow of coupling constants in SDW(z) phase, ${\bm h}\perp {\bm D}$. $g_{\rm bs}/(2\pi v)=0.23$, $J'/(2\pi v)=0.001$,
		$D=0.01J$, $h_x/D=0.1$, $h_z=0$ and $\lambda=0.2$. Here $l_{\rm inter}\simeq 6.9$, $l_\varphi\simeq2$. The dominant coupling is $g_x$ shown in red, and $g_x(\ell^*)=1$ at $\ell^*\simeq6.8$.}
	\label{fig:flow01}
\end{figure}

\begin{figure}[!tbp]
	\includegraphics[width=0.8\columnwidth]{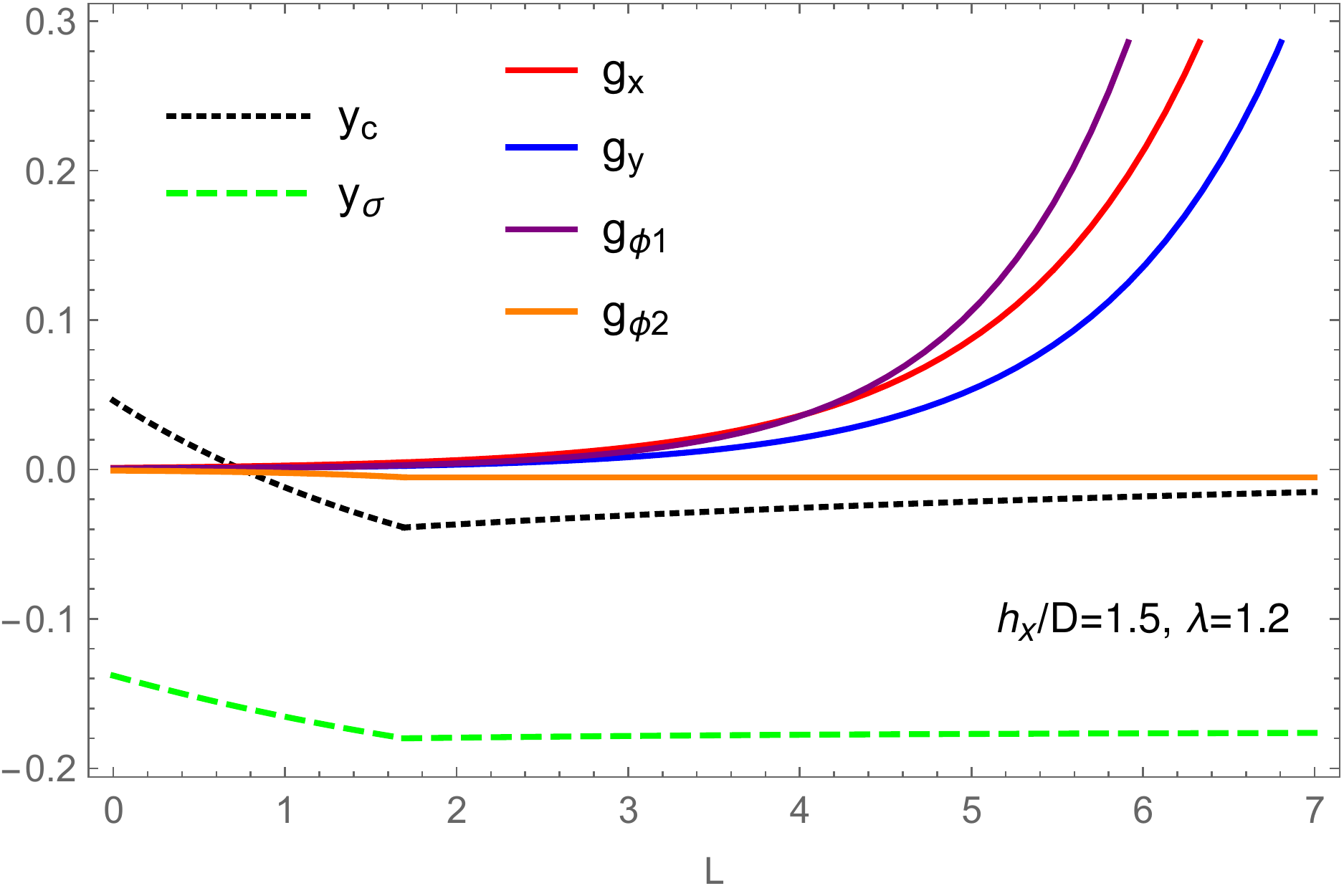}
	\caption{(Color online) Typical flow of coupling constants in distorted-cone phase, ${\bm h}\perp {\bm D}$. $g_{\rm bs}/(2\pi v)=0.23$, $J'/(2\pi v)=0.001$,
		$D=0.01J$, $h_x/D=1$, $h_z=0$ and $\lambda=1.2$. Here $l_{\rm inter}\simeq 6.9$,
		 $l_\varphi\simeq1.7$. The dominant coupling is $g_{\varphi_1}$ shown in purple, and $g_{\varphi_1}(\ell^*)=1$ at $\ell^*\simeq7.0$.}
	\label{fig:flow02}
\end{figure}

\begin{figure}[!tbp]
	\includegraphics[width=0.8\columnwidth]{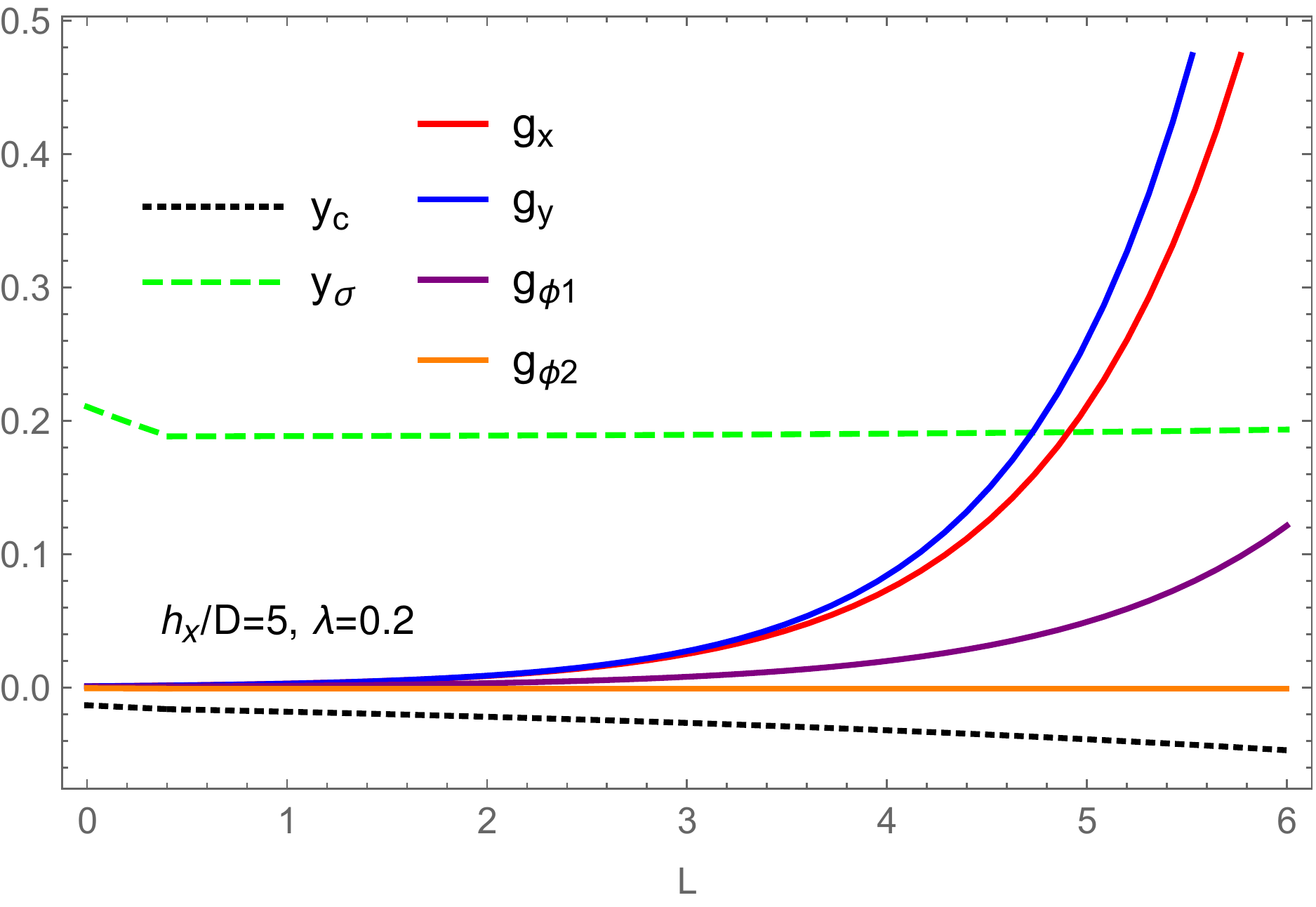}
	\caption{(Color online) Typical flow of coupling constants in SDW(y) phase, ${\bm h}\perp {\bm D}$. $g_{\rm bs}/(2\pi v)=0.23$, $J'/(2\pi v)=0.001$,
		$D=0.01J$, $h_x/D=5$, $h_z=0$ and $\lambda=0.2$. Here $l_{\rm inter}\simeq 6.9$, $l_\varphi\simeq 0.4$. The dominant coupling is $g_y$ shown in blue, $g_y(\ell^*)=1$ at $\ell^*\simeq6.2$.}
	\label{fig:flow03}
\end{figure}

Using the relation $\lambda = c' D^2/J^2$, with $c'=(2\sqrt{2} v/g_{\rm bs})^2$, 
we are now in position to calculate the physical $h-D$ phase diagram -- the result is presented in Fig.~\ref{fig:boundaryplot}. 
The boundary between SDW(y) and distorted-cone is linear with $h_x/D\simeq 1.5$, which corresponds to the vertical boundary in Fig.~\ref{fig:pd2}. The line separating SDW(z) and SDW(y) phases is determined by the condition $g_y(l)=g_x(l)$, which leads to
\begin{equation}
[\cos\theta_0^y]^2\exp[-\int_{0}^{l} dl' 2y_C(l')]=1.
\label{eq:bound2}
\end{equation}
If $D$ is small, $\cos{\theta_0^y}\sim 1$, which implies $y_C(l)<0$. Using \eqref{yyc}, Eq.~\eqref{eq:bound2} reduces to $h^2/D^2=2/\lambda$. 
Hence the critical magnetic field $h_c/J=(2\pi v/g_{\rm bs})\pi \sim 0.23\pi$ is independent of the value of $D$. 
Being quite large, this value should be considered an order-of-magnitude estimate.
(Here we have used $g_{\rm bs}\simeq 0.23\times (2\pi v)$ from Ref.~\onlinecite{eggert}.)
Typical flows of coupling constants for each of the phases in Fig.~\ref{fig:boundaryplot} are shown in Fig.~\ref{fig:flow01},~\ref{fig:flow02},~\ref{fig:flow03}. 

\section{Discussion}
\label{sec:conclusion}

Many of recent revolutionary developments in condensed matter physics, ranging from ferroelectrics \cite{Cheong2007} to spintronics  \cite{Manchon2015} to 
topological quantum phases \cite{annual-review2014,nussinov2015,Savary-Balents}, are associated with strong spin-orbit interactions. 
Even when not particularly strong, spin-orbit coupling is seen to control important aspects of low-energy physics of systems such as
$\alpha-$ and $\kappa-$phase BEDT-TTF and BEDT-TSF organic salts, which are made of light C, S, and H atoms \cite{valenti2016}.

Our study adds a new physically-motivated model to this fast growing list: a quasi-2d (or 3d) system of weakly coupled antiferromagnetic Heisenberg 
spin-$1/2$ chains subject to the uniform but \textit{staggered between chains} Dzyaloshinskii-Moriya interaction.

\subsection{Experimental implications}

The obtained $T-{\rm vs}-M(h)$ phase diagrams in Fig.~\ref{fig:Tc} and Fig.~\ref{fig:Tc_2} have striking resemblance with the experimentally determined, 
via specific heat measurements \cite{Halg2014}, phase diagrams of chain materials 
K$_2$CuSO$_4$Cl$_2$ and  K$_2$CuSO$_4$Br$_2$, respectively. The first of this is interpreted as a weak-DM material with $(D/J')_{\rm Cl} = 1.3$, see Appendix~\ref{app:j'},
in which the only magnetic order is of the standard cone type.  

\begin{figure}[t!]
	\includegraphics[width=0.8\columnwidth]{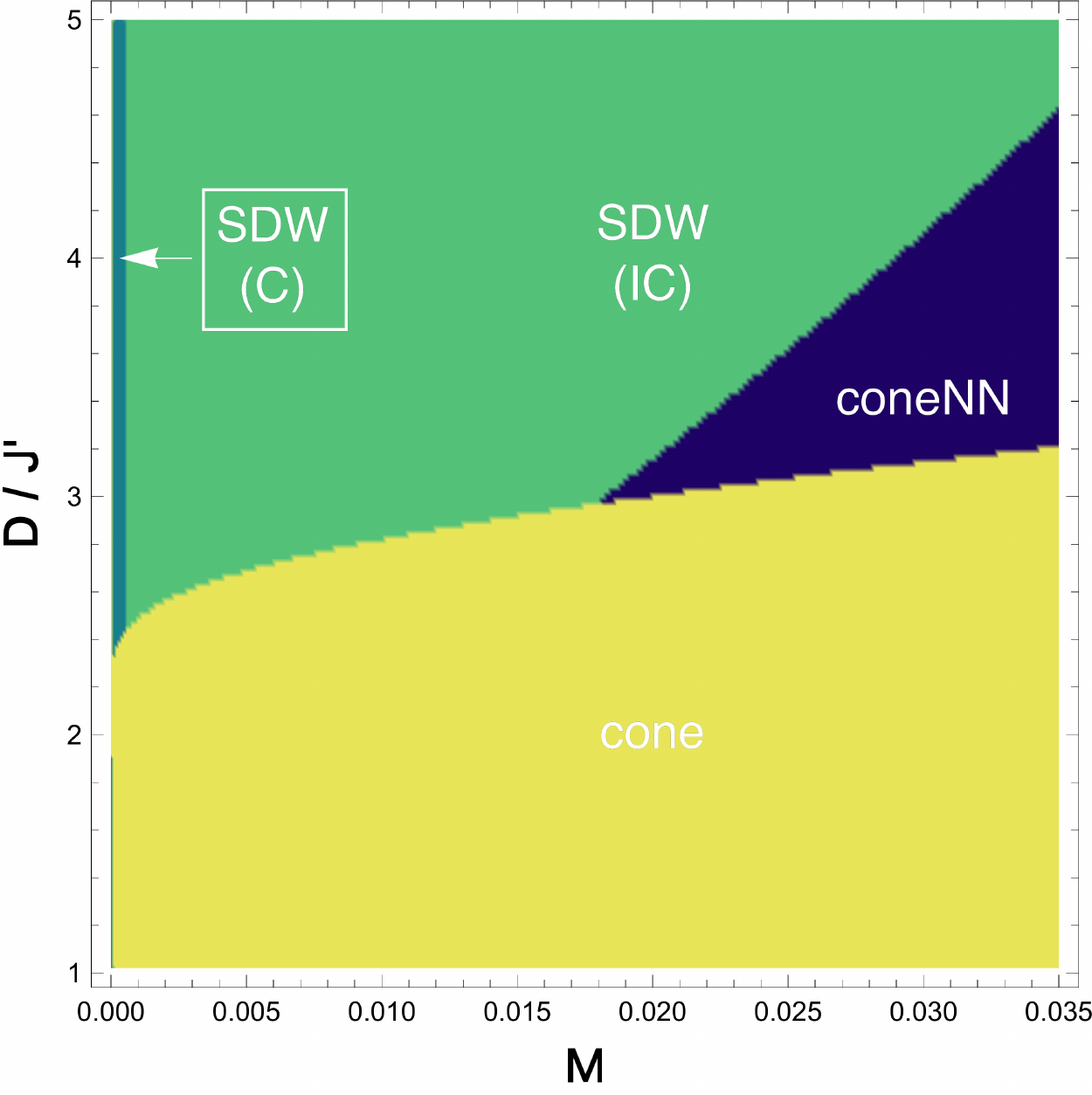}
	\caption{\label{fig:MDJp} (Color online) Small magnetization $M-D$ phase diagram for the case of ${\bm h} \parallel {\bm D}$, obtained by the CMF calculation.
		Here $J=20.5$ K, $J' = 0.0045 J = 0.09$ K. The cone phase is bounded by $D/J' \approx 4.2$ from above for all $M\in(0,0.5)$.
		See Fig.\ref{fig:mdphase} in Appendix \ref{app:j'} for the phase diagram in the 
		wider range of magnetization $0.02 < M < 0.48$.} 
\end{figure}

The Br-based material is more interesting and exhibits a low-field phase transition between
two different orders of experimentally-yet-unknown nature. Interaction parameters for this material have been estimated experimentally \cite{Halg2014} to be $J =20.5$ K, and $D=0.28$ K. Fitting zero-field $T_c$ of this material to that of the commensurate SDW order gives us $J' = 0.09$ K, see Appendix~\ref{app:j'} for more details.
Therefore $(D/J')_{\rm Br}\approx 3.1$, which places K$_2$CuSO$_4$Br$_2$ in the intermediate-DM range. 
Fig.~\ref{fig:MDJp} shows that 
$D/J' = 3.1$ is strong enough to suppress cone ordering
at small magnetic fields, but nonetheless is not sufficiently strong to prevent the cone phase from emerging at slightly greater magnetic field.
Analysis in Appendix~\ref{app:j'} shows that for this particular value of $D/J'$ one encounters {\em three} quantum phase transitions in the narrow interval
of magnetization $0 \leq M \leq 0.025$: commensurate-incommensurate SDW, incommensurate SDW to coneNN, and finally coneNN to the commensurate cone phase.
The cone gets stabilized above $M=0.025$, see Fig.~\ref{fig:Br0}. This rapid progression of phase transitions is not seen in the experiment \cite{Halg2014}.
There, rather, a single transition at $B_{\rm Br}=0.1$T is observed, although it must be said that the commensurate-incommensurate SDW may be
just too difficult to identify. Converting the observed field magnitude to energy units, via $h_{\rm Br} = g \mu_B B_{\rm Br}/k_B = 0.134$ K, 
we estimate the corresponding magnetization value as $M_{\rm Br} = h_{\rm Br}/(2\pi v) = h_{\rm Br}/(\pi^2 J_{\rm Br}) \approx 0.0007$.
This is much smaller than the critical cone magnetization $M=0.025$ estimated above.

However the present discussion, much of which is summarized graphically in Fig.~\ref{fig:MDJp}, shows that the region of $D/J' \approx 3$
is particularly tricky. Small, order of $5\% - 10\%$ changes, in $J'$ and $D$ can significantly affect the ratio $D/J'$ and lead to dramatically 
different predictions for the phase composition at small magnetization. Specifically, increasing $D/J'$ to $\simeq 4$ eliminates the cone phase 
from the competition completely as now one observes only C-IC SDW and SDW-to-coneNN transitions, in a much closer qualitative agreement
with the experiment. Given significant uncertainties in parameter values of K$_2$CuSO$_4$Br$_2$, a more quantitative description of the full experimental situation 
is not possible at the moment.

We hope that our detailed investigation will prompt further experimental studies of these interesting compounds,
in particular in the less studied so far ${\bm h} \perp {\bm D}$ configuration,
and will shed more light on the intricate interplay between the magnetic field, DM and inter-chain interactions present
in this interesting class of quasi-one-dimensional materials.
It is interesting to note that unique geometry of DM interactions makes  K$_2$CuSO$_4$Br$_2$ somewhat similar to the honeycomb iridate 
material Li$_2$IrO$_3$ an incommensurate magnetic order of
which is characterized by unusual counter-rotating spirals on neighboring sublattices \cite{kimchi2015,kimchi2016}. 

\subsection{Summary and future directions}

We have systematically investigated complicated interplay of DM interaction and external magnetic field, applied either along or perpendicular to DM vector ${\bm D}=D\hat{z}$. 
Combining techniques of bosonization, renormalizaion group and chain mean-field theory, we are able to identify the phase diagram of the system. 
In all considered cases the ground state is determined by the inter-chain interaction, which is however strongly affected by the chain backscattering, 
which in turn is very sensitive to the mutual orientation of ${\bm D}$ and ${\bm h}$.

In ${\bm h}\parallel {\bm D}$ configuration the phase diagram is strongly depended on the ratio $D/J'$.
For weak DM interaction, $D < 1.9 J'$, there is only a single cone phase, with spins spiraling in the plane perpendicular to ${\bm D}$. 
Strong DM interaction is found to promote the collinear SDW state. The basic reason for this is strong frustration of the inter-chain cone channel, 
caused by the opposite sense of rotation of spins in neighboring chains (which, in turn, is caused by the opposite directions
of the DM vectors in the neighboring chains).
As a result, the transverse cone ordering is strongly frustrated and the less-relevant SDW state gets stabilized. 
However, the SDW is the ground state only in a very low magnetic field. Increasing the magnetic field upto critical value $h_c\sim J'$, 
we find a (most likely, discontinuous) phase transition from the incommensurate SDW state to the coneNN state which is driven by the 
fluctuation-generated cone-type interaction between the 
next-neighbor (NN) chains. These RG-based arguments are fully supported by the chain mean field calculations.

For ${\bm h}\perp {\bm D}$, we find two distinct SDW states in the plane normal to the magnetic field in the experimentally relevant limit 
of not too strong DM interaction, $D\ll J$. Since none of these states is a lower-symmetry version of the other, the
phase transition between the different SDWs is of spin-flop kind, and is expected to be of the first-order. The transition field $h_c \sim 0.23\pi J$ is (almost) independent of $D$. 
In the limit of $D\sim J$ (impractical for the experiment), there is also a ``distorted-cone" state in which spins rotate in the plane
normal to vector ${\bm D}$, see Figure~\ref{fig:boundaryplot}. We have carried out two-stage RG calculations and 
determined the $\lambda-h/D$ and $h-D$ phase diagrams for this geometry numerically.

All of the obtained results are based on perturbative calculations, framed in either RG or CMF language. The complete consistency between these two techniques 
observed in our work provides strong support in favor of its validity. Nonetheless, an independent check of the presented arguments is highly desired. We hope
our work will stimulate numerical studies of this interesting problem along the lines of quantum Monte-Carlo studies in Refs. \onlinecite{Sandvik1999,Yasuda2005}.

In concluding, we would like to mention potential relevance of our model to the currently popular coupled-wire approach to (mostly chiral) spin liquids \cite{kane2002,neupert2014,sela2015}.
The essence of this approach consists in devising interchain interactions in such a way as to suppress all interchain couplings between the relevant, in RG sense, degrees of freedom
(such as staggered magnetization and dimerization). The remaining marginal interactions of current-current kind then conspire to produce gapped chiral phase with gapless 
chiral excitations on the edges. Staggered DM interactions of the kind considered here are, as we have shown, actually quite effective in removing 
$N^+_y N^{-}_{y+1}$ terms. At the same time, the remaining interchain SDW term grows progressively less relevant as magnetic field is increased towards the
saturation value. Provided that one finds way to suppress fluctuation-generated relevant coneNN like couplings between more distant chains,
described in Section~\ref{sec:coneNN}, one can hope to be able to destabilize weak SDW long-ranged magnetic order with the help of additional 
weak interactions (of yet unknown kind) and drive the system into a two-dimensional spin liquid phase.

\section*{Acknowledgement}

We would like to thank M. H\"alg, K. Povarov, A. I. Smirnov and A. Zheludev for detailed discussions of the experiments, and L. Balents for 
insightful theoretical remarks. This work is supported by the National Science Foundation grant NSF DMR-1507054.

\appendix
\section{Operator product expansion (OPE) and perturbative RG}\label{app:ope}
We have a set of operators {$O_i(x)$} in the perturbation Eq.~\eqref{H'1} and ~\eqref{H'} , with $O_i(x)=J^{a}_{R/L}(x)$ or $N^a(x)$, where $a=x,y,z$.
Product of any two operators can be replaced by a series of terms involving operators of the same set,
\begin{equation}
\lim\limits_{x\to 0}O_i(x)O_j(0)=\sum_k C_{ij}^{k}\frac{1}{|x|^{\Delta_i+\Delta_j-\Delta_k}}O_k(0).
\end{equation}
This identity is known as the \textit{operator product expansion}
(OPE)\cite{fradkin}, it tells us how different operators \textit{fuse} with another. 
In our case, the fusion rules of spin currents ${\bm J}_{R/L}$, staggered magnetization ${\bm N}$ and dimerization $\xi$ are\cite{ope},
\begin{equation}
\begin{gathered}
J_R^a(x,\tau)J_R^b(0)=\frac{\delta^{ab}}{8\pi^2(v\tau-ix)^2}+\frac{i\epsilon^{abc}J_R^c(0)}{2\pi(v\tau-ix)},\\
J_L^a(x,\tau)J_L^b(0)=\frac{\delta^{ab}}{8\pi^2(v\tau+ix)^2}+\frac{i\epsilon^{abc}J_L^c(0)}{2\pi(v\tau+ix)}.\\
J_R^a(x,\tau)N^b(0)=\frac{i\epsilon^{abc}N^c(0)-i\delta^{ab}\xi(0)}{4\pi(v\tau-ix)},\\
J_L^a(x,\tau)N^b(0)=\frac{i\epsilon^{abc}N^c(0)+i\delta^{ab}\xi(0)}{4\pi(v\tau+ix)}.\\
J_R^a(x,\tau)\xi(0)=\frac{iN^a(0)}{4\pi(v\tau-ix)},\\
J_L^a(x,\tau)\xi(0)=\frac{-iN^a(0)}{4\pi(v\tau+ix)}.
\end{gathered}
\label{app:ope1}
\end{equation}
It can be shown that the coefficients $C_{ij}^{k}$, which are known as \textit{structure constants} of the OPE, fix the quadratic terms in the RG (\textit{renormalization group}) 
flow of coupling constants, specifically,
\begin{equation}
\frac{dg_k}{dl}=(2-\Delta_k)g_k-\sum_{i,j}C_{ij}^{k}g_ig_j.
\end{equation}
$\Delta_k$ is the scaling dimension of the coupling term, which in the zero field limit is $2$ and $1$ for ${\bm J}_{yR} \cdot {\bm J}_{yL}$ and 
${\bm N}_y \cdot {\bm N}_{y+1}$ coupling terms, correspondingly.

Here, we provide an example of applying OPE and RG to $g_x N_y^xN_{y+1}^x$ term in our inter-chain Hamiltonian \eqref{H'}. In perturbative RG, there is a term,
\begin{widetext}
\begin{equation}
\begin{split}
&\frac{1}{2}(2\pi v\int \mathrm{d}x \mathrm{d}\tau g_x N_y^x(x,\tau)N_{y+1}^x(x,\tau))(2\pi v\int \mathrm{d}x'\mathrm{d}\tau'y_x M^x_{R,y}(x',\tau')M_{L,y}^x(x',\tau'))\\
=&\frac{1}{2}(2\pi v)^2\frac{1}{(4\pi)^2}\int \mathrm{d}x \mathrm{d}\tau\int \mathrm{d}X \mathrm{d}T ~g_x y_x \frac{N_{y}^x(X,T)N_{y+1}^x(X,T)}{(v\tau-ix)(v\tau+ix)}=2\pi v ~\delta g_x\int \mathrm{d}X \mathrm{d}T ~N_{y}^x(X,T)N_{y+1}^x(X,T).
\end{split}
\label{eq:egrg}
\end{equation}
\end{widetext}
Here,we have applied the OPE in the first step. In the second line $(X,vT)$ are the center of mass coordinates, while $x \to x - x'$ and $\tau \to \tau - \tau'$ are the relative ones.
The correction $\delta g_x$ is given by the integral over RG shell from $a$ to $a' = e^\delta l$,
\begin{equation}
\delta g_x=2\times 2\times\frac{1}{8}g_xy_x\int_{a}^{a'} \mathrm{d}r \frac{1}{r}=\frac{1}{2}g_xy_x\ln(\frac{a'}{a}).
\end{equation}
The first $2$ comes from two neighboring chain, the second $2$ is due to there are two equivalent term as Eq.~\eqref{eq:egrg} when one does perturbative expansion. 
This is equivalent to
\begin{equation}
\frac{dg_x}{dl}=g_x+\frac{1}{2}g_xy_x+\ldots .
\label{eq:1dg}
\end{equation}
The other two terms which give complete the RG equation~\eqref {eq:1dg} are similar as Eq.~\eqref{eq:egrg}, and they are proportional to,
\begin{equation}
\int \mathrm{d}x \mathrm{d}\tau  N_y^xN_{y+1}^x(x,\tau) \int\mathrm{d}x'\mathrm{d}\tau'y_yM^y_{R,y}M_{L,y}^y(x',\tau'),
\end{equation}
and
\begin{equation}
\int \mathrm{d}x \mathrm{d}\tau  N_y^xN_{y+1}^x(x,\tau) \int\mathrm{d}x'\mathrm{d}\tau'y_zM^z_{R,y}M_{L,y}^z(x',\tau').
\end{equation}
In the end the complete RG equations for $g_x$ is,
\begin{equation}
\frac{dg_x}{dl}=g_x+\frac{1}{2}g_xy_x-\frac{1}{2}g_xy_y-\frac{1}{2}g_xy_z .
\label{eq:cdg}
\end{equation}
The minus sign of last two terms are from the Levi-Civita epsilon in the fusion rules \eqref{app:ope1}.

Then the RG equations of all the perturbation terms in Hamiltonian \eqref{eq:15} are,
\begin{equation}
\begin{gathered}
\frac{dy_x}{dl}=y_yy_z,\;\;\frac{dy_y}{dl}=y_zy_x,\;\;
\frac{dy_z}{dl}=y_xy_y,\\
\frac{dg_x}{dl}=g_x[1+\frac{1}{2}(y_x-y_y-y_z)],\\
\frac{dg_y}{dl}=g_y[1+\frac{1}{2}(y_y-y_z-y_x)],\\
\frac{dg_z}{dl}=g_z[1+\frac{1}{2}(y_z-y_x-y_y)].\\
\end{gathered}
\label{rg1}
\end{equation}
With $y_x(0)=y_y(0)$ (see \eqref{eq:ini1}), we have $y_x(l)=y_y(l)$ and $g_x(l)=g_y(l)$. Therefore, Eq.~\eqref{rg1} reduces to,
\begin{equation}
\begin{gathered}
\frac{d y_B}{d \ell}=y_By_z,\;
\frac{d y_z}{d \ell}=y_B^2,\\
\frac{d g_\theta}{d \ell}=g_{\theta}(1-\frac{1}{2}y_z),\;
\frac{dg_z}{d\ell}=g_z[1+\frac{1}{2}(y_z-2y_B)].
\end{gathered}
\label{rg2}
\end{equation}
Here $g_\theta$ and $y_B$ are defined in Eq.~\eqref{eq:ini2}. Marginal couplings $y_{z,B}$ grows much slower than $g_{\theta,z}$, so that we can approximate \eqref{rg2}
by replacing $y_{z,B}$ with their initial values,
\begin{equation}
\begin{gathered}
\frac{d g_\theta}{d \ell}=g_{\theta}[1+\frac{g_{\rm bs}}{4\pi v}(1+\lambda)],\;
\frac{dg_z}{d\ell}=g_z[1+\frac{g_{\rm bs}}{4\pi v}(1-\lambda)].
\end{gathered}
\end{equation}
With $g_{\rm bs},\lambda >0$, we see $g_\theta$ grows faster than $g_z$.

\section{Generation of next-neighbor (NN) chain coupling}
 \label{app:nn_chain}
Starting from interaction ${\cal H}_{\rm cone}$ in Eq.~\eqref{H'} we obtain the partition function $Z_{\theta}$ as
\begin{equation}
\begin{gathered}
Z_\theta=\int D\theta e^{-S_{0}}e^{\sum_{y}\int \mathrm{d}x \mathrm{d}\tau {\cal H}_{\rm cone}}.
\end{gathered}
\end{equation}
where, $S_0$ and $Z_0$ are the action and partition function of independent spin chains. 
We expand $Z_\theta$ in power of ${\cal H}_{\rm cone}$ to the second order,
\begin{equation}
Z_\theta
=\int D\theta e^{-S_{0}}\Big\{1+\sum_{y}\int \mathrm{d}x \mathrm{d}\tau {\cal H}_{\rm cone}+S^{(2)}\Big\}.
\end{equation}
The first order term contributes nothing to the next-neighbor (NN) chain coupling.
We are interested in the second-order term which reads
\begin{equation}
S^{(2)}= \frac{1}{2}\iint \mathrm{d}x_1\mathrm{d}x_2\mathrm{d}\tau_1\mathrm{d}\tau_2 (\sum_y {\cal H}_{\rm cone})^2 .
\end{equation}
Introduce short-hand notation $A_{\mu}(y)=e^{i\mu[\sqrt{2\pi}(\tilde{\theta}_y-\tilde{\theta}_{y+1})+2t_{\theta}^{y}x]}$
in terms of which the inter-chain Hamiltonian reads
\begin{equation}
{\cal H}_{\rm cone}=\pi v A^2 g_{\theta} \sum_{\mu=\pm 1} \int \mathrm{d}x A_{\mu}(y),
\end{equation}
The terms which produce interaction between next-nearest chains can then be written as
\begin{widetext}
	\begin{equation}
	\begin{split}
	&S^{(2)}=\frac{1}{2}(\pi vA^2g_{\theta})^2\sum_{y}\sum_{\mu=\pm 1} \sum_{\nu=\pm 1}\int \mathrm{d}x_1 \mathrm{d}\tau_1\int \mathrm{d}x_2 \mathrm{d}\tau_2 A_{\mu}(y)A_{\nu}(y+1).
	\end{split}
	\label{S'}
	\end{equation}
	Rewrite the expression in the integral,
	\begin{equation}
	\begin{split}
	\sum_{\mu=\pm 1} \sum_{\nu=\pm 1}A_{\mu}(y)A_{\nu}(y+1)=&\sum_{\mu=\pm 1} \sum_{\nu=\pm 1}e^{i\mu\sqrt{2\pi}\tilde{\theta}_y(x_1)}e^{-i\nu\sqrt{2\pi}\tilde{\theta}_{y+2}(x_2)}e^{-i[\mu\tilde{\theta}_{y+1}(x_1)-\nu\tilde{\theta}_{y+1}(x_2)]}e^{i2[\mu t_{\theta}^{y}x_1+\nu t_{\theta}^{y+1}x_2]},\\
	=&\sum_{\mu=\nu}e^{i\mu\sqrt{2\pi}[\tilde{\theta}_y(x_1)-\tilde{\theta}_{y+2}(x_2)]}e^{-i\mu[\tilde{\theta}_{y+1}(x_1)-\tilde{\theta}_{y+1}(x_2)]}e^{i2\mu [t_{\theta}^{y}x_1+t_{\theta}^{y+1}x_2]}\\
	&+\sum_{\mu=-\nu}e^{i\mu\sqrt{2\pi}[\tilde{\theta}_y(x_1)+\tilde{\theta}_{y+2}(x_2)]}e^{-i\mu[\tilde{\theta}_{y+1}(x_1)+\tilde{\theta}_{y+1}(x_2)]}e^{i2\mu [t_{\theta}^{y}x_1-t_{\theta}^{y+1}x_2]}.
	\end{split}
	\end{equation}
Now we integrate out field $\tilde{\theta}_{y+1}$ from the intermediate $(y+1)$'s chain in $S^{(2)}$, only $\mu=\nu$ produces finite contribution, 
\begin{equation}
\begin{split}
S^{(2)}=\frac{1}{2}(\pi vA^2g_{\theta})^2\sum_{y}\int \mathrm{d}x_1 \mathrm{d}\tau_1\int \mathrm{d}x_2 \mathrm{d}\tau_2\sum_{\mu=\pm 1}e^{i\mu\sqrt{2\pi}[\tilde{\theta}_y(r_1)-\tilde{\theta}_{y+2}(r_2)]}
e^{i2\mu t_{\theta}^{y} [x_1-x_2]}\langle e^{-i\mu\sqrt{2\pi}[\tilde{\theta}_{y+1}(r_1)-\tilde{\theta}_{y+1}(r_2)]}\rangle .
\end{split}
\end{equation}
Here the $(y+1)$'s chain correlation function
\begin{equation}
\langle e^{-i\mu\sqrt{2\pi}[\tilde{\theta}_{y+1}(r_1)-\tilde{\theta}_{y+1}(r_2)]}\rangle =\frac{1}{|r_1-r_2|^{1/K}},
\end{equation}
where $K=2\pi/\beta^2$, $K=1$ in the absence of magnetic field, and $r_{1/2}=(x_{1/2},v\tau_{1/2})$, is the coordinates in space-time.
Switch to the center of mass and relative coordinates, $R=(r_1+r_2)/2,\; r=r_1-r_2, y=v\tau$, then $\tilde{\theta}_y(r_1)=\tilde{\theta}_y(R+r/2)\simeq \tilde{\theta}_y(R)$, and
\begin{equation}
\begin{split}
S^{(2)}=\sum_y \frac{(\pi vA^2g_{\theta})^2}{2v^2}\sum_{\mu=\pm 1}\int \mathrm{d}^2Re^{i\mu\sqrt{2\pi}[\tilde{\theta}_y(R)-\tilde{\theta}_{y+2}(R)]} \int \mathrm{d}xdy e^{i2\mu t_{\theta}^{y} x}\frac{1}{(x^2+y^2)^{\Delta_1}},\\
\end{split}
\end{equation}
here $\Delta_1=1/(2K)$ is the scaling dimension of $N^{\pm}$, which depends on magnetic field as shown in Table.~\ref{table:scaling_d}.  
The integral over relative $(x,y)$ coordinates is easy to evaluate
\begin{eqnarray}
S^{(2)}=\frac{(\pi vA^2g_{\theta})^2}{v} \pi t_{\theta}^{2\Delta_1-2}\frac{\Gamma(1-\Delta_1)}{\Gamma(\Delta_1)} \sum_{y}\int \mathrm{d}x\mathrm{d}\tau\cos[\sqrt{2\pi}(\tilde{\theta}_y(r)-\tilde{\theta}_{y+2}(r))] = -\int \mathrm{d}\tau \sum_y {\cal H}_{NN},
\label{eq:eff_S}
\end{eqnarray}
\end{widetext}
Re-exponentiating this term we obtain the desired effective action describing interaction between next-nearest chains.
Using $N^+_{y}N^-_{y+2} = A^2 \cos[\sqrt{2\pi}(\tilde{\theta}_y-\tilde{\theta}_{y+2})]$, we can read off the coupling for Eq.~\eqref{2nd neighbour},
\begin{equation}
\begin{gathered}
 2\pi v   G_{\theta}=-\frac{\pi A^2 }{4}f(\Delta_1)\frac{(J')^2}{D},\\
    f(\Delta_1)=t_{\theta}^{2\Delta_1-1}\frac{\Gamma(1-\Delta_1)}{\Gamma(\Delta_1)}.
\end{gathered}
\end{equation}
Here, $f(\Delta_1)$, as a function of $\Delta_1$, starts from 1 as the field increases from zero,
when the scaling dimension $\Delta_1$ is $1/2$.
\begin{figure}[!]
	\includegraphics[width=0.8\columnwidth]{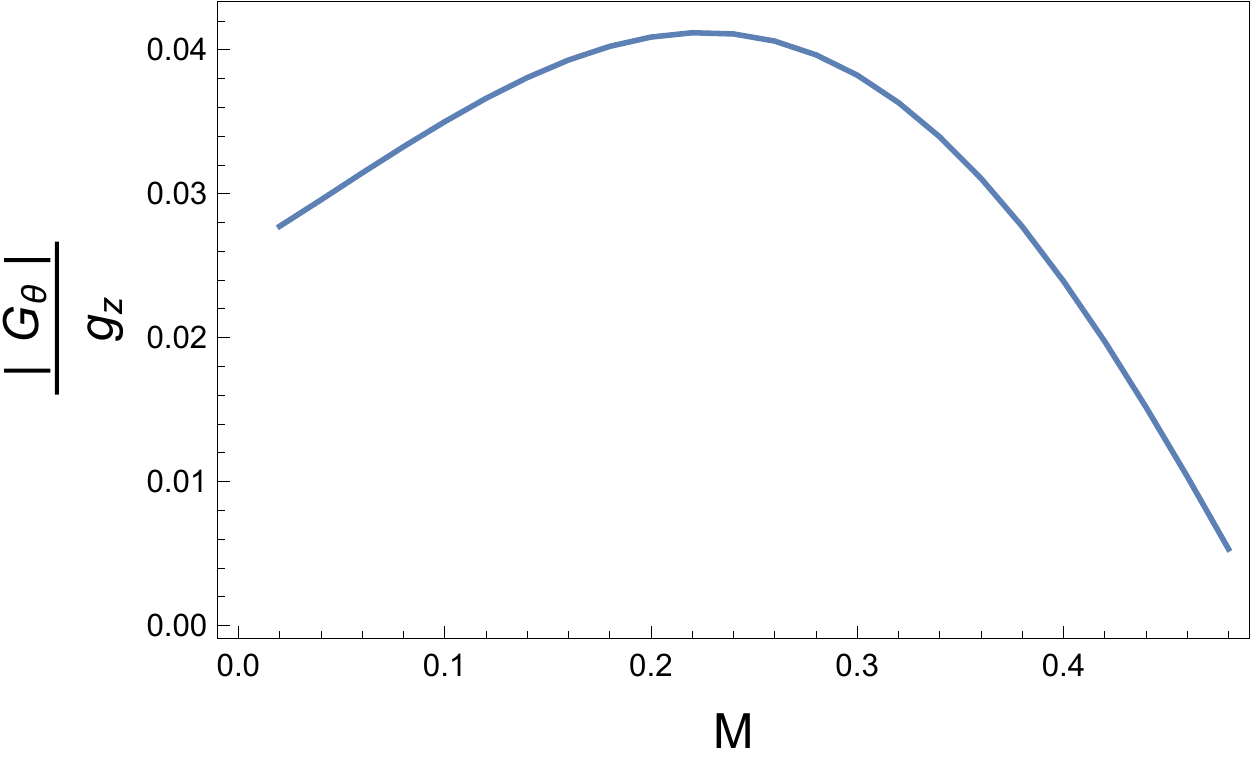}
	\caption{Coupling constant of the transverse interaction between next-nearest chains, $G_\theta$, showen as the ratio of $|G_\theta(0)|/g_z(0)$ versus magnetization M. 
		Here DM interaction is strong: $J'=0.001J$, $D/J=0.01$.}
	\label{Gnn}
\end{figure}
\section{Critical temperature by chain mean field (CMF) approximation}
\label{app:cmf}
Chain mean field (CMF) approximation consists in replacing the interchain interaction \cite{oleg_cmf}
by the self-consistent single-chain model 
\begin{equation}
- \cos(\sqrt{2\pi}\theta_{y}) \cos(\sqrt{2\pi}\theta_{y+1}) \to  -\Psi \cos(\sqrt{2\pi}\theta_{y}),
\end{equation}
where $\Psi$ stands for the expectation value of the staggered magnetization 
\begin{equation}
\Psi\equiv \langle \cos(\sqrt{2\pi}\theta_{y})\rangle.
\end{equation}
Therefore the Hamiltonian of the system reduces to the sum of independent sine-Gordon models
\begin{equation}
{\cal H}=\sum_y \int  \mathrm{d}x \frac{v}{2}[(\partial_x{\phi})^2+(\partial_x{\theta})^2] - 2c\Psi  \cos(\sqrt{2\pi}\theta_{y}),
\label{sine-gordon}
\end{equation}
where factor of $2$ arises from coupling to the two neighboring chains. To determine the critical temperature, we expand partition function corresponding to the 
Hamiltonian \eqref{sine-gordon} to the first order in $\Psi$ and arrive at the self-consistent condition for $\Psi\neq 0$, which is \cite{Giamarchi}
\begin{equation}
\begin{split}
&\frac{1}{2c}=\chi(q=0,\omega_n=0;T_c)\\
&\quad =\int \mathrm{d}x\int_{0}^{1/T_c}\mathrm{d}\tau e^{i(qx+\omega\tau)}\langle O(x,\tau)O(0,0) \rangle_0,
\end{split}
\label{app:cmf1}
\end{equation}
where $\chi(q,\omega_n;T)$ is momentum and frequency dependent susceptibility at finite temperature $T$. Depending on the type of the order we consider, the operator $O$ stands for
\begin{equation} 
O=\cos(\sqrt{4\pi\Delta_1}\;\theta)\quad \text{or}\quad O=\cos(\sqrt{4\pi\Delta_2}\;{\phi}).
\end{equation}
Scaling dimensions are listed in Table.~\ref{table:scaling_d}, $\Delta_1=\pi R^2$ and $\Delta_2=\pi/\beta^2$.
Now we examine the ordering temperatures of each interaction in Eq.\eqref{H'} and ~\eqref{2nd neighbour} individually. 
Here we follow the standard calculation in Ref.~\onlinecite{oleg_cmf} which gives the following expressions for static susceptibilities (these are Eqns. (D.55) and (D.57) of Ref.~\onlinecite{oleg_cmf}):
for SDW order
\begin{widetext}
\begin{equation}
\chi(q=0,\omega_n=0;T)
=\frac{\pi }{2 v} \Big[(2\pi T /v)^{2 \Delta -2}\frac{\Gamma(1-\Delta)\Gamma(\Delta/2)^2}{\Gamma(\Delta)\Gamma(1-\Delta/2)^2} -\frac{\Gamma(\Delta-1/2)}{\sqrt{\pi}(1-\Delta)\Gamma(\Delta)} \Big],
\label{eq:chi}
\end{equation}
and for cone order
\begin{equation}
\begin{split}
\chi(q=q_0,\omega_n=0;T)&=\frac{\pi }{2v}(2\pi T/v)^{2 \Delta -2}\frac{\Gamma(1-\Delta)}{\Gamma(\Delta)}\big| \frac{\Gamma(\Delta/2+ivq_0/4\pi T)}{\Gamma(1-\Delta/2+ivq_0/4\pi T)}\big|^2\\
&=\frac{1}{4\pi v}(2\pi T /v)^{2 \Delta -2}\frac{\Gamma(1-\Delta)}{\Gamma(\Delta)}|\Gamma(1-\Delta/2+ivq_0/4\pi T)|^4\times[\cosh(vq_0/2 T)-\cos(\pi \Delta)].
\end{split}
\label{eq:chi2}
\end{equation}
\end{widetext}
Here, $\Delta$ is either $\Delta_1$ or $\Delta_2$.
The second term in the bracket of Eq.~\eqref{eq:chi} removes the 
non-physical divergence in the limit $\Delta \to 1$ near the saturation field. A similar compensating term is not needed in Eq.~\eqref{eq:chi2} because there $\Delta \approx 1/2$.

\subsection{Cone order}
\label{app:cone2}

Consider first the cone order in finite temperature, and its Hamiltonian is given by first line in Eq.~\eqref{eq:inter-cmf},
\begin{equation}
\begin{gathered}
{\cal H}_{\rm cone}=c_1 \sum_y \int  \mathrm{d}x  \cos[\beta(\tilde{\theta}_{y}-\tilde{\theta}_{y+1})+2(-1)^yt_{\theta} x], \\
\end{gathered}
\end{equation}
with $c_1=J'A_3^2$. We apply position-dependent shift to $\tilde{\theta}$ field to remove the oscillation and change the overall sign,
\begin{equation}
\tilde{\theta}_y=\breve{\theta}_y+(-1)^y\frac{\pi}{2\beta}-(-1)^{y}\frac{t_\theta}{\beta}x,
\end{equation}
Next we apply CMF approximation
\begin{eqnarray}
{\cal H}_{\rm cone} &=& \sum_y \int \mathrm{d}x \frac{v}{2}[(\partial_x \tilde{\phi})^2+(\partial_x \breve{\theta}-(-1)^y t_\theta/\beta)^2]\nonumber\\
&& -2c_1\Psi_1 \int  \mathrm{d}x  \cos(\beta\breve{\theta}_{y}),
\label{eq:h0andcone}
\end{eqnarray}
where $\Psi_1=\langle \cos(\beta\breve{\theta}_{y})\rangle$. 
Susceptibilities of the original field $\tilde{\theta}$ and shifted field $\breve{\theta}$ are related by \cite{oleg_cmf}
\begin{equation}
\chi_{\breve{\theta}-\breve{\theta}}(q=0,\omega=0;T)=\chi_{\tilde{\theta}-\tilde{\theta}}(q_0=\frac{D}{v},\omega=0;T),
\end{equation}
Using \eqref{eq:chi2} and \eqref{app:cmf1} the ordering temperature for this cone state $T_{\rm cone}$ is obtained as
\begin{equation}
\begin{split}
&1=\eta_1\left(\frac{2\pi T_{\rm cone}}{v}\right)^{2 \Delta_1 -2}\frac{\Gamma(1-\Delta_1)}{\Gamma(\Delta_1)}|\Gamma(\Delta_1/2+iy)|^4\\
&\quad \times [\cosh(2\pi y)-\cos(\pi \Delta_1)],
\end{split}
\label{solution_cone}
\end{equation}
with $\eta_1=c_1/(2\pi v)=J'A_3^2/(2\pi v)$, and
\begin{equation*}
y=\frac{q_0 v}{4\pi T} = \frac{D}{4\pi T}, \qquad \Delta_1=\pi R^2.
\end{equation*}
Plots of $T_{\rm cone}$ for system with weak DM interaction and in the presence of magnetic field are shown as the green curves in Fig.~\ref{fig:Tc} and \ref{fig:three}.

\begin{figure}[!tbp]
 	\includegraphics[width=0.8\columnwidth]{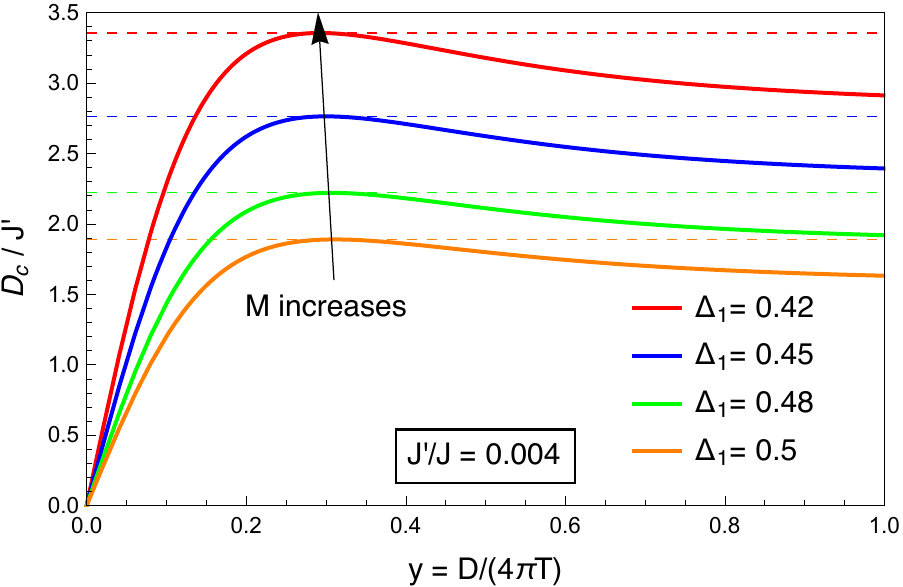}
 	\caption{(Color online) Plot of right side of Eq.~\eqref{eq:dc}, showing maximum increases when $\Delta_1$ decreases, implying the critical $D_c$ increases with field. Here, we consider low-field condition only, where field-dependence of $v$ and $A_3$ have been neglected. Horizontal dotted lines indicate critical $D/J'$ required to destroy the cone state.}
 	\label{fig:possible-ratio1}
 \end{figure}

Fig.~\ref{fig:three} shows that increasing $D$ suppresses cone state. When $D/J'$ is bigger than a critical value, the solution of $T_{\rm cone}$ starts to disappear. We can estimate critical $D/J'$ 
ratio by rearranging \eqref{solution_cone} as 
\begin{widetext}
\begin{equation}
\begin{split}
\frac{D}{J'}= 2 \left(\frac{v}{J'}\right)^{\frac{1-2\Delta_1}{2-2\Delta_1}} y
\left(\frac{A^2_3}{2\pi}\frac{\Gamma(1-\Delta_1)}{\Gamma(\Delta_1)}|\Gamma(\Delta_1/2+iy)|^4 [\cosh(2\pi y)-\cos(\pi \Delta_1)]\right)^{\frac{1}{2-2\Delta_1}}.
\end{split}
\label{eq:dc}
\end{equation}
\end{widetext}
The scaling of $D/J'$ with the $v/J'$ ratio obtained here matches that in \eqref{app:Dscaling}, which is obtained via a different, commensurate-incommensurate based,
reasoning in Appendix~\ref{app:cic}.
 The right side of Eq.~\eqref{eq:dc} for relatively low field is shown in Fig.~\ref{fig:possible-ratio1}, where we set $v=\pi J/2$, and $A_3\simeq 1/2$, 
 so that $\Delta_1$ is the only parameter dependent on field. 
The magnetization dependence of $\Delta_1=\pi R^2$ appears via $M$-dependence of the compactification radius $R$ \cite{oleg_cmf}
 \begin{equation}
 2\pi R^2=1-\frac{1}{2\ln(M_0/M)},
 \label{eq:2piR}
 \end{equation} 
 where $M_0=\sqrt{8/(\pi e)}$ and the limit of small magnetization $M$ is assumed. Therefore, Fig.~\ref{fig:possible-ratio1} shows that the critical $D$ increases with field:
 critical $D/J' \approx 1.9$ at $\Delta_1 =0.5$, which corresponds to $M=0$, but increases to $ \approx 2.75$ at $\Delta_1 =0.45$, which corresponds to $M \approx 0.0065$,
 according to \eqref{eq:2piR}. Note that this corresponds to a rather small magnetic field $h = 2\pi v M \approx \pi^2 M J = 0.064 J$ on the scale of the chain exchange $J$.
 Therefore material with $D = 2.75 J'$ will be in the longitudinal SDW phase at zero magnetic field but transitions, in a discontinuous fashion, to the commensurate cone phase
 in a small, but finite, magnetic field. This behavior seems to correspond to the case of K$_2$CuSO$_4$Br$_2$, as we describe in Appendix \ref{app:j'}.
 
 Importantly, the right-hand-side of \eqref{eq:dc} is bounded by the absolute maximum which is a weak function of the $J'/J$ ratio. For $J'/J=0.004$, chosen in Fig.~\ref{fig:possible-ratio1},
 that maximum value is approximately $6.5$. Therefore for the material with $D/J' \geq 6.5$ the cone phase does not realize at all -- the remaining competition is between the SDW phase,
 which prevails at small magnetization, and the cone-NN phase which emerges at higher $M$, as is discussed in Section~\ref{strong_DM_para}.

\subsection[sdw]{SDW order}
\begin{figure}[t]
	\includegraphics[width=0.8\columnwidth]{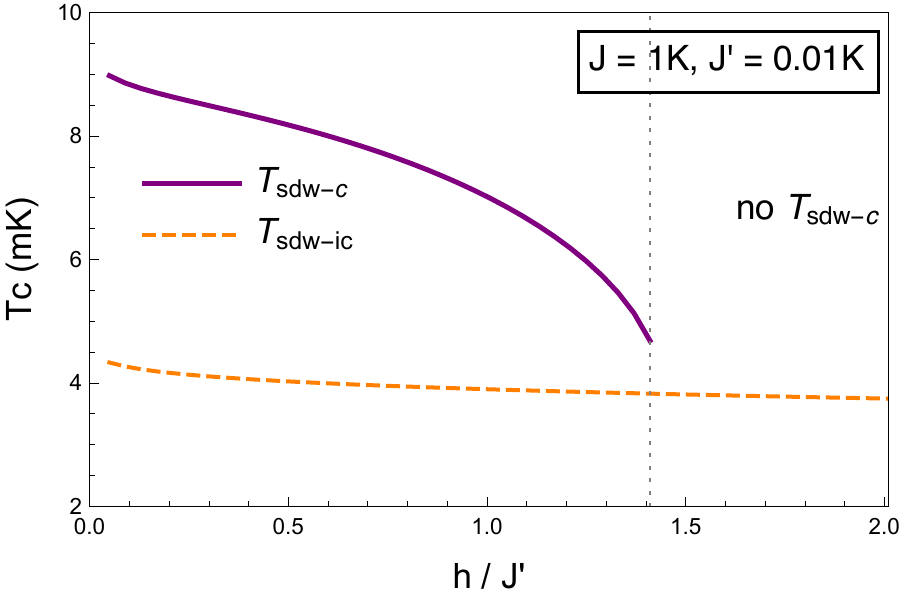}
	\caption{(Color online) Ordering temperatures of commensurate SDW ($T_{\rm sdw-c}$, purple solid line), and incommensurtate SDW ($T_{\rm sdw-ic}$, orange dashed line) 
		versus $h/J'$. Here $J=1$ K, and ${J}'=0.01$ K. Around $h/J'\sim 1.4$, longitudinal SDW order changes from the commensurate to the incommensurate one.}
	\label{fig:2sdw}
\end{figure}

As discussed in Section \ref{strong_DM_para}, the SDW order is commensurate for $h < h_{\rm c-ic}$ and becomes incommensurate in higher fields.
In the commensurate case we have 
\begin{equation}
{\cal H}_{\rm sdw}=2c_2 \sum_y \int \mathrm{d}x \sin(\frac{2\pi}{\beta}\tilde{\phi}_y+t_\phi x)\sin(\frac{2\pi}{\beta}\tilde{\phi}_{y+1}+t_\phi x),
\label{eq:sdwc}
\end{equation}
with $c_2=J'A_1^2/2$. Shifting $\tilde{\phi}$ by
\begin{equation}
\begin{gathered}
 \tilde{\phi}_y\to\breve{\phi}_y - \beta t_\phi x/2\pi - \sqrt{\pi/2} ~y
\end{gathered}
\label{eq:h2}
\end{equation}
and applying the CMF approximation, \eqref{eq:sdwc} transforms into
\begin{equation}
{\cal H}_{\rm sdw}=-4c_2{\Psi_2}\sum_y \int  \mathrm{d}x [ \cos\frac{2\pi}{\beta}\breve{\phi}_y],
\label{eq:h2-1}
\end{equation}
where ${\Psi_2}=\langle \cos\frac{2\pi}{\beta}\breve{\phi}_y\rangle$. In complete similarity with \eqref{solution_cone}, the shift produces wave vector $q_0 = t_\phi$ which strongly affects the critical temperature of
the commensurate SDW state
\begin{equation}
\begin{split}
1=&\eta_2(\frac{2\pi T_{\rm sdw-c}}{v})^{2 \Delta_2 -2}\frac{\Gamma(1-\Delta_2)}{\Gamma(\Delta_2)}|\Gamma(\frac{\Delta_2}{2}+iy)|^4 \\&\times\big[\cosh(2\pi y)-\cos(\pi \Delta_2)\big].
\end{split}
\label{eq:solution_csdw}
\end{equation}
Here $y={t_\phi v}/{(4\pi T_{\rm sdw-c})} = h/(4\pi T_{\rm sdw-c}), \quad\eta_2={c_2}/{(\pi v)} = J'A_1^2/(2\pi v).$
Similar to the case of the cone ordering, the solution of \eqref{eq:solution_csdw} exists as long as $h < h_{\rm c-ic}$. 
If one estimates the right-hand side of \eqref{eq:solution_csdw} by its $h=0$ value when $\Delta_2 = 1/2$, then one obtains
that $h_{\rm c-ic} = 1.9 J'$. This is because equations \eqref{solution_cone}
and \eqref{eq:solution_csdw} are identical in the limit of small magnetic field when $\Delta_1 = \Delta_2 = 1/2$. 
Solving \eqref{eq:solution_csdw} numerically, which accounts for the magnetic field dependence of the scaling dimension ($\Delta_2$ increases with the field, which
means that SDW order weakens),
results in a smaller critical field $h_{\rm c-ic} \approx 1.4 J'$ as Fig.~\ref{fig:2sdw} shows.

For $h > h_{\rm c-ic}$ we consider incommensurate SDW Hamiltonian of which differs from \eqref{eq:sdwc} by the absence of oscillatory term.
This, of course, is equivalent to neglecting $\tilde{g}_{\phi}$ in ${\cal H}_{\rm sdw}$ in \eqref{H'}. Therefore now 
\begin{equation}
\begin{gathered}
{\cal H}_{\rm sdw}=c_2 \sum_y \int  \mathrm{d}x  \cos\big[\frac{2\pi}{\beta}({\tilde{\phi}}_y-{\tilde{\phi}}_{y+1})\big],
\end{gathered}
\end{equation}
Here we shift $\tilde{\phi} \to \tilde{\phi}_y +\beta y/2$ which changes the sign of ${\cal H}_{\rm sdw}$. The CMF approximation then leads to
\begin{equation}
1=2c_2\chi(q=0,\omega=0;T_{\rm sdw-ic}),
\end{equation}
where the susceptibility in given by Eq.~\eqref{eq:chi}.
The ordering temperature of the incommensurate SDW order is
\begin{equation}
T_{\rm sdw-ic}=\frac{v}{2\pi}\left(\frac{\eta_2\dfrac{\Gamma(1-\Delta_2)\Gamma(\Delta_2/2)^2}{\Gamma(\Delta_2)\Gamma(1-\Delta_2/2)^2}}{1+\eta_2 \dfrac{\Gamma(\Delta_2-1/2)}
{\sqrt{\pi}(1-\Delta_2)\Gamma(\Delta_2)}}\right)^{1/(2-2\Delta_2)}.
\end{equation}
where $\eta_2= \pi c_2/v=\pi J'A_1^2/2v$. As explained below \eqref{eq:chi}, term in the denominator of the expression inside brackets in this equation removes 
divergence of the numerator in the $\Delta_2 \to 1$ limit (high-field limit).

Since the critical field $h_{\rm c-ic} \approx 1.4 J'$ is sufficiently small, we focus on the incommensurate SDW order when studying the phase transition between it 
and the cone-NN phase in Sec.~\ref{sec:CMF}.
Plots of SDW's $T_{\rm sdw}$ are shown as orange curves in Fig.~\ref{fig:Tc},~\ref{fig:three} and \ref{fig:Tc_2}.

\subsection{ConeNN order} 
When it comes to the coneNN state, the calculations are straightforward.
\begin{equation}
\begin{gathered}
{\cal H}_{\rm NN}=-c_3 \sum_y \int  \mathrm{d}x  \cos\big[\beta(\tilde{\theta}_{y}-\tilde{\theta}_{y+2})\big], \\
 c_3=\frac{\pi}{4}\frac{J'^2}{D}A_3^4t_{{\theta}}^{2\Delta-1}\frac{\Gamma(1-\Delta_1)}{\Gamma(\Delta_1)}.
\end{gathered}
\label{eq:hconenn}
\end{equation}
Note that coupling constant $c_3$ should be considered an estimate, valid up to numerical pre-factor of order $1$, 
since it is calculated via perturbative RG, see Appendix \ref{app:nn_chain}.

The ordering temperature has a simple form, due to the fact that ${\cal H}_{\rm NN}$ is free from oscillation and  $T_{\rm coneNN}$ is free from divergence ($\Delta_1\leq1/2$),
\begin{equation}
T_{\rm coneNN}=\frac{v}{2\pi}\Big[\eta_3\dfrac{\Gamma(1-\Delta_1)\Gamma(\Delta_1/2)^2}{\Gamma(\Delta_1)\Gamma(1-\Delta_1/2)^2}\Big]^{1/(2-2\Delta_1)},
\label{eq:tccone2n}
\end{equation}
where $\eta_3=\pi c_3/v$. 
The plot of $T_{\rm coneNN}$ is shown as the blue curve in Fig.~\ref{fig:Tc_2} for strong DM interaction.

\section{Mean-field treatment of the C-IC transition}
\label{app:cic}

Commensurate-incommensurate transition (CIT) appears several times in our work, both in connection with the DM-induced CIT in the cone state and with the magnetic
field induced CIT in the SDW state, see discussions in Sections \ref{sub:weak_DM_para} and \ref{strong_DM_para}, 
and calculations in Appendix \ref{app:cmf}. Here we sketch an approximate mean-field treatment of this transition at zero temperature.

As an example, let us consider ${\cal H}_{\rm cone}$ in Eq.~\eqref{eq:h0andcone} for a particular chain $y$, and suppose $y$ is even. Then,
removing all $\tilde{}$ and $\breve{}$ symbols which do not play any role in this discussion, we need to consider a single-chain Hamiltonian
\begin{eqnarray}
H_{\rm cit}&=&\int \mathrm{d}x \Big( \frac{1}{2v}(\partial_x\varphi_y)^2+\frac{v}{2}(\partial_x \theta_y)^2-\frac{D}{\sqrt{2\pi}}\partial_x\theta_y \nonumber\\ 
&&-\lambda\cos(\beta\theta_y) \Big),
\end{eqnarray}
where $\lambda=2c_1\Psi \sim J' \Psi$ depends on the self-consistently determined value of the order parameter $\Psi = \langle \cos(\beta\theta_y)\rangle$.

\begin{figure}[t]
	\includegraphics[width=0.8\columnwidth]{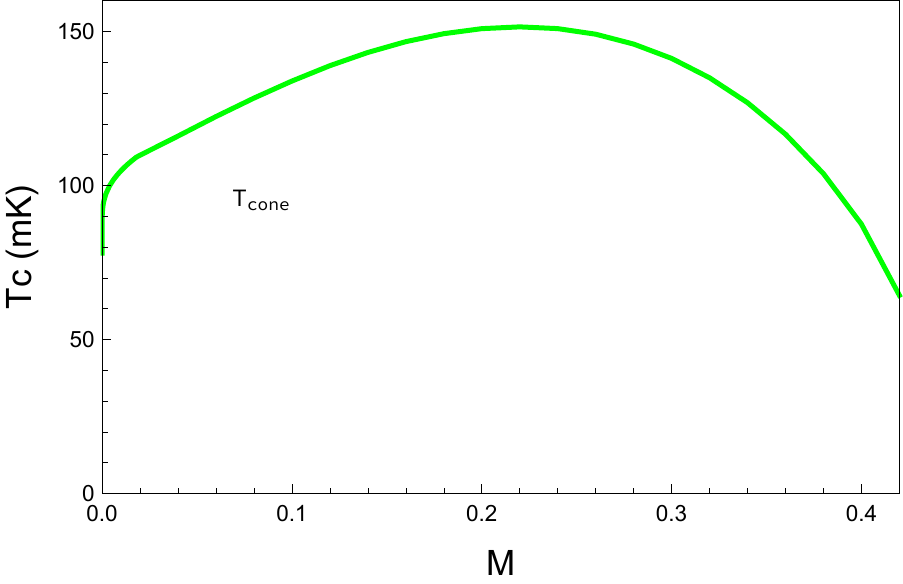}
	\caption{\label{fig:tcCl} (Color online) Critical temperatures of 
		cone ($T_{{\rm cone}}$, green solid) as a function of 
		magnetization $M$  for K$_2$CuSO$_4$Cl$_2$, with  $J_{\rm Cl}=3.1$ K, $D_{\rm Cl}=0.11$ K and ${ J}_{\rm Cl}'=0.083$ K from Eq.~\eqref{solution_cone} 
		by setting $T_{\rm cone}^{h=0}$ to $77$ mK.
		Here the phase diagram consists of a single cone phase.}
\end{figure}

According to Ref.~\onlinecite{Ru2013} (Appendix A.2), critical value $D_c$, above which ground state becomes incommensurate, scales as 
\begin{equation}
D_c \sim \sqrt{\lambda v} \Big(\frac{\lambda}{v}\Big)^{\Delta/(4 - 2 \Delta)},
\end{equation}
where $\Delta = \beta^2/(4\pi)$ is the scaling dimension of the cosine operator in $H_{\rm cit}$.
At the same time, according to Ref.~\onlinecite{oleg_cmf} (Appendix D.5) in the commensurate phase the order parameter scales as 
\begin{equation}
\Psi \sim \Big(\frac{J'}{v}\Big)^{\Delta/(2 - 2 \Delta)}.
\end{equation}
Combining the last two equation we derive that
\begin{equation}
\frac{D_c}{J'} \sim  \Big(\frac{v}{J'}\Big)^{(1-2\Delta)/(2 - 2 \Delta)}.
\label{app:Dscaling}
\end{equation}
We observe that $D_c$ is function of magnetization $M$, via dependence of $\Delta(M)$ on it. Since $\Delta(M)$ is decreasing function of magnetization, 
$\Delta(M=0) = 1/2$ while $\Delta(M=1/2) = 1/4$, critical $D_c$ is {\em smallest} at $M=0$: at this point $D_c/J' \sim 1$, in agreement with our comparison
of critical temperatures in the previous Appendix \ref{app:cmf}. As $\Delta \to 1/4$, which corresponds to the high-field limit, the critical ratio increases to
$(v/J')^{1/3} \gg 1$.

Put differently, our estimate of $D_c \approx 1.9 J'$, obtained in Appendix \ref{app:cone2}, provides the lower bound of the DM interaction magnitude $D$
required to destroy the commensurate cone state. If material is characterized by $D < D_c(M=0)$, the commensurate cone phase is stable in the whole range
of magnetization $0 \leq M \leq 1/2$.

\begin{figure}[!]
	\includegraphics[width=0.9\columnwidth]{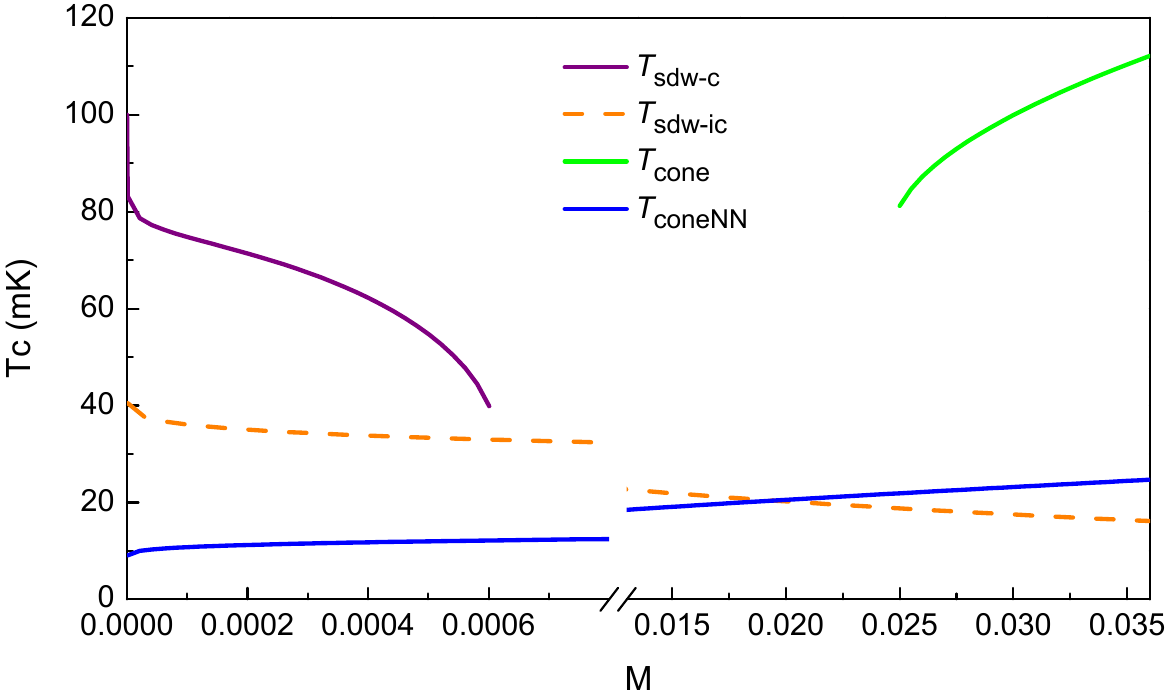}
	\caption{\label{fig:Br0} (Color online) Critical temperatures of 
		commensurate SDW ($T_{{\rm sdew-c}}$, purple solid line), incommensurate SDW ($T_{{\rm sdew-ic}}$, orange dashed line),  
		commensurate cone ($T_{{\rm cone}}$, green solid line) and coneNN ($T_{{\rm coneNN}}$, blue solid line)  as a function of 
		magnetization $M$, with  $J=20.5$ K, $J'=0.091$ K  and $D=0.28$ K. Transition between SDW(IC) and coneNN happens at $M\sim 0.018$. 
		Solution of $T_{{\rm cone}}$ appears discontinuously at $M\simeq 0.025$. Note that in order to accommodate 
		all phases in the single graph the horizontal axis is broken into two regions.}
\end{figure}

\section{Estimate of the inter-chain exchange $J'$}
\label{app:j'}
A variety of experimental techniques has been employed to characterize the parameters of K$_2$CuSO$_4$Cl$_2$ and K$_2$CuSO$_4$Br$_2$ \cite{Halg2014,smirnov}. 
The dominant intra-chain exchange $J$ has been estimated using the empirical fitting function of Ref.~\onlinecite{Johnston2000} to fit the uniform magnetic susceptibility data
as well as by fitting the inelastic neutron scattering continuum, a unique feature of the Heisenberg spin-1/2 chain,  to the M\"{u}ller ansatz \cite{Muller1981}.
DM vector $\bm D$ has been measured by electron spin resonance  (ESR) as described in Sec.~\ref{sec:esr}. However  the inter-chain exchange interaction $J'$ has been 
estimated from the chain mean-field theory fit based on Monte-Carlo improved
study in Ref.~\onlinecite{Yasuda2005}. This fit, however, completely neglects crucial for understanding of these materials DM interactions and, moreover,
assumes that spin chains form simple non-frustrated cubic structure. The second assumption is not justified as well. Inelastic neutron scattering data 
show that the interchain exchange between spin chains in the $a-b$ plane is at least an order of magnitude stronger than that along the $c$-axis, 
connecting different  $a-b$ planes. As a result, it is more appropriate to consider the current problem as two-dimensional whereby spin chains, running along the $a$-axis,
interact weakly via $J' \ll J$ directed along the $b$-axis. This is the geometry assumed in the present work. 

The inter-chain $J'$ is estimated from the value
of the zero-field critical temperature $T_{\rm c}$, which is calculated with the help of the chain mean field (CMF) approximation in Appendix~\ref{app:cmf}.
At $h=0$, and using $\Delta_1 = 1/2$ and $A_3 =1/2$, Eq.~\eqref{solution_cone} predicts 
$J' = (4\pi)^2 T_{\rm cone}^{h=0}/[|\Gamma(1/4 + i D/(4\pi T_{\rm cone}^{h=0}))|^4 \cosh(D/2T_{\rm cone}^{h=0})]$.
Here $T_{\rm cone}^{h=0}$ =77 mK is the experimentally determined transition temperature of K$_2$CuSO$_4$Cl$_2$ at zero magnetic field
and $D=0.11$ K. We obtain ${ J}'_{\rm Cl}=0.083$ K.

Fig.~\ref{fig:tcCl} shows $T_{\rm cone}$ and $T_{\rm sdw}$ for K$_2$CuSO$_4$Cl$_2$ as a function of magnetization $M$. 
It compares well to Fig. 14 in Ref.~\onlinecite{Halg2014}. 
As expected, the cone phase is the ground state of this two-dimensional system at all $M$.
The (approximately) factor of 2 difference between our result and the previous estimate in Ref.~\onlinecite{Halg2014} is caused by the assumed by us
two-dimensional geometry of the system and by the finite value of $D/J' = 1.3$ for this system, which slightly frustrates transverse inter-chain exchange.

For K$_2$CuSO$_4$Br$_2$, which is characterized by strong DM interaction, 
the value of the interchain exchange ${J}'$ can be estimated by identifying the zero-field ordering temperature $T_{\rm exp} = 0.1$ K  \cite{Halg2014} with
that of the commensurate longitudinal SDW order, Eq.\eqref{eq:solution_csdw}. For $h=0$ this gives $T_{\rm sdw-c} = A_1^2 \Gamma(1/4)^4 J'/(2\pi)^2 = 1.094 J'$, 
so that $J' \approx 0.091$ K.

\begin{figure}[!]
	\includegraphics[width=0.9\columnwidth]{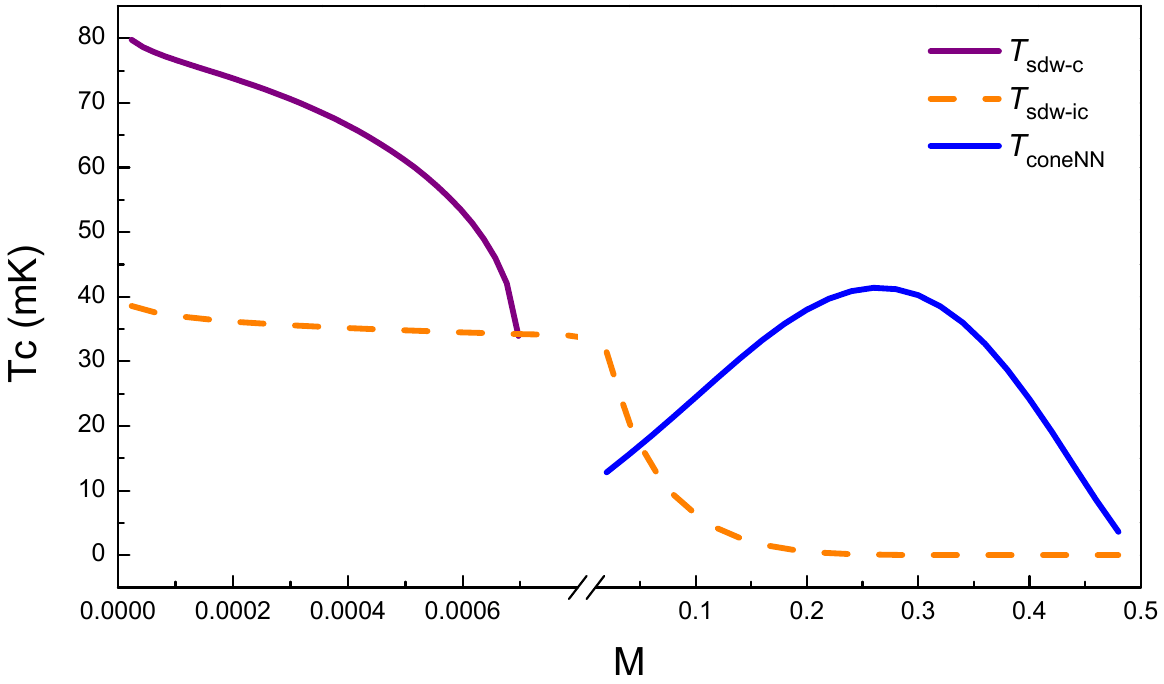}
	\caption{\label{fig:Br2} (Color online) Critical temperatures of 
		commensurate SDW ($T_{{\rm sdew-c}}$, purple solid line), incommensurate SDW ($T_{{\rm sdw}}$, orange dashed line) and 
		coneNN ($T_{{\rm coneNN}}$, blue solid line) and as a function of 
		magnetization $M$, with  $J=20.5$ K, $J'=0.091$ K  and $D=0.4$ K. Here $D$ is large enough to destroy the cone state in the full magnetization range. 
		Note that in order to accommodate 
		all three phases in the single graph the horizontal axis is broken into two regions.}
\end{figure}

\begin{table}[!t]
	{\renewcommand{\arraystretch}{1.5}%
		\begin{tabular}{c c c c c c}
			\hline\hline
			&$J (K)$ &$D (K)$ & $J'_{\rm exp} (K)$ &$\;{J'} (K)$ by CMF & $D/J'$\\
			\hline
			K$_2$CuSO$_4$Cl$_2$ & $\;\;3.1$  &$\;0.11$  & $0.031$  &$0.083$  & 1.3\\
			\hline
			K$_2$CuSO$_4$Br$_2$  &$20.5$ &$\;0.28$ &$0.034$ & $0.091$ & 3.1\\
			\hline\hline
		\end{tabular}}
		\caption{Exchange constants for K$_2$CuSO$_4$Cl$_2$ and K$_2$CuSO$_4$Br$_2$: intra-chain exchange $J$; magnitude of DM interaction $D$;
			inter-chain exchange $J'_{\rm exp}$ from Ref.~\onlinecite{Halg2014}: it is obtained by fitting experimental $T_c$ data \cite{Halg2014}  
			to the $d=3$ Heisenberg-exchange-only theory of 
			Ref.~\onlinecite{Yasuda2005}; inter-chain exchange ${J}'$ in the fifth column is obtained by fitting experimental $T_c$ data to our CMF calculations.}
		\label{app:table}
	\end{table}

Most important outcome of these calculations consists in finding significantly different estimates of the $D/J'$ ratio for the two materials, see Table~\ref{app:table}.
K$_2$CuSO$_4$Cl$_2$ is characterized by $D/J' =1.3$ which is below the critical value of $1.9$ which destroys the cone phase at $M=0$.
As a result, the phase diagram of K$_2$CuSO$_4$Cl$_2$ consists of a single cone phase.

To the contrary, K$_2$CuSO$_4$Br$_2$ has roughly two times greater value, $D/J' =3.1$, which results in a much more complex sequence of transitions
with increasing $M$, as Fig.~\ref{fig:Br0} shows.
The ground state at smallest $M \leq 0.0006$ is commensurate SDW which changes into an incommensurate SDW order for $0.0006 \leq M \leq 0.018$.
In the very narrow window $0.018 \leq M \leq 0.025$ the coneNN order takes over but then is replaced, again discontinuously, by the commensurate cone
order. Within the CMF description the coneNN-cone transition is discontinuous. 
The discontinuity in $T_c$ is significant, its value increases by a factor of about 2. This feature is not seen in the experiment and most likely indicates that
actual ratios of $D/J'$ and $J'/J$ for this interesting material are somewhat different from the values estimated by us here.

Importantly, that difference can be quite small. We find that the region of parameters with $D \approx 3 J'$ is very tricky, small changes in $D/J'$
change the outcome completely. For example, hypothetical material with slightly greater DM interaction, $D=0.4$ K so that $D/J' = 4.4$, 
turns out to be strongly DM-frustrated and does not support the cone phase at any magnetization, as Fig.~\ref{fig:Br2} shows.
Such a material would show two different transitions: first, at tiny magnetization of the order of $M=0.0007$, the commensurate SDW order 
changes to the incommensurate one. Then, at much higher magnetization of about $M=0.09$, there is a first order transition 
from the incommensurate SDW to the coneNN phase. This time there is no discontinuity in the $T_c(M)$ but the derivative $dT_c/dM$ is discontinuous still.

\begin{figure}[t!]
	\includegraphics[width=0.8\columnwidth]{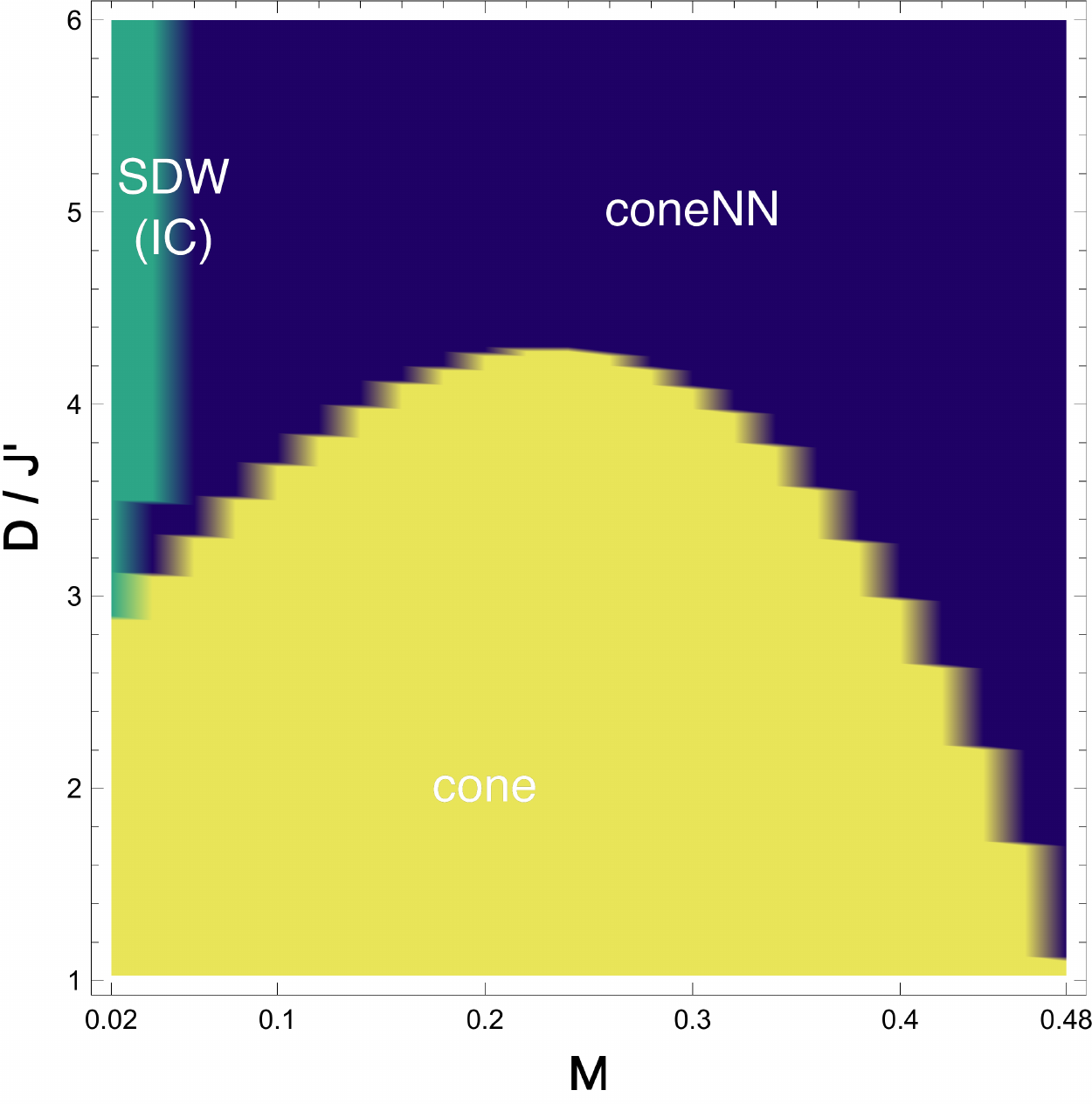}
	\caption{\label{fig:mdphase} (Color online) $M-D$ phase diagram for the case of ${\bm h} \parallel {\bm D}$, obtained by the CMF calculation.
		Here $J=20.5$ K, $J' = 0.0045J$. Cone phase is suppressed by large $D/J'$, and large field/magnetization.}
\end{figure}

The multitude of possible behaviors is summarized by phase diagrams in Fig.~\ref{fig:MDJp}, which focuses on the small $M$ range, and Fig.~\ref{fig:mdphase}, in
which the full range of $M$ is explored. In numerically calculating $T_c$'s for these diagrams we set $J=20.5$ K  and $J'\simeq 0.0045 J = 0.09$ K.
Being restricted to small values of $M$, Fig.~\ref{fig:MDJp} is calculated by keeping parameters $v$ and $A_{1,3}$ at their $M=0$ values but taking the variation
of the scaling dimensions with $M$ via Eq.\eqref{eq:2piR}. The commensurate-incommensurate transition between the two SDW phases happens at very small magnetization,
as has already been seen in Fig.~\ref{fig:Br0}. The ``triple point" where three phases intersect is at $M\simeq0.02$ and $D/J'\simeq 3$.

Figure~\ref{fig:mdphase} accounts for the $M$ dependence of all parameters that appear in the expressions for various $T_c$'s.
This is done with the help of numerical data from Ref.~\onlinecite{Essler2003} in which the smallest magnetization value is $0.02$.
This, as our discussion above shows, is too big a magnetization for the commensurate SDW state which therefore is absent from Fig.~\ref{fig:mdphase}.
As discussed previously, the cone order is first enhanced by $M$, due to the decrease of the corresponding scaling dimension, and then gets
suppressed at large magnetization, basically due to the Zeeman effect. It should be noted that our one-dimensional CMF calculations are not valid near the satuation,
$M\to 0.5$, where the velocity $v$ of chain spin excitations vanishes to zero. This shortcoming has already been discussed in Ref.~\onlinecite{oleg_cmf}.

Once again, Fig.~\ref{fig:mdphase} shows that SDW phase is restricted to low magnetization values. Staggered between chains DM interaction 
is effective in suppressing the commensurate cone phase for all $M$.  
For material with strong DM interaction such as  $D/J'>4.2$ (for example the $D=0.4$ K material in Fig.~\ref{fig:Br2}), 
the commensurate cone phase is entirely avoided as one increases $M$ from zero to saturation.
\begin{figure}[t]
	\includegraphics[width=0.8\columnwidth]{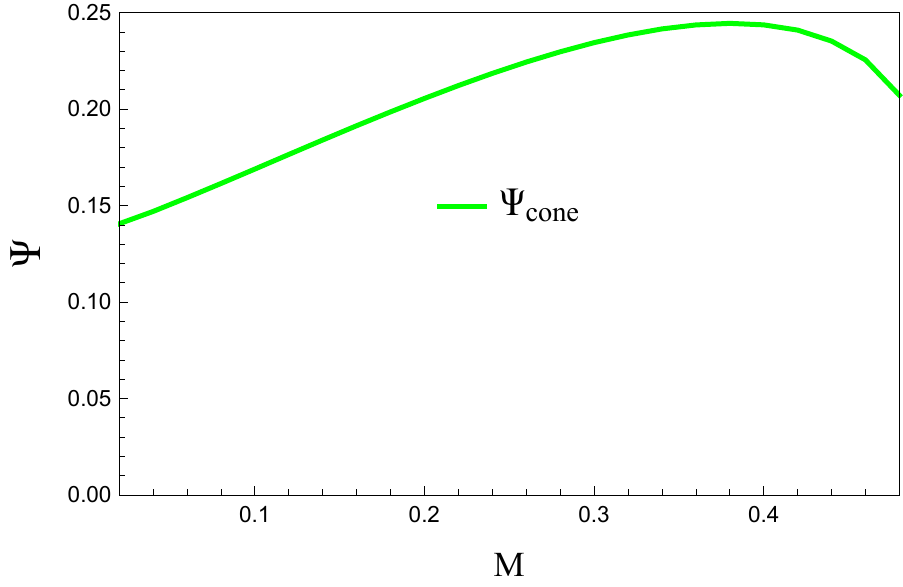}
	\caption{\label{fig:orderCl} Order parameter of cone ($\Psi_{\rm cone}$, green solid line) in K$_2$CuSO$_4$Cl$_2$, where ${J}'/J=0.027$ and $D/J'=1.3$. Note that $\Psi_{\rm cone}$ is enhanced by field.}
\end{figure}
\section{Order parameter at $T=0$ by CMF}
\label{app:order_parameter} 

Here we propose to study the magnetic orders in more details by calculating the associate order parameters, 
even though experimental attempts to measure them, via neutron scattering and muon-spin spectroscopy, remain inconclusive for now  ~\cite{Halg'sphdthesis}.
Our calculation of the order parameters is based on the CMF approximation in Sec.~\ref{app:cmf}, where the effective Hamiltonian reduces to a sine-Gordon model\cite{fradkin,Gogolin} 
as in Eq.~\eqref{sine-gordon}, its action reads
\begin{equation}
S_{\rm{sG}} = \int dx dy \Big(\frac{1}{2}(\partial_x \theta)^2 + \frac{1}{2}(\partial_y \theta)^2 - 2 \mu
\cos[\beta\theta] \Big) .
\label{eq:action-sg}
\end{equation}
Here, $\mu=c \langle \cos\beta\theta\rangle/v$, and $\tau=y/v$. 
According to Refs.~\onlinecite{Lukyanov1997,oleg_cmf}, expression for ${\Psi}\equiv  \langle \cos\beta\theta\rangle$ as a function of magnetization $M$ reads
\begin{equation}
{\Psi}(M)=\Big[(\frac{c}{v})^{\beta'^2}\sigma'(M)^{1-\beta'^2}\Big]^{1/(1-2\beta'^2)},
\label{eq:psi'}
\end{equation}
where $\beta' = \beta/\sqrt{8\pi}$, and
\begin{eqnarray}
\label{eq:sg4}
&&\sigma'(M) =\frac{\tan[\pi \xi/2]}{2\pi (1-\beta'^2)} \left[\frac{\Gamma(\frac{\xi}{2})}{\Gamma(\frac{1+\xi}{2})}\right]^2
\Big[\frac{\pi \Gamma(1-\beta'^2)}{\Gamma(\beta'^2)}\Big]^{1/(1-\beta'^2)},\nonumber\\
&&\quad\xi = \frac{\beta'^2}{1-\beta'^2}=\frac{\beta^2}{8\pi - \beta^2}.
\end{eqnarray}
Eq.~\eqref{eq:psi'} is a general form of order parameter for sine-Gordon model.
The three interactions in consideration are Eq.~\eqref{eq:h0andcone}, \eqref{eq:h2-1} and \eqref{eq:hconenn},
with $\beta=2\pi R$, and their corresponding parameters $\beta'$ are 
\begin{equation}
\beta_{1,3}'=\Delta_1/2,\quad
\beta_{2}'=\Delta_2/2,
\end{equation}
where $\beta'_{1,2,3}$ are associated with ${\Psi}_{1,2,3}$, and $\Psi_1=\langle \cos(\beta\breve{\theta}_{y})\rangle$ (defined below Eq.~\eqref{eq:h0andcone}), ${\Psi_2}=\langle \cos\frac{2\pi}{\beta}\breve{\phi}_y\rangle$ (defined below Eq.~\eqref{eq:h2-1}) and ${\Psi_3}=\langle \cos{\beta}\tilde{\theta}_y\rangle$.

\begin{figure}[t]
	\includegraphics[width=0.8\columnwidth]{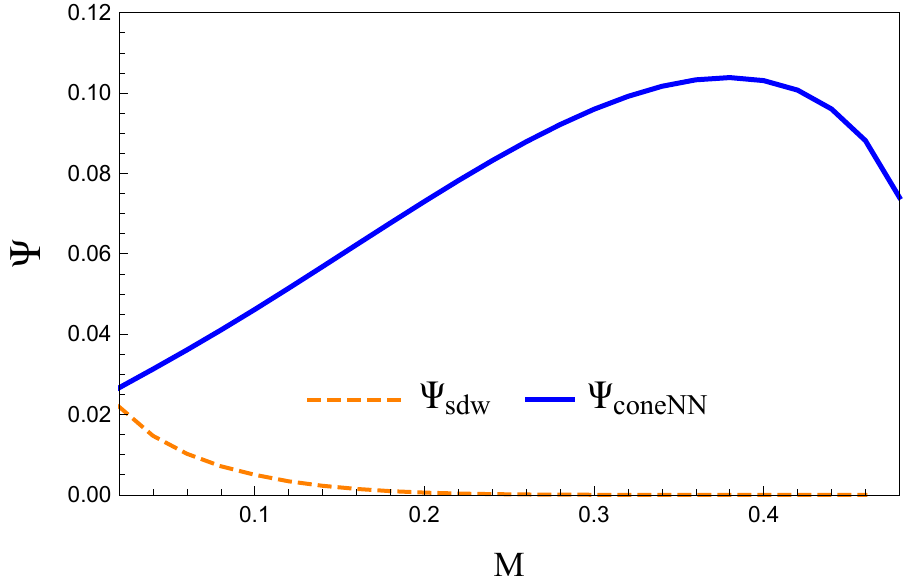}
	\caption{\label{fig:orderBr} Order parameters of SDW ($\Psi_{\rm sdw}$, orange dashed line) and coneNN ($\Psi_{\rm coneNN}$, blue solid line) in Br-compound, 
		where ${J}'/J=0.004$ and $D/J'=3.1$. Note that magnetic field enhances the coneNN order but suppresses the SDW one.}
\end{figure}

Now we can compute the order parameters for two materials K$_2$CuSO$_4$Cl$_2$ and K$_2$CuSO$_4$Br$_2$ exchange constants of which are estimated in Table~\ref{app:table}. 
For K$_2$CuSO$_4$Cl$_2$ the only phase to be considered is the cone. Its order parameter $\Psi_{\rm cone}$
\begin{equation}
\Psi_{\rm cone}=A_3\Big[(\frac{c_1}{v})^{\Delta_1}\sigma'(M)^{2-\Delta_1}\Big]^{1/(2-2\Delta_1)}
\end{equation}
is shown in Fig.~\ref{fig:orderCl}. For K$_2$CuSO$_4$Br$_2$ two order parameters need to be considered,
\begin{eqnarray}
&&\Psi_{\rm sdw}=A_1\Big[(\frac{c_2}{v})^{\Delta_2}\sigma'(M)^{2-\Delta_2}\Big]^{1/(2-2\Delta_2)},\nonumber\\
&&\Psi_{\rm coneNN}=A_3\Big[(\frac{c_3}{v})^{\Delta_1}\sigma'(M)^{2-\Delta_1}\Big]^{1/(2-2\Delta_1)}
\end{eqnarray}
and they are shown in Fig.~\ref{fig:orderBr}. Observe that the scaling of $\Psi$'s with $J'/v$ follows the RG prediction \eqref{eq:orderparam}.

Comparing Figures \ref{fig:orderCl} and \ref{fig:orderBr}, we notice the order parameters has smaller magnitude in Br-compound, 
due to its stronger DM interaction which frustrates the system more. Also, cone-type orders are enhanced by magnetic field,  while the SDW order is suppressed by it.

\bibliography{reference}

\end{document}